\documentclass[prb,aps,twocolumn,reprint,floatfix,showpacs]{revtex4-2}
\usepackage{booktabs}
\usepackage{array}
\newcommand{\PreserveBackslash}[1]{\let\temp=\\#1\let\\=\temp}
\newcolumntype{C}[1]{>{\PreserveBackslash\centering}p{#1}}
\newcolumntype{R}[1]{>{\PreserveBackslash\raggedleft}p{#1}}
\newcolumntype{L}[1]{>{\PreserveBackslash\raggedright}p{#1}}

\usepackage{verbatim}

\usepackage{dcolumn}
\usepackage{bm}

\usepackage{mathrsfs}
\usepackage{latexsym}
\usepackage{amssymb}
\usepackage{amsmath}
\usepackage{amsfonts}
\usepackage{wasysym}

\usepackage{mathtools}
\usepackage{braket}
\usepackage[T1]{fontenc}
\usepackage[utf8]{inputenc}

\usepackage[toc,page]{appendix}

\usepackage{titlesec}
\usepackage{graphicx}
\graphicspath{{Figures/}}
\usepackage[caption=false]{subfig}

\titleformat{\paragraph}
{\normalfont\normalsize\bfseries}{\theparagraph}{1em}{}
\titlespacing*{\paragraph}{0pt}{3.25ex plus 1ex minus .2ex}{1.5ex plus .2ex}

\renewcommand\arraystretch{1.25}

\allowdisplaybreaks

\usepackage{color}
\usepackage[colorlinks,bookmarks=false,citecolor=blue,linkcolor=blue,urlcolor=blue,pdfstartview=FitH]{hyperref}

\definecolor{darkred}{rgb}{0.7,0.0,0.0}

\definecolor{light-gray}{gray}{0.5}
\def\cgray{\color{light-gray}}

\def\cbl{\color{blue}}
\definecolor{darkblue}{rgb}{0,0.02,0.45}

\definecolor{darkgreen}{rgb}{0.02,0.45,0.0}

\definecolor{violet}{rgb}{0.8,0.2,0.6}

\newcommand{\be}{\begin{equation}}
\newcommand{\ee}{\end{equation}}
\newcommand{\bea}{\begin{eqnarray}}
\newcommand{\eea}{\end{eqnarray}}
\newcommand{\sbe}{\small\begin{equation}}
\newcommand{\see}{\end{equation}\normalsize}
\newcommand{\sbea}{\small\begin{eqnarray}}
\newcommand{\seea}{\end{eqnarray}\normalsize}
\newcommand{\mi}{\mathrm{i}}

\def\bs{\boldsymbol}

\def\mc{\mathcal}

\DeclareMathAlphabet\mathbfcal{OMS}{cmsy}{b}{n}

\usepackage{times}

\bibliography{refs} 

\begin{document}

\title{Multi-$Q$ magnetic phases from frustration and chiral interactions}

\author{Marios Georgiou}
\affiliation{Department of Physics, Loughborough University, Loughborough LE11 3TU, UK}

\author{Ioannis Rousochatzakis}
\affiliation{Department of Physics, Loughborough University, Loughborough LE11 3TU, UK}

\author{Joseph J. Betouras}
\affiliation{Department of Physics, Loughborough University, Loughborough LE11 3TU, UK}

\date{\today}

\begin{abstract}
We investigate the effect of Dzyaloshinskii-Moriya (DM) interactions in the planar pyrochlore (checkerboard) antiferromagnet, one of the paradigmatic models of spin frustration, and establish the classical phase diagram using a combination of analytical and numerical approaches. While anisotropic interactions generally tend to remove the frustration, here we show that a high degree of frustration survives in a large region of the phase diagram. In conjunction with the fixed handedness introduced by the DM anisotropy, this spawns a cascade of incommensurate and double-twisted multi-domain phases, consisting of spatially intertwined domains of the underlying competing phases of the isotropic frustrated point. The results underpin a novel mechanism of generating multi-$Q$ phases in systems that combine high degree of frustration and chiral interactions.
\end{abstract}

\pacs{}

\maketitle

\section{Introduction}
The search for novel phases of matter arising from the interplay of strong correlations and competing interactions has been a central theme in condensed matter physics research for many years.
Highly frustrated magnets have been a major platform in this search, owing to the vast parameter space (combining spin and orbital degrees of freedom, Coulomb repulsion, crystal-field effects and spin-orbit coupling), the tremendous advances in experimental technology and the strong impetus from the discovery of appropriate materials~\cite{HFMBook,DiepBook,RamirezBook}.  
By now, there is a wealth of prominent correlated systems -- 
from Heisenberg and Ising antiferromagnets on geometrically frustrated networks  
to the more recent anisotropic Kitaev magnets --
where competing interactions tend to preempt long-range magnetic ordering and give rise to classical or quantum spin liquids~\cite{Anderson1973,Balents2010,Savary2016,Trebst2017,Zhou2017,Winter2017,Hermanns2017,Hermanns2017,Knolle2019}.

In practice, such correlated orders require a certain degree of fine-tuning of the microscopic interactions, and various perturbations that are inevitably present in real materials can remove the frustration completely and stabilize one of the competing magnetic orders at low enough temperatures.

Here we report on a model system where a large degree of frustration can survive even in the presence of appreciable anisotropic perturbations. 
The system of interest lacks an inversion center, 
i.e., it is a system where symmetry-allowed perturbations can introduce a fixed handedness-chirality.
As we show below, the combination of these two ingredients -- robust frustration and fixed handedness -- gives rise to rich physics, already at the classical level: 
i) In a conventional magnetically ordered system, with a unique order parameter, the introduction of a chiral perturbation can induce a long-wavelength twisting of the primary order parameter field and may even lead to nolinear soliton-like modulations, such as 1D domain-wall lattices, skyrmions, and vortex lattices of various kinds~\cite{Dz64,IZYUMOV,Bogdanov1989,
Bogdanov1989,Bogdanov1994,
Roessler2006,Oleg2014, 
muehlbauer2009,yu2010,tonomura2012,
Seki2012, 
Shiba:1983js,Muhlbauer2011,Yukawa2012,Okubo2011,Okubo2012,Kamiya2014,Z2vortices
}.
Unlike conventional magnets, however, highly-frustrated magnets have multiple (competing) `order parameter' fields, which opens the door to `multi-{\bf Q}' states featuring more than one fields.
ii) The frustration spawns a cascade of closely spaced incommensurate and commensurate, multi-domain phases, made of spatially intertwined fields of the parent, frustrated phase.

\begin{figure}[!b]
\subfloat[][]{\includegraphics[width=0.58\linewidth]{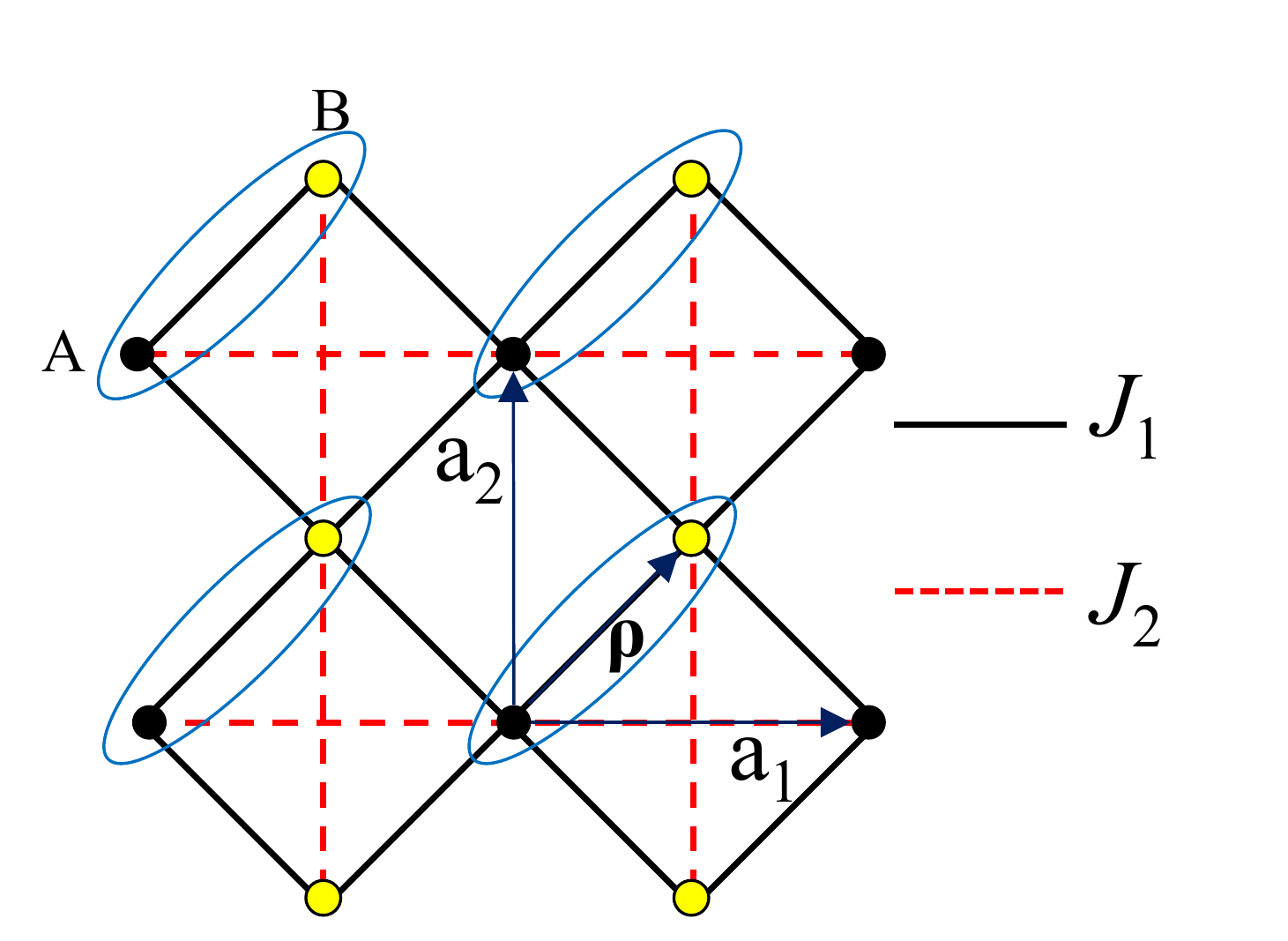}\label{fig:checkerboard-lattice-a}}
\subfloat[][]{\includegraphics[width=0.42\linewidth]{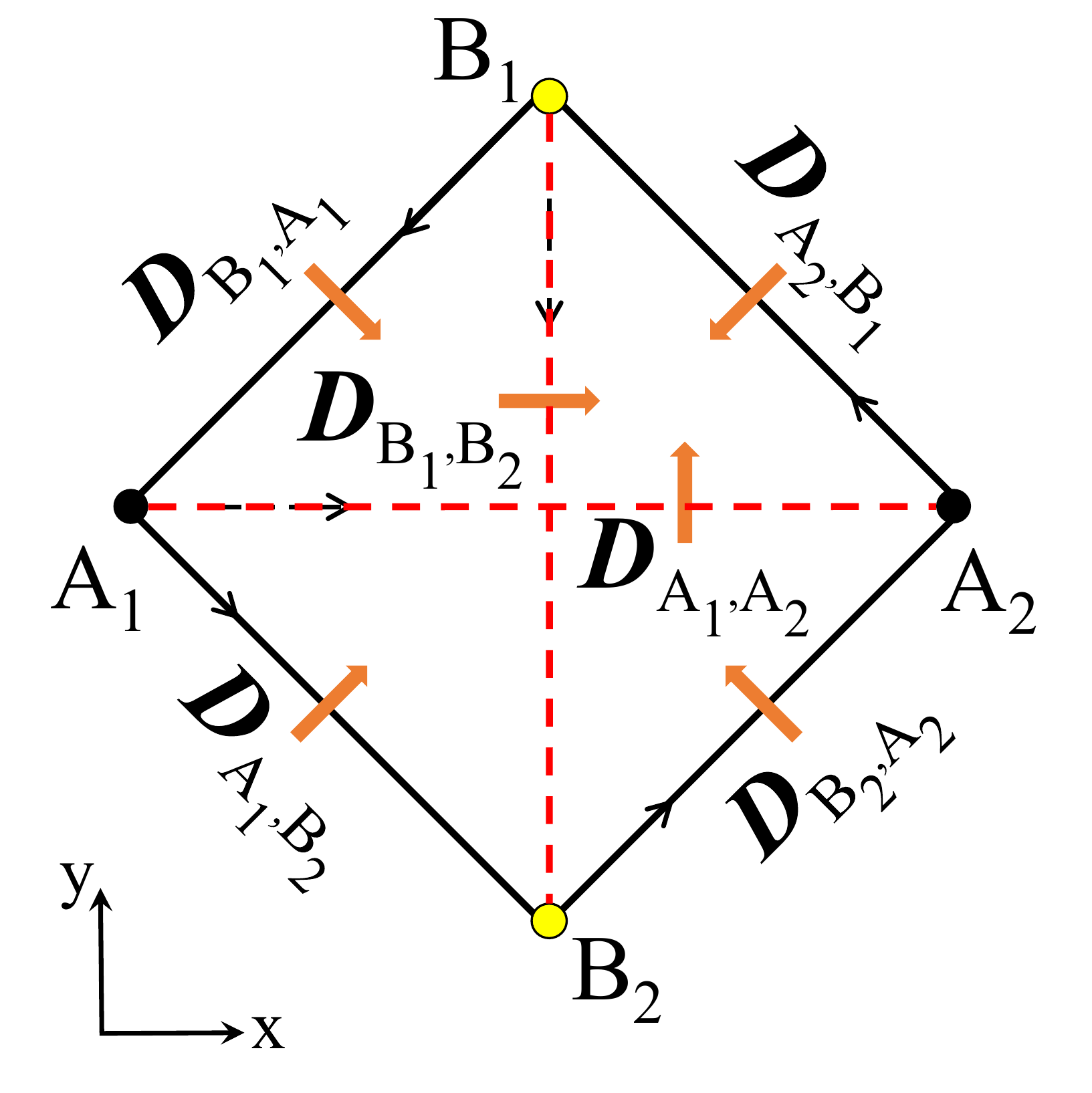}\label{fig:checkerboard-lattice-b}}
\caption{(Color online) (a) The checkerboard lattice is a square Bravais lattice (with primitive translations ${\bf a}_{1}=(a,0)$ and ${\bf a}_{2}=(0,a)$, where $a$ is the lattice constant) and a basis of two atoms A, B (the vector connecting these atoms is $\bs{\rho}= \frac{a}{2}({\bf x}+{\bf y})$). 
Solid (black) and dashed (red) lines show the topology of the couplings (Heisenberg and DM) considered in our model. 
(b) The topology of the (in-plane) DM vectors inside the crossed plaquettes.
}\label{fig:checkerboard_lattice}
\end{figure}

More specifically, we consider the impact of Dzyalozinskii-Moriya interactions~\cite{Dz1958,Moriya1960} on the planar pyrochlore (checkerboard) Heisenberg antiferromagnet (see Fig.~\ref{fig:checkerboard_lattice}), one of the prominent playgrounds of geometric frustration in two spatial dimensions~\cite{MoessnerChalker1998,Singh1998,Lieb1999,Elhajal2001,CanalsGaranin2002,Canals2002,BrenigHonecker2002,Fouet2003,Moessner2004,Bernier2004,Chan2011,Bishop2012}. This model has attracted much interest due its close relationship to the three-dimensional pyrochlore antiferromagnet, as both are networks of corner sharing tetrahedra (projected tetrahedra in the 2D case)~\cite{MoessnerChalker1998,Elhajal2001,Moessner2004}. 
In these lattices, the Heisenberg Hamiltonian $\mc{H}_{0}$ can be re-written (modulo an overall global constant) as~\cite{MoessnerChalker1998}
\be
\mc{H}_0=\frac{J}{2}\sum\nolimits_t {\bf S}_t^2\,,
\ee
where the sum runs over all tetrahedra $t$ and ${\bf S}_t$ is the total spin of the $t$-th tetrahedron. 

\begin{figure}[!t]
\includegraphics[width=0.8\linewidth]{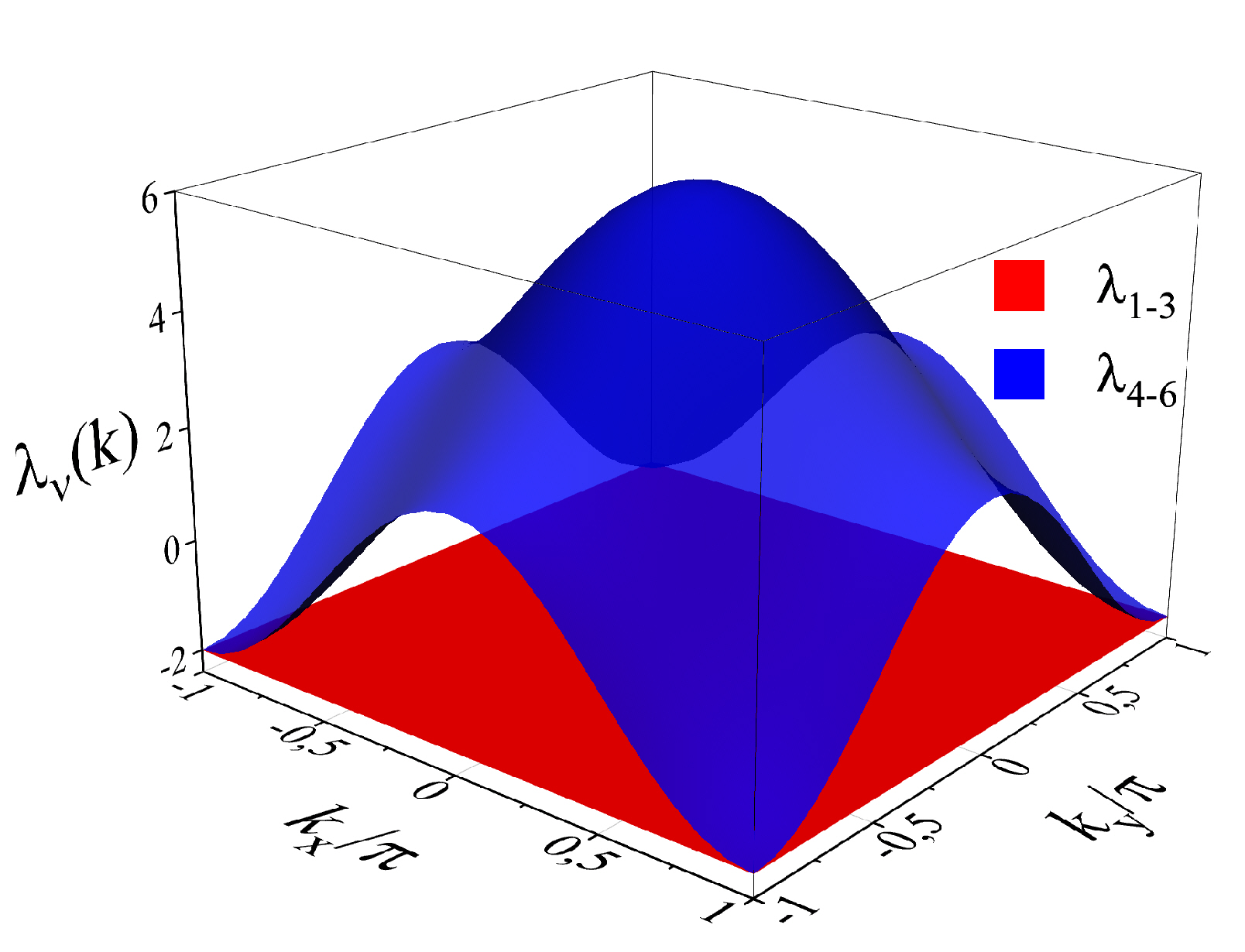}
\caption{(Color online) Dispersion of the six eigenvalues of the coupling matrix $\bs{\Lambda}({\bf k})$ [Eq.~(\ref{eq:LambdaFullLattice}) for $D_{1,2}\!=\!0$]  of the isotropic planar pyrochlore antiferromagnet, $\lambda_{1-3}\!=\!-2$ (red) and $\lambda_{4-6}\!=\!2(1\!+\!\cos k_x \!+\!\cos k_y)$ (blue).}
\label{fig:LT-isotropic-bands}
\end{figure}

At the classical level, and with $J\!>\!0$, these systems feature a vast manifold of ground states, encompassing all possible configurations that satisfy the local constraint ${\bf S}_t\!=\!0$ for all $t$.
This infinite degeneracy, which is a hallmark of classical highly-frustrated magnets, is reflected in the fact that three out of six bands of the Fourier-transformed interaction matrix $\bs{\Lambda}_0({\bf k})$ [defined in Eq.~(\ref{eq:LambdaFullLattice}) for $D_{1,2}\to0$] are completely flat in ${\bf k}$-space, see Fig.~\ref{fig:LT-isotropic-bands} and Refs.~\cite{CanalsGaranin2002,Canals2002}.
Moreover, out of the twelve symmetry-resolved fields that can be constructed locally on each tetrahedron (see Sec.~\ref{sec:SymmetryResolvedFields}), the classical ground state manifold of $\mc{H}_0$ can host up to nine  such fields (and any combinations thereof, which satisfy the spin length constraints).

The introduction of symmetry-allowed DM interactions $D_1$ and $D_2$ between first and second neighbours generally lifts the ground state degeneracy of $\mc{H}_0$ everywhere, except along the special line $D_2=\sqrt{2}D_1$, where the three flat bands of $\bs{\Lambda}_0({\bf k})$ remain flat up to first order in $D/J$ (see App.~\ref{App:LT-full-model}).
As the ratio $D/J$ is of the order of 10\% in typical materials~\cite{Moriya1960}, the broader vicinity of the line $D_2=\sqrt{2}D_1$ therefore marks an extended region in parameter space with an infinite number of nearly degenerate states. In conjunction with the lack of inversion symmetry, this spawns the cascade of incommensurate and commensurate phases shown in Fig.~\ref{fig:phasediag} (and analyzed in detail below), which are otherwise completely unexpected from the physics of a single tetrahedron (Fig.~\ref{fig:single-tetra-phase}).

This study therefore presents a novel mechanism of generating multi-$Q$ states by combining frustration and chiral interactions. 
Our results can be relevant to layered checkerboard antiferromagnets, such as the family of oxychalcogenide compounds A$_{2}$F$_{2}$Fe$_{2}$OQ$_{2}$ (A= Sr,Ba; Q=S,Se), La$_2$O$_2$Fe$_2$O(Se,S)$_2$~\cite{Si10, Freelon19}, (LaO)$_2$Mn$_2$Se$_2$O and (BaF)$_2$Mn$_2$Se$_2$O~\cite{Chen1}, Na$_2$Fe$_2$Se$_2$O~\cite{Chen2}, BaFe$_2$Se$_2$O~\cite{Kageyama16}, Sr$_2$F$_2$Fe$_2$OS$_2$~\cite{Morosan13}, (Ce,Nd)$_2$O$_2$Fe$_2$OSe$_2$~\cite{McCabe14}, and Pr$_2$O$_2$(Mn,Fe)$_2$OSe$_2$~\cite{McCabe17}. 
This study can also drive new insights to the 3D pyrochlore extension of the model, for which some work on the effect of DM anisotropy has been known for some time~\cite{Elhajal05}. This platform offers a broader range of available materials (such as Cd$_2$Os$_2$O$_7$~\cite{Bogdanov13, Sohn17,Zhang21}, FeF$_3$~\cite{Amirabbasi19},  Sm$_2$Ir$_2$O$_7$~\cite{McMorrow16} and Y$_2$Ir$_2$O$_7$~\cite{Son19,Hwang20}), and, moreover, can naturally satisfy the requirement of frustration (equal Heisenberg exchange on all bonds of a tetrahedron) by symmetry, unlike the 2D case, which requires fine tuning. 

The remaining part of the paper is organised as follows. In Sec.~\ref{sec:model} we introduce the model and give a detailed symmetry analysis. In Sec.~\ref{sec:phase-diag} we summarize the main results and present the phase diagram of the model. Our critical analysis begins with a close examination of the physics of a single tetrahedron in Sec.~\ref{sec:single-tetra}. The discussion of the full lattice model is split into two main parts, the first (Sec.~\ref{sec:full-model-phasesI}) on the incommensurate 1D modulated phases, and the second (Sec.~\ref{sec:full-model-phasesII}) on the multi-domain phases. 
Our conclusions and a discussion is given in Sec.~\ref{sec:summary}, while a body of technical details and auxiliary information are relegated to the Appendices (\ref{sec:single-tetra-app}-\ref{sec:DM-along-z}).

\section{Model \& preliminary insights from symmetry}\label{sec:model}

\subsection{Model and symmetries}
We consider classical Heisenberg spins ${\bf S}_i$ of length $S=1$, residing at the vertices of a two-dimensional checkerboard lattice, see Fig.~\ref{fig:checkerboard-lattice-a}. This can be thought of as a square Bravais lattice with a basis of two atoms, labeled $A$ and $B$, or, equivalently, as a network of corner-sharing crossed plaquettes (projected tetrahedra), see Fig.~\ref{fig:checkerboard-lattice-b}.
The vertices $i$ of the lattice can be labeled as 
\be
i \mapsto ({\bf R},\nu),~~
{\bf R}=n_1{\bf a}_1+n_2 {\bf a}_2,~~
\nu=\text{A or B}
\ee
where ${\bf a}_{1,2}$ are two primitive translations of the underlying Bravais lattice (see Fig.~\ref{fig:checkerboard-lattice-a}) and $n_{1,2}$ are integers. The physical positions ${\bf r}$ of the spins are given by
\be
{\bf r}_{{\bf R},\text{A}}={\bf R},~~~
{\bf r}_{{\bf R},\text{B}}={\bf R}+\rho\,,
\ee
where $\rho=\frac{a}{2}({\bf x}+{\bf y})$ (see Fig.~\ref{fig:checkerboard-lattice-a}).

The spins interact with each other with Heisenberg and Dzyaloshinskii-Moriya (DM) exchange couplings, and the model is described by the Hamiltonian
\be\label{eq:Ham}
\mc{H} = \sum_{i<j} J_{ij} {\bf S}_i\cdot{\bf S}_j + {\bf D}_{ij} \cdot {\bf S}_i\times{\bf S}_j\,,
\ee
where $J_{ij}$ denotes the Heisenberg exchange between sites $i$ and $j$, and ${\bf D}_{ij}$ the corresponding DM vectors. We include nearest-neighbour (NN) couplings $J_1$ and ${\bf D}_1$ [solid (black) lines in Fig.~\ref{fig:checkerboard_lattice}] and second neighbour couplings $J_2$ and ${\bf D}_2$ [dashed (red) lines in Fig.~\ref{fig:checkerboard_lattice}].
As we are interested in the impact of the DM couplings on a highly frustrated magnet, we shall limit ourselves to the special case 
\be
J_1=J_2=J\,,
\ee
and, unless otherwise specified, we shall measure energies in units of $J\!=\!1$ in the following.

Turning to the DM interactions, we shall consider systems with $C_{4\mathrm{v}}$ symmetry around the center of the crossed plaquettes.
This symmetry group is generated by the fourfold axis $C_{4z}$ and the reflection operation $\sigma_{xz}$, which map the four spins of the plaquette as follows (see site labelling of Fig.~\ref{fig:checkerboard-lattice-b}):
\be
\renewcommand{\arraystretch}{1.25}
\!\!\!\!\!\!\!\begin{array}{| c | c |}
\hline
C_{4z} & \sigma_{xz}\\
\hline
{\bf A}_1\!\mapsto\!\big(B_2^y,-B_2^x,B_2^z\big) & {\bf A}_1 \!\mapsto\! ( -A_1^x, A_1^y, -A_1^z)\\
{\bf A}_2\!\mapsto\!\big(B_1^y,-B_1^x,B_1^z\big) & {\bf A}_2 \!\mapsto\! (-A_2^x, A_2^y, -A_2^z)\\
{\bf B}_1\!\mapsto\!\big(A_1^y,-A_1^x,A_1^z\big) & {\bf B}_1 \!\mapsto\! ( -B_2^x, B_2^y, -B_2^z)\\
{\bf B}_2\!\mapsto\!\big(A_2^y,-A_2^x,A_2^z\big) & {\bf B}_2 \!\mapsto\! ( -B_1^x, B_1^y, -B_1^z)\\
\hline
\end{array}\,.
\ee
The in-plane components of the DM vectors that are allowed by this symmetry group are as follows (using the site labelling and reference frame of Fig.~\ref{fig:checkerboard-lattice-b}) 
\bea\label{eq:DMvecs}
\renewcommand{\arraystretch}{1.4}
\begin{array}{c}
{\bf D}_{\text{A}_1, \text{B}_2}= - {\bf D}_{\text{A}_2, \text{B}_1} = D_1~\frac{{\bf x}+{\bf y}}{\sqrt{2}}\,,\\
{\bf D}_{\text{B}_1, \text{A}_1} = -{\bf D}_{\text{B}_2, \text{A}_2}= D_1 ~\frac{{\bf x}-{\bf y}}{\sqrt{2}}\,,\\
{\bf D}_{\text{A}_1, \text{A}_2}= D_2 ~{\bf y}\,,~~~ {\bf D}_{\text{B}_1, \text{B}_2}= D_2 ~{\bf x} \,,
\end{array}
\eea
where ${\bf x}$ and ${\bf y}$ are unit vectors along the $x$- and $y$-axis, respectively, and the constants $D_1$ and $D_2$ are two independent coupling constants of our model.
The $C_{4\mathrm{v}}$ symmetry allows, in addition, a uniform, out-of-plane component $D_{1}'{\bf z}$ for the NN DM vectors. In the following we shall disregard this component (see however discussion in  App.~\ref{sec:DM-along-z}) and focus on the most interesting effects arising from the interplay of $D_1$ and $D_2$,  which, unlike $D_1'$, are the only couplings that break explicitly the real space inversion around the center of the crossed plaquettes. As such, they can give rise to so-called Lifshitz invariants (linear spatial derivatives of order parameter fields)~\cite{Dz64,IZYUMOV,Bogdanov1989}, which are known to stabilize multi-${\bf Q}$ phases with soliton-like modulations of primary order parameter fields~\cite{
Abrikosov1957,
Wright89,
Bogdanov1989,Bogdanov1994,Bogdanov1995,Roessler2006,Oleg2014, 
muehlbauer2009,yu2010,tonomura2012,
Seki2012, 
Shiba:1983js,Muhlbauer2011,Yukawa2012,Okubo2012,Kamiya2014,Z2vortices
}.

\subsection{Duality of the model}
Apart from the global symmetries (translations, $C_{4z}$, $\sigma_{xz}$, and time reversal $\mc{T}$), the model that we study has, in addition, a useful {\it duality}. Indeed, as the DM vectors considered here all lie in the $xy$-plane, the mirror operation $\mc{M}_{xy}=\mc{T}\sigma_{xy}$, which maps
\be
\label{eq: duality-trans}
\mc{M}_{xy}:~~~{\bf S}_i ~\mapsto~ (S_i^x, S_i^y, -S_i^z)\,,
\ee
effectively flips the signs of $D_1$ and $D_2$, i.e., it maps the point $(D_1,D_2)$ of the phase diagram to the point $(-D_1,-D_2)$. This duality allows to restrict our study to half of the phase diagram, e.g., the region with $D_2\ge 0$.

\begin{figure*}[!t]
\!\!\!\subfloat{\includegraphics[width=0.49\textwidth]{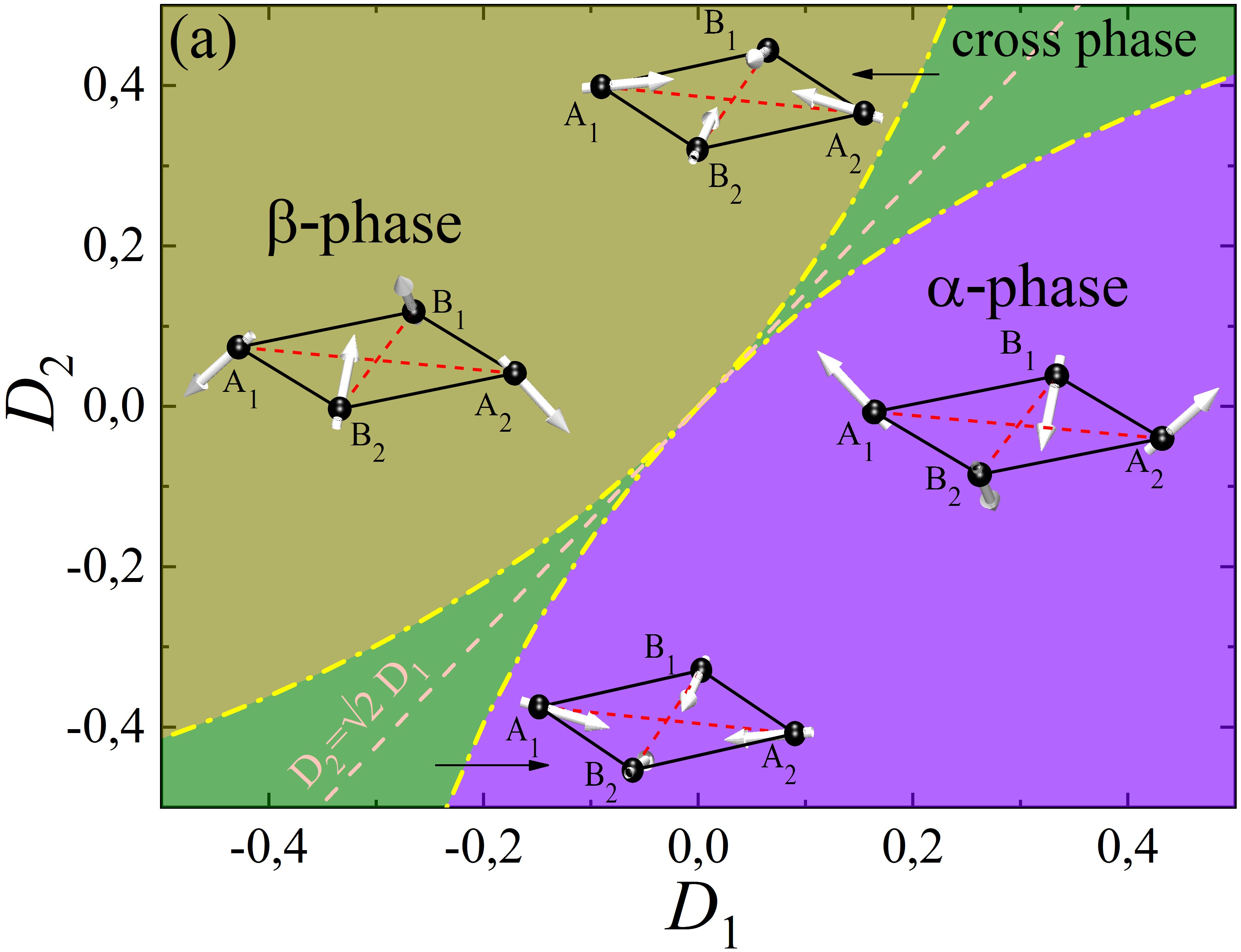}\label{fig:single-tetra-phase}}
~~~~
\subfloat{\includegraphics[width=0.51\textwidth]{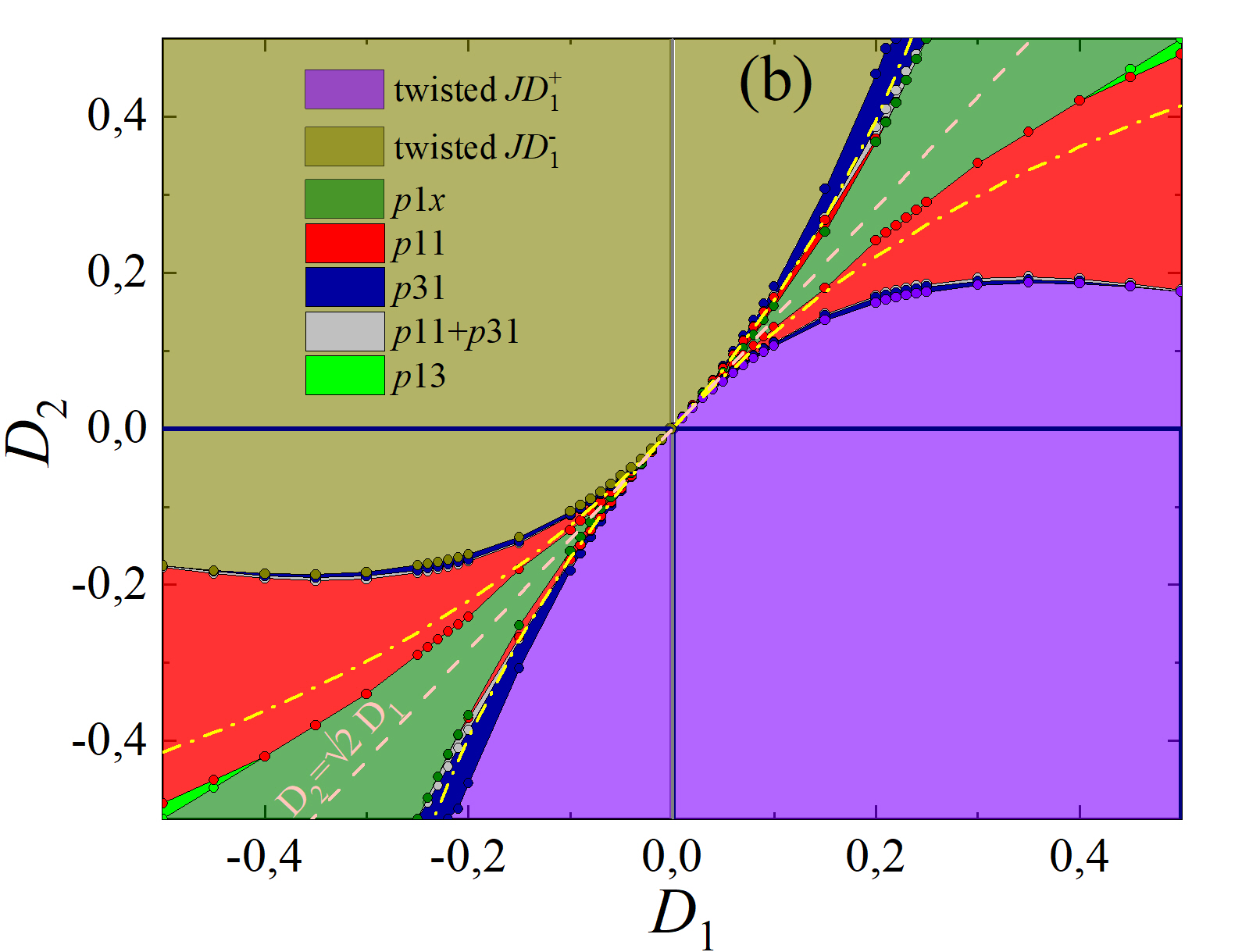}\label{fig:phasediag}}
\caption{(Color online) 
(a) The phase diagram of the Hamiltonian (\ref{eq:Ham}) for a single tetrahedron (in units of $J=1$), as obtained from the analytical Luttinger-Tisza method (see App.~\ref{app:LTsingleTetrahedron}), contains three phases, the $\alpha$-, $\beta$- and the cross-phase. 
The corresponding configurations are given in Eqs.~(\ref{eq:single-tetra-spins-a}), (\ref{eq:single-tetra-spins-b}) and (\ref{eq:single-tetra-spins-cp}), (\ref{eq:single-tetra-spins-cn}), respectively.
The boundary line between the cross- and the $\alpha$-phase is $D_2\!=\!\frac{\sqrt{2}D_{1}}{J+\sqrt{2}D_{1}}$, while that between the cross- and the $\beta$-phase is $D_2\!=\!\frac{\sqrt{2}D_{1}}{J-\sqrt{2}D_{1}}$.
(b) The classical phase diagram of the full lattice model of Eq.~(\ref{eq:Ham}), as obtained by a numerical iterative energy minimization method. 
The phases at opposite points in the diagram(i.e., $(D_1,D_2)$ and $-(D_1,D_2)$) are dual to each other, where the duality refers to the mirror operation $\mc{M}_{xy}$, see main text.
For comparison, we have superimposed the boundary lines (in yellow) of the single tetrahedron phase diagram of panel (a).}\label{fig:PhaseDiagrams}
\end{figure*}

\subsection{Symmetry-resolved fields}\label{sec:SymmetryResolvedFields}
It is useful at this point to re-write the energy in terms of symmetry-resolved order parameter fields of isolated tetrahedra. 
Following the analysis of App.~\ref{sec:symmetry-analysis}, the $12\times12$ representation $\bs{\Gamma}$ of the point group $C_{4\mathrm{v}}$ formed by the (normalized) spin configurations $\bs{\mc{S}}$ of the four spins of the tetrahedron (see site labelling in Fig.~\ref{fig:checkerboard-lattice-b}), 
\be
\!\bs{\mc{S}}\!=\!\frac{1}{2}(A_1^x,\!A_1^y,\!A_1^z,\!A_2^x,\!A_2^y,\!A_2^z,\!B_1^x,\!B_1^y,\!B_1^z,\!B_2^x,\!B_2^y,\!B_2^z),~~
\ee
decomposes into irreducible representations (irreps) of $C_{4\mathrm{v}}$ as
\be
\bs{\Gamma}= \mc{A}_1 \oplus 2\mc{A}_2 \oplus 2 \mc{B}_1 \oplus \mc{B}_2 \oplus 3 \mc{E}\,,
\ee
where $\mc{A}_{1}$, $\mc{A}_{2}$, $\mc{B}_{1}$ and $\mc{B}_{2}$ are one-dimensional and $\mc{E}$ is two-dimensional.  
A list of the associated symmetrized basis vectors ${\bf v}_i$ ($i\!=\!1$-$12$) are given in Eq.~(\ref{eq:irrep-basis-vecs}), and $\bs{\mc{S}}$ decomposes as
\be\label{eq:spinor-in-basis}
\bs{\mc{S}}= \sum\nolimits_i
m_{i}~\mathbf{v}_{i}\,,
\ee
where the `fields' $m_i$ (normalized so that their maximum value equals 1) are given by 
\small
\be
\label{eq:order-params}
\renewcommand\arraystretch{1.5}
\!\!\!\!\begin{array}{l}
m_{\mc{A}_{1}}\!=\!\frac{1}{4}(A_1^y-A_2^y+B_1^x-B_2^x),
\\
m_{\mc{A}_{2a}}\!=\!\frac{1}{4}(A_1^x - A_2^x - B_1^y + B_2^y),
\\
m_{\mc{A}_{2b}}\!=\!\frac{1}{4}(A_1^z\!+\!A_2^z\!+\!B_1^z\!+\!B_2^z)=\frac{1}{4}S_t^z,
\\
m_{\mc{B}_{1a}}\!=\!\frac{1}{4\sqrt{2}}(A_1^x\!-\!A_1^z \!-\! A_2^x\!-\!A_2^z \!+\! B_1^y\!+\! B_1^z \!-\! B_2^y\!+\!B_2^z),
\\
m_{\mc{B}_{1b}}\!=\!\frac{1}{4\sqrt{2}}(A_1^x\!+\!A_1^z \!-\! A_2^x\!+\!A_2^z \!+\! B_1^y\!-\! B_1^z \!-\! B_2^y\!-\!B_2^z),
\\
m_{\mc{B}_{2}}\!=\!\frac{1}{4}(A_1^y\!-\!A_2^y\!-\!B_1^x\!+\!B_2^x),
\\
\Big[\!\begin{array}{l}
m_{\mc{E}_a,{1}}\\
m_{\mc{E}_a,{2}}
\end{array}\!\Big]
\!\!=\!\!\frac{1}{4\sqrt{2}}\Big[\!\begin{array}{l}
A_1^x\!+\!A_1^y\!+\!A_2^x\!+\!A_2^y\!+\!B_1^x\!+\!B_1^y\!+\!B_2^x\!+\!B_2^y\\
A_1^x\!-\!A_1^y\!+\!A_2^x\!-\!A_2^y\!+\!B_1^x\!-\!B_1^y\!+\!B_2^x\!-\! B_2^y
\end{array}\!\Big],
\\
\Big[\!\begin{array}{l}
m_{\mc{E}_{b'},{1}}\\
m_{\mc{E}_{b'},{2}}
\end{array}\!\Big]
\!\!=\!\!\frac{1}{4\sqrt{2}}\Big[\!\begin{array}{l}
A_1^x\!-\!A_1^y\!+\!A_2^x\!-\!A_2^y\!-\!B_1^x\!+\!B_1^y\!-\!B_2^x\!+\!B_2^y\\
A_1^x\!+\!A_1^y\!+\!A_2^x\!+\!A_2^y\!-\! B_1^x\!-\!B_1^y\!-\!B_2^x\!-\!B_2^y
\end{array}\!\Big],
\\
\Big[\!\begin{array}{l}
m_{\mc{E}_{c'},{1}}\\
m_{\mc{E}_{c'},{2}}
\end{array}\!\Big]
\!=\!\frac{1}{4}\Big[\!\begin{array}{l}
A_1^z\!-\!A_2^z\!-\!B_1^z\!+\!B_2^z\\
A_1^z\!-\!A_2^z\!+\!B_1^z\!-\!B_2^z
\end{array}\!\Big],
\end{array}~~
\ee
\normalsize
in the above $\{m_{\mc{A}_{2a}}, m_{\mc{A}_{2b}}\}$ are two different fields that transform like $\mc{A}_2$, $\{m_{\mc{B}_{1a}},m_{\mc{B}_{1b}}\}$ are the two fields that transform like $\mc{B}_1$, and the last three lines denote the three
two-component fields $\{{\bf m}_{\mc{E}_a},{\bf m}_{\mc{E}_{b'}},{\bf m}_{\mc{E}_{c'}}\}$ that transform like $\mc{E}$. 
To simplify the form of the Hamiltonian we shall use the fields
\be\label{eq:bpcp}
{\bf m}_{\mc{E}_{b}} \!\equiv\!\frac{1}{\sqrt{2}} ({\bf m}_{\mc{E}_{b'}}+{\bf m}_{\mc{E}_{c'}}),~~
{\bf m}_{\mc{E}_{c}}\!\equiv\!\frac{1}{\sqrt{2}} ({\bf m}_{\mc{E}_{b'}}-{\bf m}_{\mc{E}_{c'}})\,,
\ee 
instead of ${\bf m}_{\mc{E}_{b'}}$ and ${\bf m}_{\mc{E}_{c'}}$. 
We note in passing that the spin length constraints translate into constraints for the fields, and one of these constraints is the relation $\sum\nolimits_i m_i^2=1$.  

The single-tetrahedron Hamiltonian $\mc{H}_t$ can be written in terms of the fields $m_i$ as
\begin{align}
\label{eq:crosscouplings}
&\mc{H}_{t}/2=
3 J (m_{\mc{A}_{2b}}^2 
+{\bf m}_{\mc{E}_a}^2)
\nonumber\\
&~~
-J(m_{\mc{A}_1}^2 
\!+\!m_{\mc{A}_{2a}}^2  
\!+\!m_{\mc{B}_{1a}}^2
\!+\!m_{\mc{B}_{1b}}^2 
\!+\!m_{\mc{B}_2}^2 
\!+\!{\bf m}_{\mc{E}_b}^2 
\!+\!{\bf m}_{\mc{E}_c}^2 )
\nonumber\\
&~~
-d_- [m_{\mc{B}_{1a}}^2
-m_{\mc{B}_{1b}}^2 
+\frac{1}{\sqrt{2}}({\bf m}_{\mc{E}_{b}}^2
-{\bf m}_{\mc{E}_{c}}^2)]
\nonumber\\
&~~
- d_+ [
2~ m_{\mc{A}_{2a}} m_{\mc{A}_{2b}}
-{\bf m}_{\mc{E}_a}\cdot ({\bf m}_{\mc{E}_b}-{\bf m}_{\mc{E}_c})]
\,,
\end{align}
where 
\be
d_\pm \equiv \sqrt{2}D_1\pm D_2\,.
\ee
Eq.~(\ref{eq:crosscouplings}) provides the following insights:

i) At the isotropic point ($d_\pm\!=\!0$), the energy can be minimized by any of the nine (out of twelve) fields that belong to
\be
\mc{A}_1,~
\mc{A}_{2a},~ 
\mc{B}_{1a},~ 
\mc{B}_{1b},~ 
\mc{B}_2,~ 
\bs{\mc{E}}_b,~
\bs{\mc{E}}_c\,,
\ee
representations or any combinations thereof, that satisfy the spin-length constraints, 
as these are the ones that appear with a negative pre-factor in Eq.~(\ref{eq:crosscouplings}) for $d_\pm\!=\!0$. 
As expected, the corresponding basis vectors ${\bf v}_i$ satisfy the condition ${\bf S}_t\!=\!0$, see Eq.~(\ref{eq:irrep-basis-vecs}). 

ii) Among the terms generated by the DM couplings, the terms proportional to $d_-$ are special in that they involve only fields ( $m_{\mc{B}_{1a}}$, $m_{\mc{B}_{1b}}$, ${\bf m}_{\mc{E}_b}$ and ${\bf m}_{\mc{E}_c}$) which are present in the ground state manifold of the isotropic point. By contrast, the two cross-coupling terms proportional to $d_+$ each involve a field ($m_{\mc{A}_{2b}}$ or ${\bf m}_{\mc{E}_a}$) that costs a nonzero Heisenberg energy. As a result, switching on the DM couplings, only the terms proportional to $d_-$ can lift the classical ground state degeneracy of the isotropic point already at first order in the DM couplings. 
If, however, we move along the line $D_2=\sqrt{2}D_1$ the coefficient $d_-$ vanishes and the lifting of the degeneracy will be quadratic in $d_+/J$. This insight arises also from an analysis of the flat bands of the coupling matrix, see App.~\ref{sec:special-line-D2-sq2D1}.

iii) The terms involving $m_{\mc{A}_{2a}}$ and $m_{\mc{A}_{2b}}$ can be grouped together into 
\be\label{eq:crosscouplingA2ab}
\gamma_- (c_\phi~ m_{\mc{A}_{2a}}\!+\!s_\phi~m_{\mc{A}_{2b}})^2+\gamma_+ (s_\phi~m_{\mc{A}_{2a}}\!-\!c_\phi~m_{\mc{A}_{2b}})^2\,,
\ee
where $c_\phi\!\equiv\!\cos\phi$, $s_\phi\!\equiv\!\sin\phi$, $\gamma_{\pm}\!\equiv\!1\pm2/\cos(2\phi)$, and
\be\label{eq:tan2phi}
\tan(2\phi)\!=\!d_+/(2J)\,.
\ee
Therefore, a positive (negative) $d_+$ promotes a nonzero mixing $c_\phi~m_{\mc{A}_{2a}}\!+\!s_\phi~m_{\mc{A}_{2b}}$ (respectively, $c_\phi~m_{\mc{A}_{2a}}\!-\!s_\phi~m_{\mc{A}_{2b}}$) of the two fields. According to Eq.~(\ref{eq:order-params}) a nonzero $m_{\mc{A}_{2b}}$ comes with an out-of-plane ferromagnetic canting $S_t^z\!=\!4\sin\phi$, with 
\be
S_t^z \sim d_+/J ~~\text{for weak}~d_+/J\,.
\ee 
This is also manifest in the full lattice model, as shown below.

iv) The cross-coupling term ${\bf m}_{\mc{E}_a}\cdot ({\bf m}_{\mc{E}_b}-{\bf m}_{\mc{E}_c})$ admixes a nonzero component of the field ${\bf m}_{\mc{E}_a}$ on top of the combination ${\bf m}_{\mc{E}_{c'}}\!=\!\frac{1}{\sqrt{2}}({\bf m}_{\mc{E}_b}\!-\!{\bf m}_{\mc{E}_c})$. This term is relevant for the `twisted all-in/all-out' phase of the full lattice model, as we discuss below.  

v) Finally, under the duality transformation $\mc{M}_{xy}$, the fields 
\be
m_{\mc{A}_{2b}} \to 
-m_{\mc{A}_{2b}}\,,~~
m_{\mc{B}_{1a}} \leftrightarrow  m_{\mc{B}_{1b}}\,,~~
{\bf m}_{\mc{E}_{b}} \leftrightarrow 
{\bf m}_{\mc{E}_{c}}\,,
\ee
while the remaining fields remain invariant. These relations account for the differences between dual phases in the phase diagram, i.e., phases connected by $D_{1,2}\mapsto -D_{1,2}$.

\section{Summary of main results: Classical phase diagram}\label{sec:phase-diag}

Figures~\ref{fig:single-tetra-phase} and \ref{fig:phasediag} show the classical phase diagrams of the single-tetrahedron problem and that of the full lattice model, respectively, in the region $|D_{1,2}|\leq 0.5 J$. The first has been obtained analytically using the Luttinger-Tisza method (see App.~\ref{app:LTsingleTetrahedron}) and the second by a numerical iterative minimization method (which is a variation of the iterative method discussed, e.g., in Refs.~\cite{LapaHenley2012,SklanHenley2012}, see App.~\ref{app:num-method}) applied to finite-size clusters with periodic boundary conditions. For the full model, the Luttinger-Tisza method works only at the isotropic point and the special line $D_2=0$, see App.~\ref{App:LT-full-model}. 
The phases of the first and second quadrant of this plane map to the phases of the third and fourth quadrant, respectively, by virtue of the duality transformation $\mc{M}_{xy}$ mentioned above, which maps $S^z\mapsto -S^z$. So, it suffices to discuss the phases of the first two quadrants only. 

In the parameter region shown, the phase diagram of a single tetrahedron features only two qualitatively different phases, denoted by `$\alpha$' and `cross' (the `$\beta$' phase is the dual of $\alpha$). The main  characteristics of these phases, which are analysed in more detail in Sec.~\ref{sec:single-tetra}, can be seen in the insets of Fig.~\ref{fig:single-tetra-phase}.
The $\alpha$-phase belongs to the irrep $\mc{B}_1$ of the $C_{4\mathrm{v}}$ group (it particular, it saturates the mode $m_{\mc{B}_{1a}}$, whereas the $\beta$-phase saturates the mode $m_{\mc{B}_{1b}}$) and is a member of the ground state manifold of the isotropic point.
By contrast, the cross phase is a mixture of the fields $\mc{A}_{2a}$ and $\mc{A}_{2b}$, which is generated by the cross-coupling term $\propto d_+ \mc{A}_{2a} \mc{A}_{2b}$. As discussed after Eq.~(\ref{eq:crosscouplingA2ab}) above, the admixture of $\mc{A}_{2b}$, involves a nonzero Heisenberg energy cost, associated with a uniform out-of-plane canting of the spins, which grows monotonously with increasing $d_+/J$. 
Otherwise, the in-plane components of the spins resemble the `all-in' (or `all-out') state of the pyrochlore spin ices~\cite{Elhajal05}, where the spins point towards (or away from) the center of the tetrahedron.

Turning to the phase diagram of the lattice model (Fig.~\ref{fig:phasediag}), we find close similarities but also some very striking differences, especially in the region around the special line $D_2=\sqrt{2}D_1$ where several more phases appear.
Indeed, besides the more extended phases termed `twisted $JD_1^\pm$' and `twisted all-in/all-out' (or $p1x$), which appear {\it locally} similar to the $\alpha$ (or $\beta$) and `cross' phases, respectively (see Figs.~\ref{fig:twisted-JD1-plus-1} and \ref{fig:p1x-panel}), we find a number of new phases around the line $D_2\!=\!\sqrt{2}D_1$ that have no analogue in the single-tetrahedron problem. These include: 
i) the phase termed `$p11$', which occupies a large part of the phase diagram and consists of a spatial intertwining of equal `domains' of the neighbrouging `$p1x$' and `twisted $JD_1^\pm$' phases, see Fig.~\ref{fig:p-phases-panel}\,(a). 
ii) The much narrower phase `$p31$' which consists of a spatial intertwining of {\it unequal} `domains' of the `$p1x$' and `twisted $JD_1^\pm$' phases, see Fig.~\ref{fig:p-phases-panel}\,(b).  
iii) The phase `$p11+p31$' which appears in a very narrow region on the border of `$p31$' phase, and consists of alternating `domains' of $p11$ and $p31$, see Fig.~\ref{fig:p-phases-panel}\,(c). 
iv) The narrow phase `$p13$', see Fig.~\ref{fig:p-phases-panel}\,(d), which onsets at relatively large $D/J$ ratios close to the special line $D_2=\sqrt{2}D_1$.

Further important aspects of the phase diagram are as follows.
First, the axis $D_2\!=\!0$ hosts a one-parameter family of degenerate ground states, which are coplanar and have a characteristic, one-dimensional, period-4 modulation, with NN spins rotating successively by 90$^\circ$ along the NN bond directions (${\bf x}$+${\bf y}$ or ${\bf x}$-${\bf y}$), see characteristic spin pattern in Fig.~\ref{fig:JD1p}.

Second, the twisted $JD_1^\pm$ phases are incommensurate phases that are spawn out of the ground state manifold of the $D_2\!=\!0$ line. Specifically, the twisted $JD_1^+$ (twisted $JD_1^-$) phase is a long-wavelength modulation made of members of the degenerate manifold of the positive (negative) $D_1$ semi-axis. The long-wavelength modulation comes with a characteristic canting of the spins which gives rise to two spirals, one for the spins of sublattice A (rotating in a fixed plane) and the other for the spins of sublattice B (rotating in a another fixed plane). The characteristic spin pattern is shown in Fig.~\ref{fig:twisted-JD1-plus-1}. The angle between the planes of the two spirals goes to zero for $D_2\to 0$, see Fig.~\ref{fig:lambdaJD1p}.
Furthermore, the twisted phases feature a very weak ferromagnetic moment per tetrahedron (with typical values of $|{\bf S}_t|$ not exceeding 0.1), showing the proximity of these phases to the ground state manifold of the isotropic point. 

Third, the phase denoted by $p1x$  in Fig.~\ref{fig:phasediag} also describes an incommensurate, one-dimensional modulation along the NN bond directions, ${\bf x}$+${\bf y}$ or ${\bf x}$-${\bf y}$, see Fig.~\ref{fig:p1x-panel}, with corresponding wavevectors very close to $(\pi,\pi)$. 
This phase can be thought of as a long-wavelength modulation that is spawn out of a one-parameter family of $(\pi,\pi)$ states, with in-plane components in the all-in/all-out configuration, which locally resembles the cross phase.
In addition, and unlike the twisted $JD_1^\pm$ phases, the $p1x$ phase features appreciable values of $|{\bf S}_t|$, which increases systematically as we depart from the isotropic point.

Fourth, the remaining phases of the phase diagram of Fig.~\ref{fig:phasediag}, i.e., the ones denoted by $p11$, $p31$, $p31+p11$ and $p13$, and shown in Fig.~\ref{fig:p-phases-panel}, feature a two-dimensional modulation, and can be thought of as multi-domain phases made of the underlying competing fields of the isotropic point. In particular, the local structure and symmetry content of each domain seem to be characteristic of the twisted $JD^\pm$ and the all-in/all-out structure. These phases show a characteristic multi-${\bf Q}$ pattern in the static spin structure factors.

In the following we turn to a more detailed analysis of these results, starting with the exact results on the single-tetrahedron problem.

\section{Single tetrahedron problem: Exact results} \label{sec:single-tetra}

For the single tetrahedron system, i.e., the 4-site cluster of Fig.~\ref{fig:checkerboard-lattice-b}, the Luttinger-Tisza minimization method delivers the exact form of the minimum energy configurations, see App.~\ref{app:LTsingleTetrahedron}. As shown in Fig.~\ref{fig:single-tetra-phase} and mentioned above, the phase diagram consists of three phases: the $\alpha$-phase, the $\beta$-phase, and the cross phase. The former two phases belong to the ground state manifold of the isotropic point, whereas the cross phase involves a nonzero Heisenberg exchange energy cost (i.e., ${\bf S}_t\neq 0$).

{\cbl $\alpha$-phase:} This phase is stabilised in a large parameter region, mainly of large and negative $D_2$, see Fig.~\ref{fig:single-tetra-phase}. Here the four spins of the crossed plaquette point along the following fixed directions (or their time-reversed directions),
\bea\label{eq:single-tetra-spins-a}
\renewcommand{\arraystretch}{1.25}
\!\!\!\begin{array}{ll}
{\bf A}_1=\frac{\eta}{\sqrt{2}}(1,0,-1)\,,   &{\bf B}_1 =\frac{\eta}{\sqrt{2}}(0,1,1)\,, \\
{\bf A}_2=\frac{\eta}{\sqrt{2}}(-1,0,-1)\,,   & {\bf B}_2 =\frac{\eta}{\sqrt{2}}(0,-1,1)\,,
\end{array}
\eea
with $\eta=\pm1$ (the $\eta=-1$ choice is shown in the respective inset in Fig.~\ref{fig:single-tetra-phase}), $m_{\mc{B}_{1a}}\!=\!\eta$ and ${\bf S}_t=0$.  
Furthermore, for $D_2<0$, this configuration saturates the lower energy bound $-|D_2|$ of the $D_2$ interactions, as the spins of the same sublattice are perpendicular to each other, ${\bf A}_1\perp{\bf A}_2$ and ${\bf B}_1\perp{\bf B}_2$.
By contrast, the energy from each $D_1$ coupling is $-D_1/\sqrt{2}$, as the angle between NN spins is $\pi/4$.
The energy per site, $E_\alpha/N$, for the $\alpha$ state is given by:
\be\label{eq:single-tetra-Ea}
E_{\alpha}/N = 
-(J+d_-)/2\,.
\ee
We note that the $C_{4\mathrm{v}}$ symmetry does not deliver any more classical ground states beyond the ones described by Eq.~(\ref{eq:single-tetra-spins-a}) and its time-reversed version.

{\cbl $\beta$-phase:} This configuration is stabilized in a wide region, mainly of large and positive $D_2$, see Fig.~\ref{fig:single-tetra-phase}. It is the dual of the $\alpha$-phase, where the duality here refers to the operation $\mc{M}_{xy}$ mentioned above.
The spins point along the following directions (or their time-reversed directions)
\bea\label{eq:single-tetra-spins-b}
\renewcommand{\arraystretch}{1.25}
\!\!\!\begin{array}{ll}
{\bf A}_1 =\frac{\eta}{\sqrt{2}}(1,0,1)\,,  &{\bf B}_1 =\frac{\eta}{\sqrt{2}}(0,1,-1)\,, \\
{\bf A}_2 =\frac{\eta}{\sqrt{2}}(-1,0,1)\,,  &{\bf B}_2 =\frac{\eta}{\sqrt{2}}(0,-1,-1)\,,
\end{array}
\eea
with $\eta=\pm1$ (the $\eta=-1$ choice is shown in the respective inset in Fig.~\ref{fig:single-tetra-phase}),  $m_{\mc{B}_{1b}}\!=\!\eta$ and ${\bf S}_t=0$.
For $D_2>0$, the $\beta$-phase saturates the lower energy bound $-|D_2|$ of the $D_2$ interactions, as in the $\alpha$-phase.
In addition, the energy per spin for the $\beta$ state is given by
\be\label{eq:single-tetra-Eb}
E_{\beta}/N = -(J-d_-)/2\,
\ee
which results from $E_\alpha/N$ by $D_{1,2}\!\to\!-D_{1,2}$. 

{\cbl Cross-phase for $D_{1,2}>0$:} This phase is stabilised in an extended parameter region surrounding the line $D_2=\sqrt{2}D_1$ for positive $D_1$ and $D_2$. Here the four spins point along the following directions (or, their time-reversed directions)
\bea\label{eq:single-tetra-spins-cp}
\renewcommand{\arraystretch}{1.25}
\!\!\!\begin{array}{ll}
{\bf A}_1\!=\! \eta(\cos \phi,0,\sin \phi),   
& {\bf A}_2\!=\!\eta(-\cos \phi,0,\sin \phi),\\
{\bf B}_1\!=\!\eta(0,-\cos \phi, \sin \phi), 
&{\bf B}_2\!=\!\eta(0,\cos \phi, \sin \phi),
\end{array}
\eea
with $\eta=\pm1$ (in the respective inset in Fig.~\ref{fig:single-tetra-phase} the $\eta=1$ case is shown), the angle $\phi$ given by Eq.~(\ref{eq:tan2phi}), and the energy per site is 
\be\label{eq:single-tetra-Ecross}
E_{\text{cross}}/N=\big(J-\sqrt{4J^2+d_+^2} ~\big)/2~.
\ee
Unlike the $\alpha$- and $\beta$-phase, the cross phase is not a member of the ground state manifold of the Heisenberg point. Moreover, it belongs to the irrep $\mc{A}_{2}$ of $C_{4\mathrm{v}}$. For this reason, the cross-$\alpha$ and cross-$\beta$ boundary lines are first order transition lines, where, e.g., the total moment $S_t^z$ drops to zero discontinuously at the $\alpha$- and $\beta$-phase boundary. Moreover, in the cross phase,
\be
m_{\mc{A}_{2a}} = \eta \cos\phi,~~~
m_{\mc{A}_{2b}} = \eta \sin\phi,
\ee
and therefore this state saturates the combination 
\be
\cos\phi ~m_{\mc{A}_{2a}} +\sin\phi~ 
m_{\mc{A}_{2b}}
\ee
that arises from the cross-coupling term $\propto d_+  m_{\mc{A}_{2a}}m_{\mc{A}_{2b}}$, see discussion after  Eq.~(\ref{eq:crosscouplingA2ab}).
The in-plane components of the spins resemble the `all-in' state of the pyrochlores~\cite{Elhajal05} (or all-out for the time-reversed configuration), where the spins point towards (or away from) the centers of the tetrahedra.
Besides the in-plane components, however, we also have a uniform out-of-plane moment ${\bf S}_t =4\eta\sin\phi~{\bf z}$, reflecting the mixing of the field $m_{\mc{A}_{2b}}\propto \sin\phi$, and which grows monotonously with $|d_+|/J$. 
Note however that the magnitude of the total moment per spin $|{\bf S}_t|/4$ remains small (relatively to the spin length) even for relatively large $d_+/J$ (for example, $|{\bf S}_t|/4\sim 0.12$ for $d_+/J=0.5$).
It is worth noting that  the point-group symmetry does not deliver any more ground states other than the one described by Eq.~(\ref{eq:single-tetra-spins-cp}) and its time-reversed configuration, since the state is invariant under $C_{4z}$, while the action of $\sigma_{xz}$ is equivalent with that of $\mc{T}$.

{\cbl Cross-phase for $D_{1,2}<0$:} This phase is stabilised in an extended region surrounding the line $D_2=\sqrt{2}D_1$ for negative $D_1$ and $D_2$, and can be obtained from the positive $D_{1,2}$ cross phase by the duality $\mc{M}_{xy}$. The four spins point along the following directions (or their time-reversed directions)
\be\label{eq:single-tetra-spins-cn}
\renewcommand{\arraystretch}{1.25}
\!\!\!\!\!\begin{array}{ll}
{\bf A}_1\!=\! \eta(\cos \phi,0,-\sin \phi),  & {\bf B}_1\!=\!-\eta(0,\cos \phi,\sin \phi)\,, \\
{\bf A}_2\!=\! -\eta(\cos \phi,0,\sin \phi),  & {\bf B}_2\!=\!\eta(0,\cos \phi, -\sin \phi)\,,
\end{array}
\ee
where $\eta=\pm1$, $\tan(2\phi) = |d_+|/(2 J)$, and the energy is again given by Eq.~(\ref{eq:single-tetra-Ecross}).

\section{Full lattice model, I: 1D modulated phases}\label{sec:full-model-phasesI}

Having established the phase diagram for the case of the single tetrahedron (Fig.~\ref{fig:single-tetra-phase}), we now turn to the study of the full lattice model.
Importantly, none of the ground states of the single tetrahedron can be directly `tiled' in the lattice without an energy cost. Nevertheless, the three largest phases of the phase diagram of Fig.~\ref{fig:phasediag} -- the twisted $JD_1^\pm$, and the  $p1x$ phase -- are 1D modulations of single-tetrahedron states.  
Here we shall focus on the main characteristics of these phases. The double-twisted phases that are sandwiched between the one-dimensional phases around the special line $D_2=\sqrt{2}D_1$, have no direct analogue in the single-tetrahedron problem and will be discussed in Sec.~\ref{sec:full-model-phasesII}.

\begin{figure}[!b]
\includegraphics[width=0.9\linewidth]{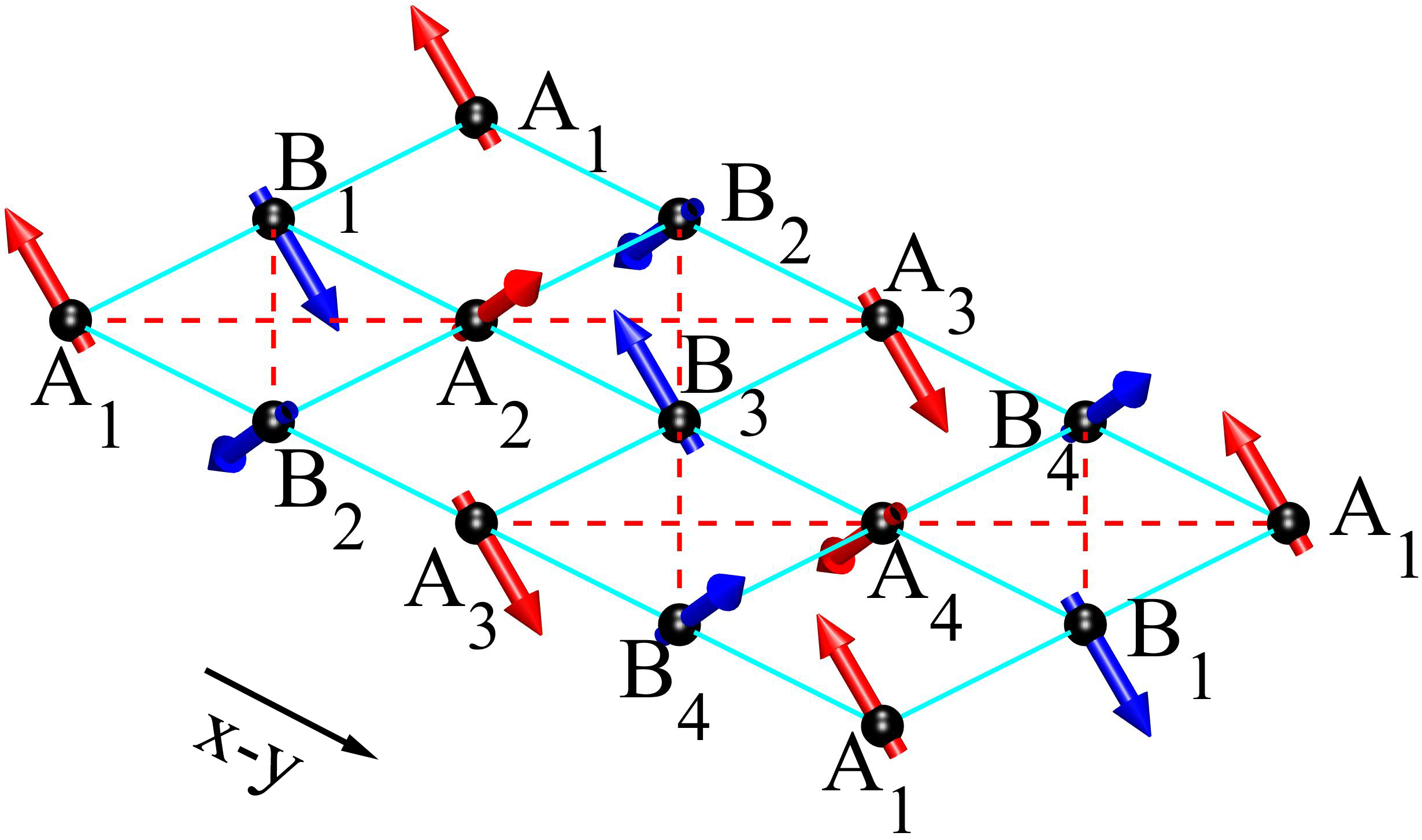}
\caption{(Color online) Magnetic unit cell of a representative member of the $JD_1^+$ manifold of Eq.~(\ref{eq:JD1pos}) [with $\varphi=5\pi/4$ and the ${\bf A}_1$ site corresponding to ${\bf r}=0$ in Eq.~(\ref{eq:JD1pos})].}\label{fig:JD1p}
\end{figure}

\subsection{Special line $D_2=0$: The parent $JD_1^\pm$ manifolds}\label{sec:JD1-state}
We begin with the phases along the special line $D_2=0$, where the Luttinger-Tisza method provides exact results for the classical ground states (see App.~\ref{App:LTJD1pm}).
The line $D_2=0$ features two continuous manifolds of degenerate ground states (denoted by $JD_1^+$ and $JD_1^-$ for $D_{1}>0$ and $D_{1}<0$, respectively), with the `twisted $JD_1^+$' and `twisted $JD_1^-$' phases of Fig.~\ref{fig:phasediag} emerging precisely from these special manifolds. 
The members of these manifolds are all {\it coplanar}, with spins rotating in either the $(x-y,z)$ or $(x+y,z)$ plane, featuring a characteristic, one-dimensional, period-4 modulation with propagation vector ${\bf Q}_0\!=\!(\pi/2,-\pi/2)$ or ${\bf Q}_0'\!=\!(\pi/2,\pi/2)$, respectively.

{\cbl Case $D_1>0$:} 
In the positive $D_1$ case, the $JD_1^+$ manifold consists of two one-parameter families of coplanar  states, each characterised by an angle $\varphi$. 
The two families are related to each other by the $C_{4z}$ symmetry. A member of the first family is shown in Fig.~\ref{fig:JD1p}. Here, the spins rotate in the $(x-y,z)$ plane, in successive steps of 90$^\circ$ as we move along ${\bf x}$-${\bf y}$, which is one of the two NN bond directions. The ordering wavevector of this orthogonal, period-4 state is ${\bf Q}_0\!=\!(\pi/2,-\pi/2)$. 
More explicitly, the first type of ground states take the general form 
\be\label{eq:JD1pos}
\renewcommand\arraystretch{1.5}
\begin{array}{ll}
{\bf S}_{{\bf R},A} =
\cos({\bf Q}_{0}\cdot{\bf R}+\varphi)\frac{{\bf x}-{\bf y}}{\sqrt{2}} - \sin({\bf Q}_{0}\cdot {\bf R}+\varphi) {\bf z}\,, \\
{\bf S}_{{\bf R},B} = -{\bf S}_{{\bf R},A}\,,
\end{array}
\ee
where the angle $\varphi$ does not affect the energy, since
\be\label{eqn:EoNJD1pos}
E/N = -J - D_1\,.
\ee
This degeneracy is accidental, as there is no $SO(2)$ symmetry around the axis $\frac{{\bf x}+{\bf y}}{\sqrt{2}}$ in the presence of $D_1$ interactions)
This phase is described by a combination of the four fields $m_{\mc{A}_1}$, $m_{\mc{B}_{1a}}$, $m_{\mc{B}_{1b}}$, and $m_{\mc{E}_b,1}$ with 
\be\label{eq:JD1pfields}
\renewcommand\arraystretch{1.5}
\begin{array}{l}
m_{\mc{A}_1}({\bf R})=-\frac{1}{2}\sin({\bf Q}_0\cdot{\bf R}+\varphi+\frac{\pi}{4}),\\
m_{\mc{B}_{1a}}({\bf R})=
\frac{1+\sqrt{2}}{2\sqrt{2}}\sin({\bf Q}_0\cdot{\bf R}+\varphi+\frac{\pi}{4}),
\\
m_{\mc{B}_{1b}}({\bf R})=
\frac{1-\sqrt{2}}{2\sqrt{2}}\sin({\bf Q}_0\cdot{\bf R}+\varphi+\frac{\pi}{4}),\\
m_{\mc{E}_b,1}({\bf R})\!=\!
\cos({\bf Q}_0\cdot{\bf R}+\varphi+\frac{\pi}{4})\,,
\end{array}
\ee
where ${\bf R}=n_1 {\bf a}_1+ n_2 {\bf a}_2$ specifies the position of the A$_1$ site of the tetrahedron (see site labeling convention in Fig.~\ref{fig:checkerboard-lattice-b}). 
For later reference, the mean squared fields $\langle m_i^2\rangle$ over the lattice are given by:

\be\label{eq:JD1pfieldsquared}
\renewcommand\arraystretch{1.5}
\begin{array}{l}
\langle m_{\mc{A}_1}^2\rangle = 1/8,\\
\langle m_{\mc{B}_{1a}}^2\rangle =\frac{3+2\sqrt{2}}{16}\simeq 0.36,\\
\langle m_{\mc{B}_{1b}}^2\rangle =\frac{3-2\sqrt{2}}{16}\simeq 0.01,\\
\langle m_{\mc{E}_{b},1}^2\rangle =1/2\,,
\end{array}
\ee
which are independent of $\varphi$. 

The physical origin of these states can be undestood by noting that they saturate the DM energy on half of the NN  bonds (in particular the ones running along ${\bf x}$-${\bf y}$), and, at the same time, they satisfy the condition ${\bf S}_t\!=\!0$ that minimizes the Heisenberg energy.
Another characteristic of these states is that the in-plane projections of the spins point along the directions of the NN bonds, unlike the states of the single-tetrahedron where the in-plane projections point toward (or away from) the centers of the tetrahedra.

The second family of ground states results from the first by the application of the $C_{4z}$ symmetry. The spins, in this case, rotate in the $(x+y,z)$ plane and the propagation vector is ${\bf Q}_0'=(\pi/2,\pi/2)$, i.e., spins rotate by 90$^\circ$ as we move along the NN bond direction ${\bf x}$+${\bf y}$.

{\cbl Case $D_1<0$:} The classical ground state manifold along the negative $D_1$ axis can be obtained from that of the positive $D_1$ axis by applying the duality transformation $\mc{M}_{xy}$, i.e., by flipping the $z$-component of the spins in Eq.~(\ref{eq:JD1pos}): \be\label{eq:JD1neg}
\renewcommand\arraystretch{1.5}
\begin{array}{ll}
{\bf S}_{{\bf R},A} = 
\cos({\bf Q}_{0}\cdot{\bf R}+\varphi)\frac{{\bf x}-{\bf y}}{\sqrt{2}} + \sin({\bf Q}_{0}\cdot {\bf R} +\varphi) {\bf z}\,, \\
{\bf S}_{{\bf R},B}=-{\bf S}_{{\bf R},A}\,.
\end{array}
\ee
This phase is written as a combination of the four fields $m_{\mc{A}_1}$, $m_{\mc{B}_{1a}}$, $m_{\mc{B}_{1b}}$, and $m_{\mc{E}_c,1}$, with 
\be\label{eq:JD1mfields}
\renewcommand\arraystretch{1.5}
\begin{array}{l}
m_{\mc{A}_1}({\bf R})=-\frac{1}{2}\sin({\bf Q}_0\cdot{\bf R}+\varphi+\frac{\pi}{4}),\\
m_{\mc{B}_{1a}}({\bf R})=
\frac{1-\sqrt{2}}{2\sqrt{2}}\sin({\bf Q}_0\cdot{\bf R}+\varphi+\frac{\pi}{4}),
\\
m_{\mc{B}_{1b}}({\bf R})=
\frac{1+\sqrt{2}}{2\sqrt{2}}\sin({\bf Q}_0\cdot{\bf R}+\varphi+\frac{\pi}{4}),\\
m_{\mc{E}_c,1}({\bf R})\!=\!
\cos({\bf Q}_0\cdot{\bf R}+\varphi+\frac{\pi}{4})\,.
\end{array}
\ee
Furthermore, ${\bf S}_t=0$ on each crossed plaquette, as in the $JD_1^+$ manifold, and the energy per site is 
\be\label{eqn:EoNJD1neg}
E/N = -J-|D_1|\,.
\ee

\subsection{The twisted $JD_1^\pm$ phases}
\label{sec:twisted-JD1pm-states}
\subsubsection{Numerical results}

We now discuss how the extended `twisted $JD_1^\pm$' phases of Fig.~\ref{fig:phasediag} emerge from the degenerate $JD_1^\pm$ manifolds.
We shall focus on the twisted $JD_1^+$ phase only (shown by violet color in Fig.~\ref{fig:phasediag}), as the twisted $JD_1^-$ phase is its dual phase. 
Figure~\ref{fig:twisted-JD1-plus-1} summarizes the main features of the twisted $JD_1^+$ phase for a representative point $(D_1,D_2)=(0.3,-0.4)$, as found on a finite-size cluster with spanning vectors ${\bf T}_1=81({\bf a}_1-{\bf a}_2)$ and ${\bf T}_2=81({\bf a}_1+{\bf a}_2)$, and $N\!=\!26244$ spins. 
This cluster accommodates the closest commensurate approximate of the (incommensurate) `twisted $JD_1^+$' phase at the given parameter point. (The way we find these clusters is discussed in  Sec.~\ref{sec:LinearAnsatzTwistedJD1p} below.) 
The main features of the representative state of Figure~\ref{fig:twisted-JD1-plus-1} are as follows:

\begin{figure*}[!t]
\includegraphics[width=\textwidth]{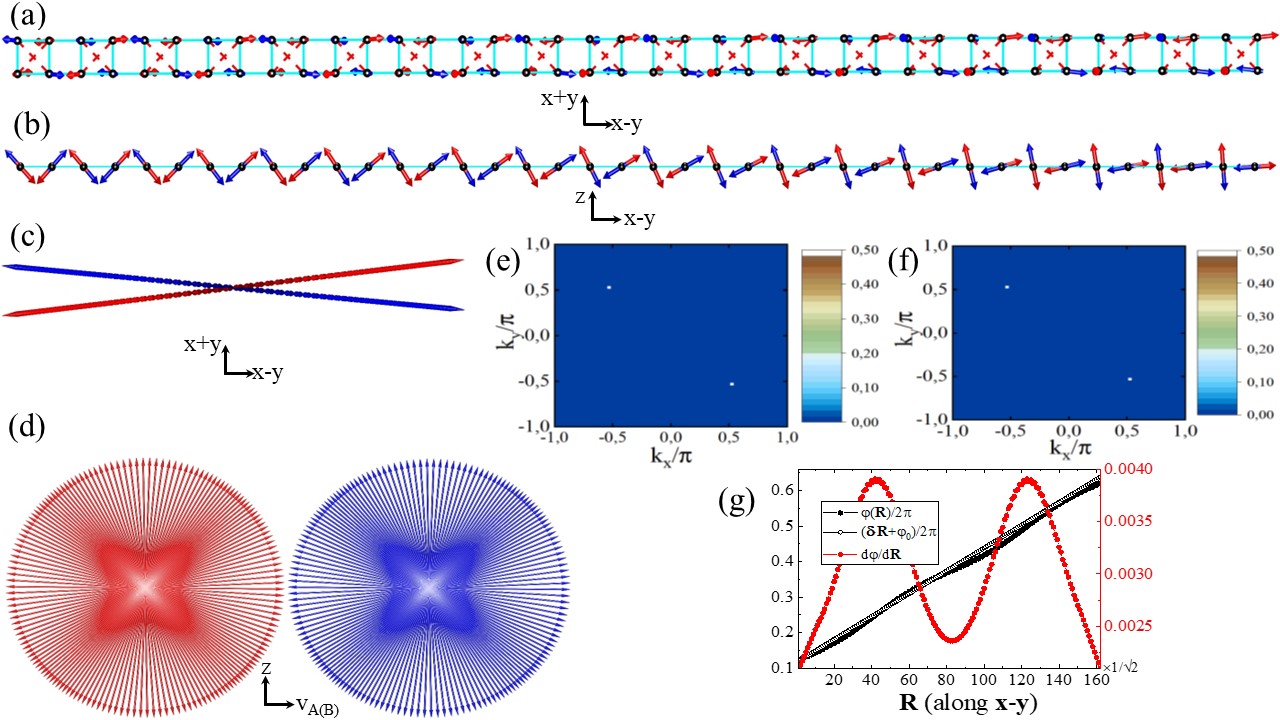}
\caption{(Color online) Key aspects of the twisted $JD_1^+$ state for a representative point $(D_1,D_2)=(0.3,-0.4)$, as found on a finite-size cluster of $N=26244$ spins, with spanning vectors ${\bf T}_1=81({\bf x}-{\bf y})$ and ${\bf T}_2=81({\bf x}+{\bf y})$. 
(a-b) $(x-y,x+y)$ and $(x-y,z)$ views of approximately 1/4 of a period of the modulation.  
(c) $(x-y,x+y)$ view of common-origin plot of the spin configuration, with A and B sublattice spins shown by red and blue arrows, respectively. 
(d) $({\bf v}^{A(B)}_{2},{\bf v}^{A(B)}_{3})$ view of the common origin plot of the A (red) and B (blue) sublattices.
(e-f) Fourier transform intensities $|{\bf S}_{\mathbf{k},A}|^2$ and $|{\bf S}_{\mathbf{k},B}|^2$ in the first Brillouin zone. 
(g) Black filled symbols: Rotation of the angle $\varphi$ of Eq.~(\ref{eq:twistedJD1posAnsatz}) along the propagation direction ${\bf x}$-${\bf y}$, as extracted from numerics. Black open symbols: Fit to the linear approximation $\varphi=\bs{\delta}\cdot{\bf R}+\varphi_0$, where $\bs{\delta}=(\delta,-\delta)$ [see Eq.~(\ref{eq:twisted-JD1plus-phi-angle})]. Red filled symbols: $\mathrm{d}\varphi/\mathrm{d}\mathbf{R}$ , along $\mathbf{x}-\mathbf{y}$, as extracted from numerics, showing the rounded plateaus in $\varphi({\bf R})$.
}\label{fig:twisted-JD1-plus-1}
\end{figure*}

\begin{figure}[!h]
\includegraphics[width=0.93\linewidth,angle=0]{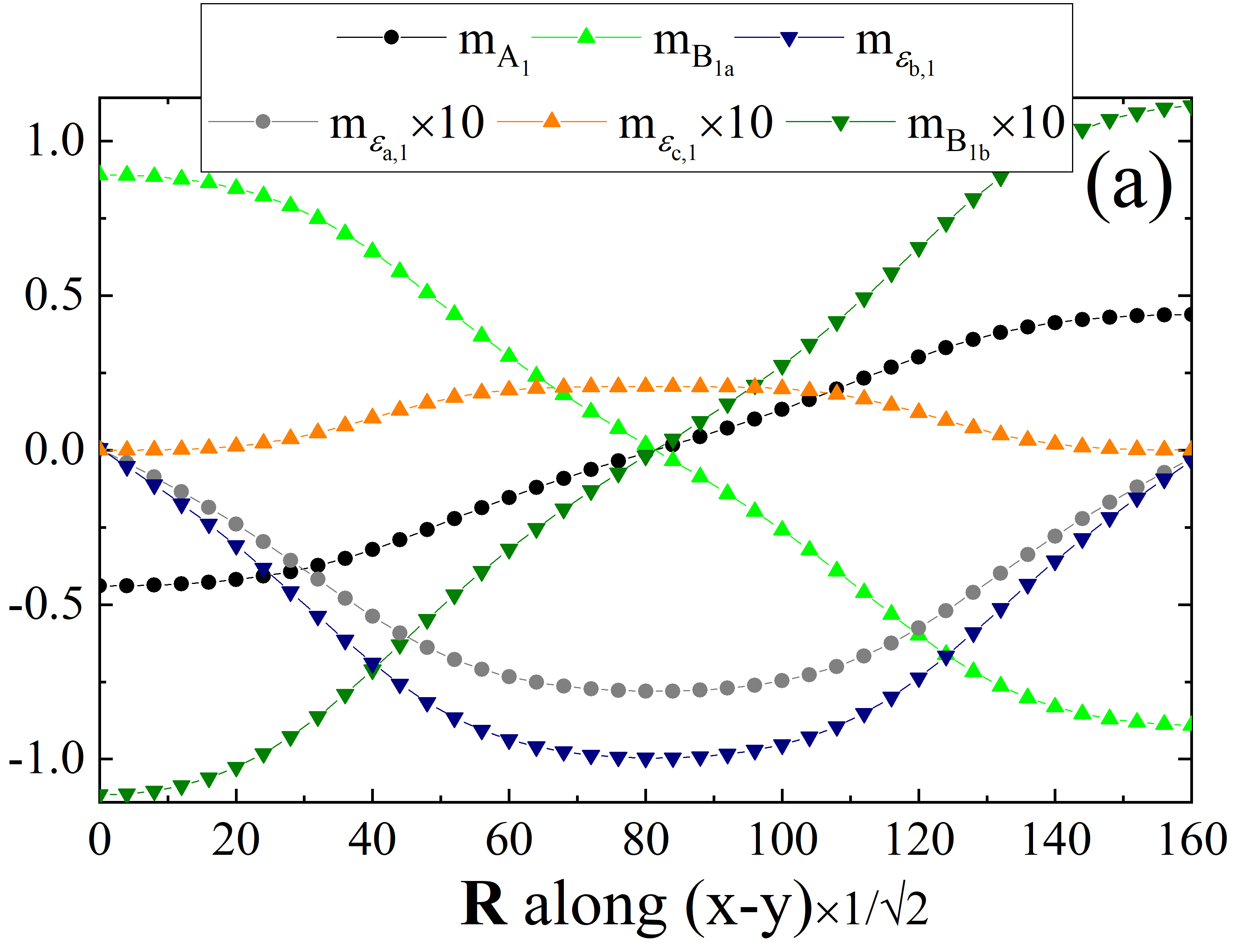}\\
\includegraphics[width=0.9\linewidth,angle=0]{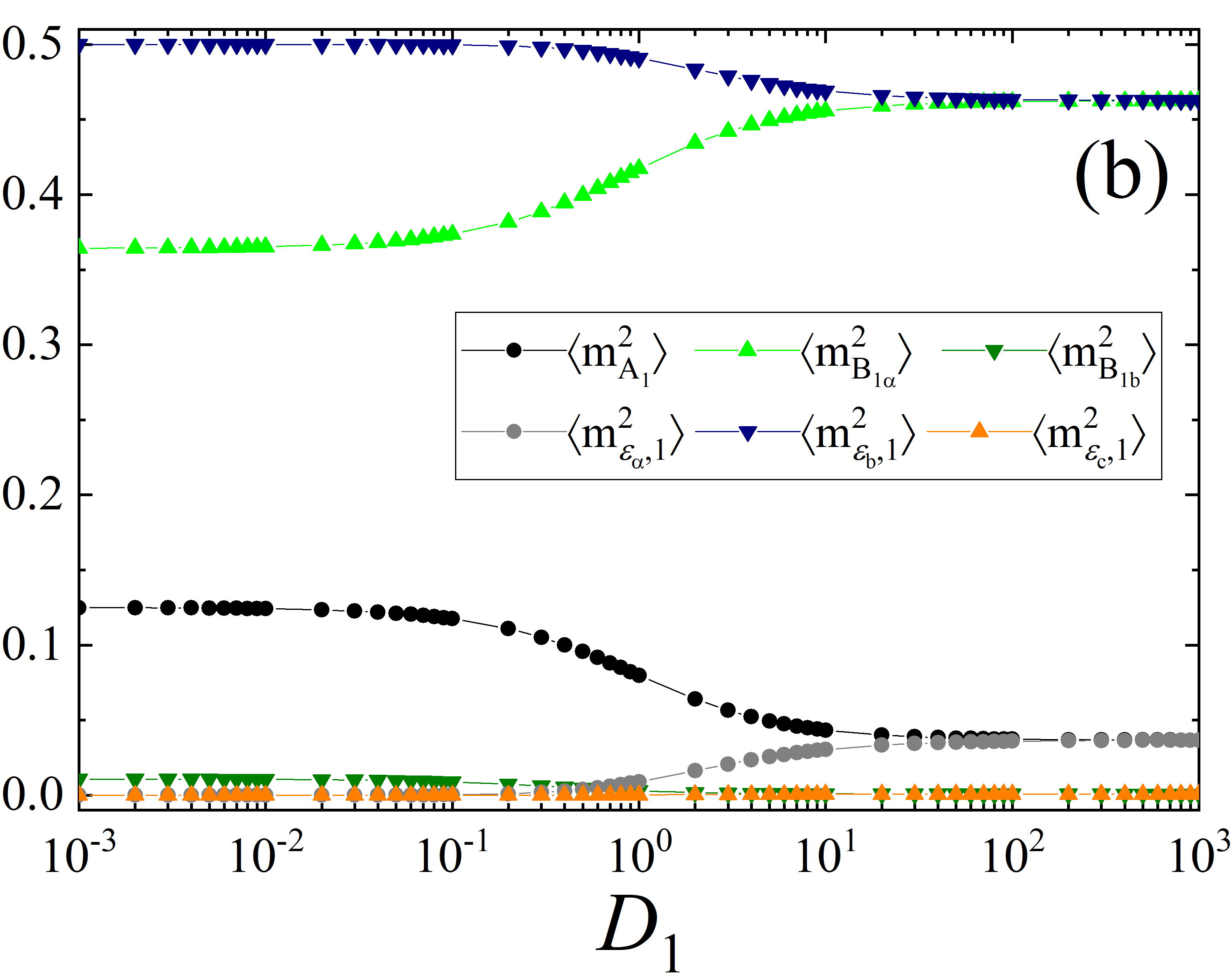}
\caption{(Color online) (a) Spatial evolution of the symmetry-resolved fields of the twisted $JD^{+}_{1}$ phase of Fig.~\ref{fig:twisted-JD1-plus-1} on every four lattice sites along the propagation direction [see also Eq.~(\ref{eq:twisted-JD1p-fields})], as found from numerics. 
(b) Evolution of the mean squared values $\langle m_i^2\rangle$ (over all lattice sites) with $D_1$ along the line $D_2\!=\!-D_{1}$, as found by the linear ansatz predictions of Eq.~(\ref{eq:JD1pAnsatzfieldsquared}), with $\delta$ and $\psi$ determined by minimizing Eq.~(\ref{eq:twisted-JD1+-Euc})}.
\label{fig:miJD1p}
\end{figure}

{\cbl{\it i) Local structure:}}
Panel (a) shows the view in the $(x-y,x+y)$ plane of about one fourth of a period of the modulation. 
The $(x-y,z)$ view shown in panel (b) reveals the close similarity of the local structure of this configuration with the period-4 modulation of the parent $JD_1^+$ states. 
This similarity is further evidenced by the fact that the dominant fields $m_i$ of the twisted $JD_1^+$ phase are precisely the ones listed in Eq.~(\ref{eq:JD1pfields}) for the $JD_1^+$ phase (albeit they now acquire a spatial dependence mirroring the  long-wavelength twisting). Moreover, as shown in Fig.~\ref{fig:miJD1p}\,(b), the corresponding mean squared values of these `parent' fields $\langle m_i^2\rangle$ approach the precise numbers listed in Eq.~(\ref{eq:JD1pfieldsquared}) for $D_1\!\to\!0$. 
Additionally, the `twisted-$JD_1^+$' phase features a very weak ($<10^{-2}$) component of $m_{\mc{E}_a,1}$ and $m_{\mc{E}_c,1}$, which are not present in the parent phase, see Fig.~\ref{fig:miJD1p}\,(b).

{\cbl{\it ii) Global two-plane structure:}}
Figure~\ref{fig:twisted-JD1-plus-1}\,(c) shows that the sublattice-A spins (blue arrows) rotate in one plane and the B spins (red arrows) rotate in another plane, and that the two planes are very close to (and symmetrically away from) the characteristic $(x-y,z)$ plane of the parent $JD_1^+$ manifold. 
Moreover, the opposite sense of rotation of the planes of the A- and B-spins is consistent with the corresponding directions of the $D_2$ vectors along the two diagonals of the tetrahedra. Indeed, the $D_2$ vector between A spins is along ${\bf y}$ and therefore the $D_2$ coupling tends to rotate the plane of the A-spins from the $(x-y,z)$ plane at $D_2\!=\!0$ toward the plane $(x,z)$. Similarly, the $D_2$ vector between B spins is along ${\bf x}$ and therefore the $D_2$ coupling tends to rotate the plane of the B-spins from the $(x-y,z)$ plane at $D_2\!=\!0$ toward the plane $(y,z)$.

The fact that the A-spins rotate in one plane and the B-spins in another plane can also be seen by the numerical calculation of the so-called {\it spin-inertia tensors} of the two separate sublattices, $\mc{I}^{\text{A}}$ and $\mc{I}^{\text{B}}$, defined as
\begin{align}
\label{eq:inertial-tensor}
\mathcal{I}^{\mu}_{\alpha\beta}&= \frac{1}{N/2} \sum_{i \in \mu} S_{i}^\alpha~ S_{i}^\beta,~~~\mu=\text{A, B},~~~
\alpha,\beta=x,y,z\,.
\end{align}
These matrices are semi-positive definite, and their trace equals one due to the spin length constraints~\cite{Henley1984,SklanHenley2012,LapaHenley2012}.
The eigenvalues $\lambda^\mu_{\ell}$ and eigenvectors ${\bf v}^\mu_{\ell}$ ($\mu=$A, B and $\ell=1,2,3$) give valuable insights on global aspects of the spin configuration.
For example, for collinear configurations,  $\lambda_1\!=\!\lambda_2\!=\!0$,  $\lambda_3\!=\!1$, and ${\bf v}_3$ gives the preferred line in spin space. 
Likewise, for coplanar configurations, $\lambda_1\!=\!0$, $\lambda_2\!+\!\lambda_3\!=\!1$, ${\bf v}_1$ gives the direction perpendicular to the spin plane, and the ratio $\lambda_2/\lambda_3$ gives information on the way the spins `cover' the given plane. 
Finally, for non-coplanar configurations, all eigenvalues are nonzero, and their relative ratios encode information about the way spins `cover' the Bloch sphere.
Additionally, the eigenspectrum of the spin-inertia tensors is also useful in building approximate ansatz configurations, as shown below.

For the representative state shown in Fig.~\ref{fig:twisted-JD1-plus-1}, we find that one of the eigenvalues is nearly zero, 
\be
\label{eq:twisted-JD1p-1}
(\lambda^\mu_{1},\lambda^\mu_{2},\lambda^\mu_{3}) = (2.2\mathrm{e}{-7}, 0.494, 0.506)\,,
\ee
for both $\mu=$A, B, meaning that the A-spins are almost coplanar and the same is true for the B-spins. This aspect holds everywhere inside the twisted $JD_1^+$ phase. 

Turning to the eigenvectors of $\mc{I}^{\text{A}/\text{B}}$, these take the form 
\be
\begin{aligned}\label{eq:evecsIAIBTwistedJD1p}
{\bf v}^{A}_{1}&\!=\!-(c_\chi, s_\chi,0), &{\bf v}^{A}_{2}&\!=\!(s_\chi,-c_\chi,0),  & {\bf v}^{A}_{3}&={\bf z}, \\
{\bf v}^{B}_{1}&=-(s_\chi,c_\chi,0), & {\bf v}^{B}_{2}&=(c_\chi, -s_\chi, 0), &{\bf v}^{B}_{3}&={\bf z},
\end{aligned}
\ee
where $c_\chi\!=\!\cos\chi$, $s_\chi\!=\!\sin\chi$, with $\chi$ staying close to $\pi/4$ inside the twisted $JD_1^+$ phase (for the state shown in Fig.~\ref{fig:twisted-JD1-plus-1},  $\chi\simeq 0.286\pi$).
The eigenvectors ${\bf v}_1^{\text{A}/\text{B}}$  corresponding to the lowest eigenvalue ($\lambda_1^{\text{A}/\text{B}}$) specify the planes of the two sublattices. For the state shown in Fig.~\ref{fig:twisted-JD1-plus-1}, 
\bea
\label{eq:twisted-JD1p-2}
{\bf v}_1^A=-(0.623,0.782,0)\,,~~
{\bf v}_1^B=(0.782,0.623,0)\,,
\eea
i.e., the two planes are almost perpendicular to ${\bf x}+{\bf y}$, and are slightly and symmetrically rotated away from the $(x-y,z)$ plane, consistent with Fig.~\ref{fig:twisted-JD1-plus-1}\,(c). 

Finally, we note that, for 
\be
D_2\!\to\!0:~
\chi\!\to\!\pi/4,~
(\lambda^\mu_{1},\lambda^\mu_{2},\lambda^\mu_{3}) \to (0, \frac{1}{2}, \frac{1}{2})\,,
\ee
for both $\mu=$A, B, which is the expected result for the parent $JD_1^+$ states.

{\cbl {\it iii) Ferromagnetic canting:}}
The canting of the two spin sublattices comes with a density-wave-like modulation of the ferromagnetic moment ${\bf S}_t$ along the direction ${\bf x}+{\bf y}$. However, the magnitude $|{\bf S}_t|$ does not exceed a few percent of the classical spin length, showing that it correlates with the very small angle between the planes of the two sublattices, as seen in Fig.~\ref{fig:twisted-JD1-plus-1}\,(c).

{\cbl{\it iv) Almost perfect linearity:}}
The fact that $\lambda_{2}^\mu/\lambda_3^\mu \approx 1$ (for both $\mu=$A, B) shows that the spins of each sublattice `cover' their respective plane in an almost uniform way, i.e., the spins form almost perfect linear helices.
This is illustrated in the common origin plots of Fig.~\ref{fig:twisted-JD1-plus-1}\,(d), but can also be seen more directly in Fig.~\ref{fig:twisted-JD1-plus-1}\,(g), which shows that the spatial evolution of the angle $\varphi(\mathbf{r})$ that parametrizes the linear ansatz discussed below [see Eqs.~(\ref{eq:twistedJD1posAnsatz})-(\ref{eq:twisted-JD1plus-phi-angle})] is almost that of a straight line. 
The weak non-linearity is addressed in
Sec.~\ref{sec:solitonsJD1p}. 

{\cbl{\it v) 1D character of the modulation:}}
Finally, the one-dimensional character of the modulation and the proximity to the parent, period-4 $JD_1^+$ state, is further evidenced in the squared spin structure factors $|{\bf S}_{{\bf k},A}|^2$ and $|{\bf S}_{{\bf k},B}|^2$ of the two sublattices, defined as (see App.~\ref{App:LT-full-model})
\be
{\bf S}_{{\bf k},\mu} =\frac{1}{N/2} \sum_{{\bf r}\in \mu} e^{-i {\bf k}\cdot{\bf r}}~{\bf S}_{{\bf r},\mu}\,,~\mu=A,B\,.~~
\ee
(where $N$ is the total number of spins), and which are shown in Fig.~\ref{fig:twisted-JD1-plus-1}\,(e-f). 
The structure factors show two dominant peaks at wavevectors $\pm(\pi/2+\delta,-\pi/2-\delta)$ with  $\delta\!=\!0.0105\pi$ (corresponding to a wavelength of $\lambda\!\simeq\!190$) for the particular state shown in Fig.~\ref{fig:twisted-JD1-plus-1}). 
We should clarify here that the calculated structure factors feature higher harmonic peaks whose intensities are too weak (about 3\% of the total intensity) to be visible in Fig.~\ref{fig:twisted-JD1-plus-1}\,(e-f). 
Such peaks ensure local stability and the spin length constraint on each site. We have further checked that the intensity of these peaks remains very weak (and therefore $\varphi({\bf R})$ remains linear to a very good approximation) throughout the stability region of the twisted $JD_1^+$ phase.

\subsubsection{The linear ansatz approximation}\label{sec:LinearAnsatzTwistedJD1p}
The above features suggest the following {\it approximate} ansatz for the twisted $JD_1^+$ phase:
\be\label{eq:twistedJD1posAnsatz}
\renewcommand\arraystretch{1.5}
\begin{array}{l}
{\bf S}_{{\bf R},\text{A}}\!\simeq\!\cos\big({\bf Q}_0\!\cdot\!{\bf R}\!+\!\varphi({\bf R})\big) {\bf v}_A \!-\! \sin\big({\bf Q}_0\!\cdot\!{\bf R}\!+\!\varphi({\bf R})\big) {\bf z} \,,
\\
{\bf S}_{{\bf R},\text{B}}\!\simeq\!-\cos \big({\bf Q}_0\!\cdot\!{\bf R}\!+\!\varphi({\bf R})\big) {\bf v}_B\!+\! \sin\big({\bf Q}_0\!\cdot\!{\bf R}\!+\!\varphi({\bf R})\big) {\bf z}\,,
\end{array}
\ee
where
\be
\label{eq:twisted-JD1plus-va-vb-vectors}
\renewcommand\arraystretch{1.5}
\begin{array}{l}
{\bf v}_{\text{A}} = \cos\frac{\psi}{2}~ \frac{{\bf x}-{\bf y}}{\sqrt{2}}-\sin\frac{\psi}{2}~\frac{{\bf x}+{\bf y}}{\sqrt{2}}\,,
\\
{\bf v}_{\text{B}} = \cos\frac{\psi}{2}~\frac{{\bf x}-{\bf y}}{\sqrt{2}}+\sin\frac{\psi}{2}~\frac{{\bf x}+{\bf y}}{\sqrt{2}}\,.
\end{array}
\ee
Here $\psi$ denotes the angle between the spin plane of the A- and B-sublattices, and $\varphi({\bf R})$ describes the long-wavelength rotation of the structure along the propagation direction ${\bf x}-{\bf y}$. Note that ${\bf v}_{\text{A}}$ and ${\bf v}_{\text{B}}$ of Eq.~(\ref{eq:twisted-JD1plus-va-vb-vectors}) have the form of ${\bf v}_2^{\text{A}}$ and ${\bf v}_2^{\text{B}}$ of Eq.~(\ref{eq:evecsIAIBTwistedJD1p}), respectively, with $\psi=\pi/2-2\chi$.

We can constrain the above ansatz further by using the numerical insights that, for states away from the phase boundaries, the angle $\varphi({\bf R})$ varies almost linearly in space [see comparison made in Fig.~\ref{fig:twisted-JD1-plus-1}\,(g)]. So, we can approximate
\be\label{eq:twisted-JD1plus-phi-angle}
\varphi({\bf R})\simeq  \bs{\delta}\cdot{\bf R} + \varphi_0\,,
\ee
where $\varphi_0$ is a constant and $\bs{\delta}=(\delta,-\delta)$, as suggested by the numerical results for the structure factor.

The linear ansatz encapsulates all of qualitative features found numerically.
In particular, the total moment ${\bf S}_t$, 
\be\label{eq:twisted-JD1p-St}
{\bf S}_t({\bf R})=- 4\sin\frac{\psi}{2} \cos\frac{2\delta+\pi}{4} \cos\zeta_{\bf R}~\frac{{\bf x}+{\bf y}}{\sqrt{2}}\,,
\ee
where $\zeta_{\bf R}\!=\!({\bf Q}_0\!+\!\bs{\delta})\!\cdot\!{\bf R}+\varphi_0+\frac{2\delta\!+\!\pi}{4}$, shows a `density-wave' modulation along the axis ${\bf x}+{\bf y}$, with a weak  magnitude $\propto\sin\frac{\psi}{2}$ that correlates with the small angle $\psi$ between the planes of the two spin sublattices. 
Moreover, the phase features the following fields $m_i({\bf R})$, 
\be\label{eq:twisted-JD1p-fields}
\renewcommand\arraystretch{1.5}
\!\!\!\begin{array}{l}
m_{\mc{A}_1}\!=\!- \sin\frac{2\delta+\pi}{4} \sin\frac{2\psi+\pi}{4}\sin\zeta_{\bf R},
\\
m_{\mc{B}_{1a}}\!=\!\frac{1}{2 \sqrt{2}} [\cos\frac{\delta+\psi}{2}\!+\!\sin\frac{\delta-\psi}{2}\!+\!2 \cos\frac{2\delta+\pi}{4}]\sin\zeta_{\bf R},
\\
m_{\mc{B}_{1b}}\!=\!\frac{1}{2 \sqrt{2}} [\cos\frac{\delta+\psi}{2}\!+\!\sin\frac{\delta-\psi}{2}\!-\!2 \cos\frac{2\delta+\pi}{4}]\sin\zeta_{\bf R},
\\
m_{\mc{E}_b,1}\!=\!\frac{1}{\sqrt{2}} [\cos \frac{\psi}{2}\cos\frac{2\delta+\pi}{4}+\sin\frac{2\delta+\pi}{4}]\cos \zeta_{\bf R},
\\
\hline
m_{\mc{E}_a,1}\!=\!-\sin\frac{\psi}{2}\cos\frac{2\delta+\pi}{4}\cos\zeta_{\bf R},
\\
m_{\mc{E}_c,1}\!=\!\frac{1}{\sqrt{2}} [\cos \frac{\psi}{2}\cos\frac{2\delta+\pi}{4}-\sin\frac{2\delta+\pi}{4}]\cos \zeta_{\bf R},
\end{array}
\ee
from which $m_{\mc{E}_a,1}$ and $m_{\mc{E}_c,1}$ are not present in the parent state and have much weaker magnitudes, as discussed above.

The mean squared values of these fields over the lattice are 
\be\label{eq:JD1pAnsatzfieldsquared}
\renewcommand\arraystretch{1.5}
\begin{array}{l}
\langle m_{\mc{A}_1}^2\rangle = \frac{1}{2}
\sin^2\frac{2\delta+\pi}{4} \sin^2\frac{2\psi+\pi}{4}\,,\\
\langle m_{\mc{B}_{1a}}^2\rangle = \frac{1}{16} [\cos\frac{\delta+\psi}{2}\!+\!\sin\frac{\delta-\psi}{2}\!+\!2 \cos\frac{2\delta+\pi}{4}]^2\,,
\\
\langle m_{\mc{B}_{1b}}^2\rangle = \frac{1}{16} [\cos\frac{\delta+\psi}{2}\!+\!\sin\frac{\delta-\psi}{2}\!-\!2 \cos\frac{2\delta+\pi}{4}]^2\,,
\\
\langle m_{\mc{E}_{b},1}^2\rangle = \frac{1}{4} [\cos \frac{\psi}{2}\cos\frac{2\delta+\pi}{4}+\sin\frac{2\delta+\pi}{4}]^2\,,
\\
\hline
\langle m_{\mc{E}_{a},1}^2\rangle =\frac{1}{2}\sin^2\frac{\psi}{2}\cos^2\frac{2\delta+\pi}{4} \,,
\\
\langle m_{\mc{E}_{c},1}^2\rangle =\frac{1}{4} [\cos \frac{\psi}{2}\cos\frac{2\delta+\pi}{4}-\sin\frac{2\delta+\pi}{4}]^2 \,.
\end{array}
\ee
Figure~\ref{fig:miJD1p}\,(b) shows the evolution of these values with $D_1$ along the line $D_2\!=\!-D_1$, with $\delta$ and $\psi$ determined by minimizing Eq.~(\ref{eq:twisted-JD1+-Euc}) below. 
As expected, the mean squared fields tend to the corresponding values of Eq.~(\ref{eq:JD1pfieldsquared}) for the parent phases in the limit of $\delta\to0$ and $\psi\to0$.

We remark that the ansatz does not otherwise satisfy local stability exactly, as can be checked explicitly by calculating the local torques $\bs{\tau}_i={\bf S}_i\times{\bf h}_i$ (where ${\bf h}_i$ is the local field exerted on the $i$-th spin site from its neighbours). For local stability, the torques should vanish identically for all spins, but in the above ansatz, they are generally nonzero, with $|\bs{\tau}_i|_{\text{max}} \sim10^{-2}$. This is also reflected in the higher harmonic structure factor peaks mentioned above.

{\cbl {\it The long-wavelength regime:}}
According to the above, the twisted $JD_1^+$ state emerges from the parent $JD_1^+$ state, and therefore the linear ansatz of Eqs.~(\ref{eq:twistedJD1posAnsatz}) and (\ref{eq:twisted-JD1plus-phi-angle}) must reduce to Eq.~(\ref{eq:JD1pos}) in the limit of $D_2\to0$, meaning that $\psi$ and $\delta$ must vanish in this limit. The wavelength $\lambda$ of the modulation is therefore expected to grow monotonously with decreasing $D_2$ and diverge at $D_2\to 0$. 
As a result, the weak $D_2$ regime poses a number of numerical challenges: To capture long-wavelength modulations one needs: i) to have a good estimate of $\lambda$ so that we choose appropriate cluster sizes to work with, and ii) to be able to efficiently minimize the energy over such large clusters. The iterative single-site update method (see App.~\ref{app:num-method}) becomes inefficient for cluster sizes of the order of $n\sim 10^2$ or larger, where $n$ defines the spanning vector ${\bf T}_1=n({\bf a}_1-{\bf a}_2)$ along the propagation direction [the size $m$ in the orthogonal direction, ${\bf T}_2=m ({\bf a}_1+{\bf a}_2)$, does not need to be large given the 1D modulation of the present phase].

The linear ansatz helps to alleviate these problems as follows. Using  Eqs.~(\ref{eq:twistedJD1posAnsatz}) and (\ref{eq:twisted-JD1plus-phi-angle}), we can  determine the energy per unit cell, $E_{uc}$, analytically in the thermodynamic limit by summing over ${\bf R}$.
The final expression depends on $D_1$, $D_2$, $\delta$ and $\psi$, but not on $\varphi_0$ (which is a free parameter characterising a global shift of the IC phase along the propagation direction):
\bea\label{eq:twisted-JD1+-Euc}
E_{uc}\!&=&\!-1\!-\!\sin\delta \!+\!(\sin\delta\!-\!1)\cos\psi \!+\!\cos\delta \big\{
\nonumber\\
\!&-&\!2D_1\cos(\psi/2)\!+\!\sqrt{2} D_2 [\cos(\psi/2)\!-\!\sin(\psi/2)] \big\}.~~~
\eea
Minimizing $E_{uc}$ with respect to $\psi$ and $\delta$ for given $D_{1,2}$ delivers $\psi$ and $\delta$. The latter gives the optimal cluster size $n$, using the relation 
\be
n=\frac{\pi}{\delta+\pi/2}k,~~k \in \mathbb{Z}\,,
\ee
with $k$ chosen so that $n$ is very close to an integer.
Next, as the linear ansatz is approximate, we use the predicted values of $\psi$, $\delta$ and $n$ to construct an initial state that we then update numerically using the iterative minimization procedure discussed in App.~\ref{app:num-method}. And, finally, we explore other nearby cluster sizes and choose the one that delivers the lowest energy. The configuration of Fig. \ref{fig:twisted-JD1-plus-1} results from such a procedure.

\begin{figure}[!t]
~~\includegraphics[width=0.9\linewidth]{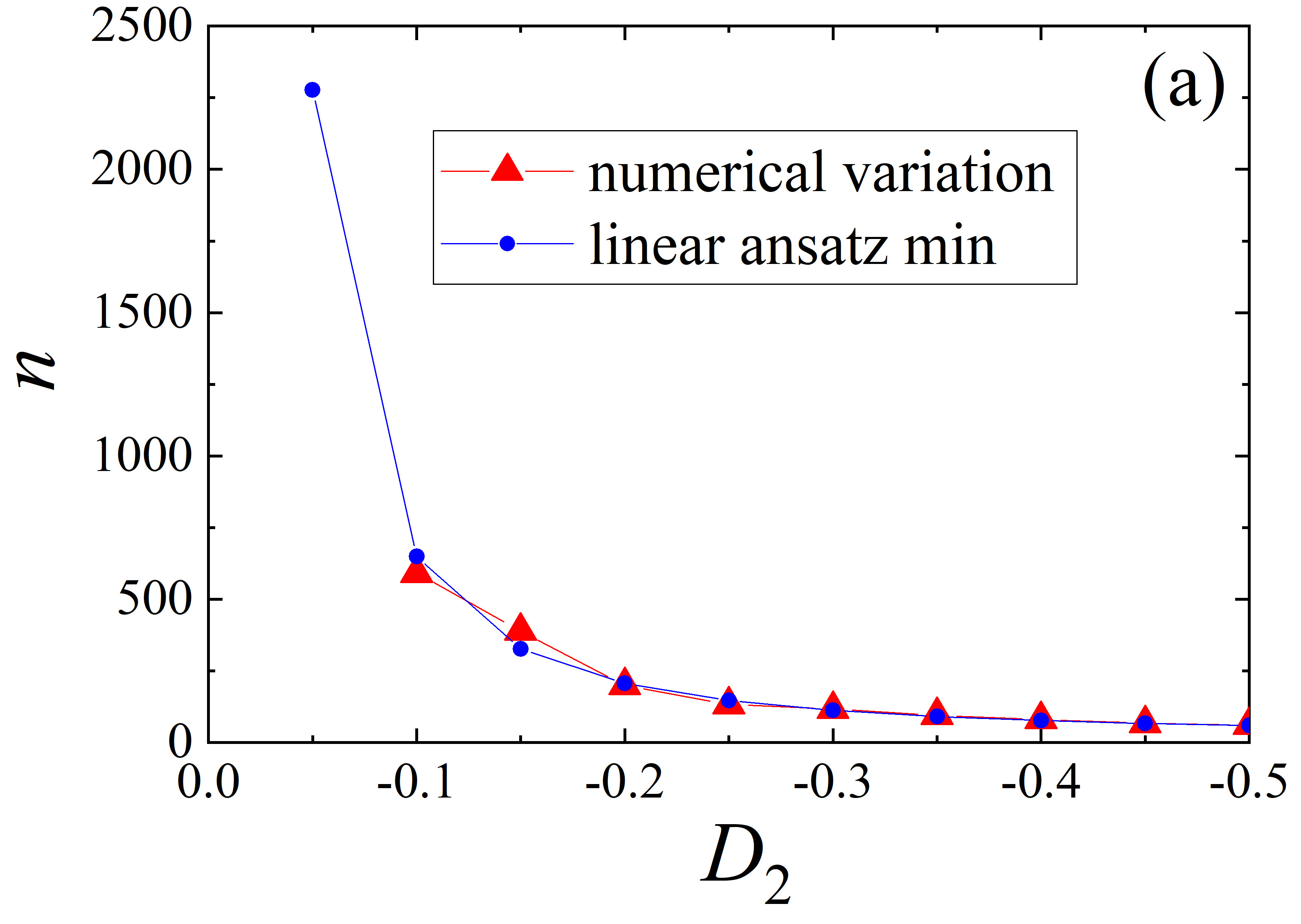}
\includegraphics[width=0.9\linewidth]{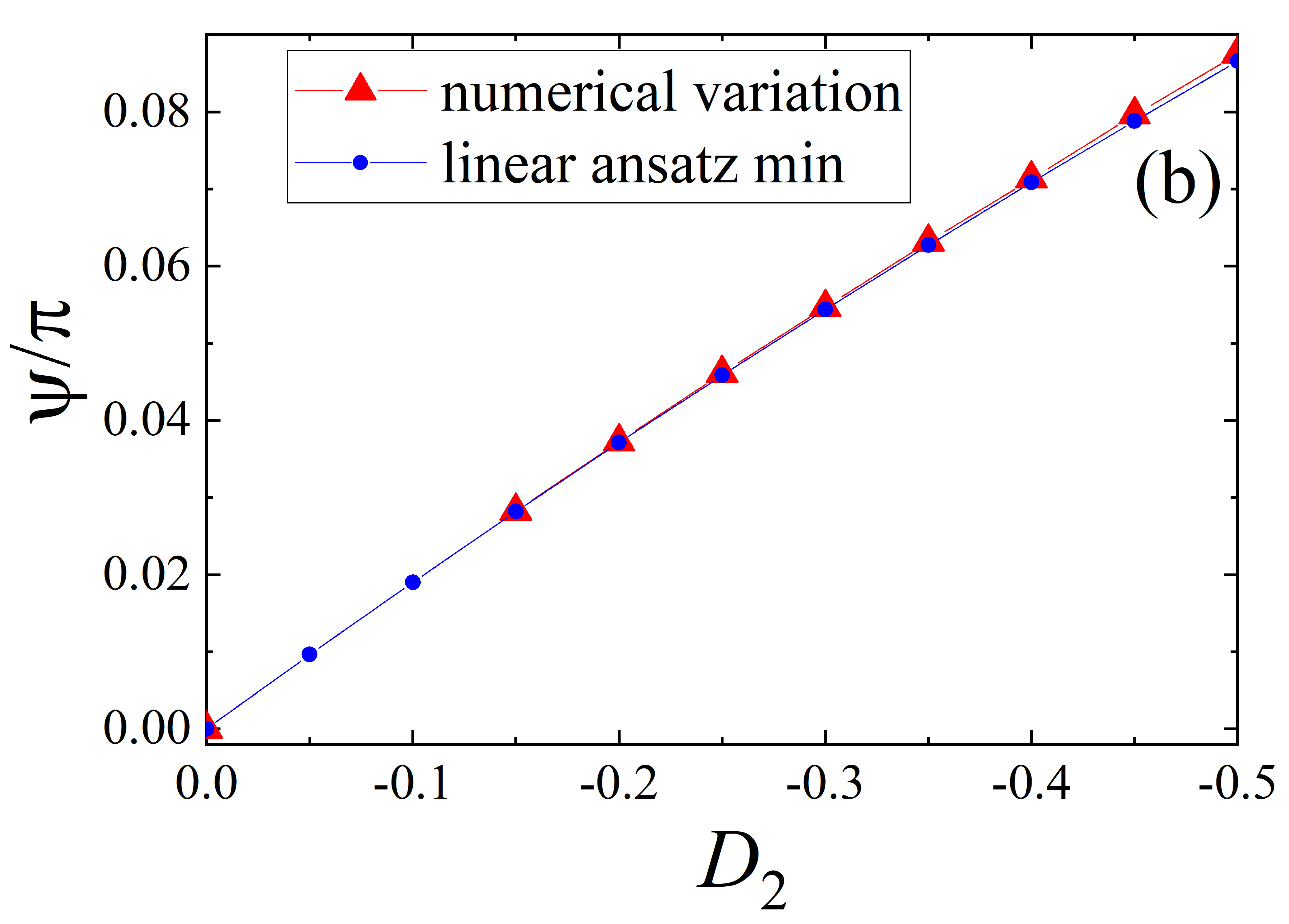}
\caption{Optimal system size $n$ (a) and the angle $\psi$ between the planes of A and B spins (b) as a function of $D_2$ for $D_1\!=\!0.3$. The data shown by triangles correspond to the values obtained numerically, whereas data shown by circles are obtained by the minimization of energy computed within the linear, Eq.~(\ref{eq:twisted-JD1+-Euc}).}
\label{fig:lambdaJD1p}
\end{figure}

Figure~\ref{fig:lambdaJD1p} shows a comparison of the optimal cluster size $n$ and the angle $\psi$ predicted by the linear ansatz with the corresponding values obtained numerically using the procedure described above. The comparison is made along the line $D_1\!=\!0.3$ inside the twisted $JD_1^+$ phase for negative $D_2$.
As expected, the difference between the two values of $n$ (Fig.~\ref{fig:lambdaJD1p}\,a) is more pronounced as $D_2$ approaches zero. On the other hand, the two sets of values of $\psi$ (Fig.~\ref{fig:lambdaJD1p}\,b) are practically everywhere the same, showing that this parameter is determined by local energetics.
Altogether, the main trends in these two panels confirm the general expectation that $\psi\to0$ and $\lambda\to\infty$ in the limit $D_2\to0$, and therefore the twisted $JD_1^+$ state is a long-wavelength modulation of the $JD_1^+$ states.

\subsubsection{C-I transition for the twisted $JD_1^\pm$ phases}\label{sec:solitonsJD1p}
As mentioned above, the linear ansatz provides a very good description of the local physics, but it misses nonlinear aspects, which are evident in a number of results: 
i) the slight deviation of $\lambda_{2}^\mu/\lambda_3^\mu$ from $1$. 
ii) the common origin plots of Fig.~\ref{fig:twisted-JD1-plus-1}\,(d) show a slightly larger spin density along four orthogonal directions in the plane of rotation. 
iii) the derivative $\frac{d\varphi}{dR}$ along the propagation direction, shown in Fig.~\ref{fig:twisted-JD1-plus-1}\,(g), reveals two rounded plateaus in $\varphi({\bf R})$.
iv) the spatial evolution of the symmetry-resolved fields $m_i({\bf R})$ shown in Fig.~\ref{fig:miJD1p}\,(a) reveals a clear deviation from the harmonic behaviour, whose position coincides with that of the plateaus in $\frac{d\varphi}{dR}$.

By continuity, it is expected that this nonlinearity becomes more pronounced as we approach the limit $D_2=0$.
Indeed, in this scenario, the ansatz of Eq.~(\ref{eq:twistedJD1posAnsatz}) would still be valid, but the linear approximation of Eq.~(\ref{eq:twisted-JD1plus-phi-angle}) would break down and the angle $\varphi$ would show a behaviour similar to that of a soliton lattice. Namely, that of a piece-wise constant $\varphi$ inside `domains' of the parent period-4 phase, separated by $\pi/2$-jumps of $\varphi$,  which play the role of `domain walls' (or `discommensurations') with a negative energy density $\propto D_2$. The distance between the domain walls would then diverge in the limit $D_2=0$, leaving a single domain of the parent phase. This {\it commensurate-incommensurate} type of transition at $D_2\!=\!0$ is usually driven by {\it Lifshitz invariants}, linear spatial derivative terms in the long-wavelength expansion of the energy~\cite{Dz64,IZYUMOV,Bogdanov1989}. 

As it turns out, checking this scenario explicitly requires studying cluster sizes that are far too large for an efficient minimization, according to Fig.~\ref{fig:lambdaJD1p}\,(a). We note, in particular, that the state shown in Fig.~\ref{fig:twisted-JD1-plus-1} corresponds to $D_2\!=\!-0.4$ and has been obtained on a cluster with $N\!=\!26244$ spins. Results on similar clusters at smaller values of $D_2$ do not show more pronounced nonlinearities, implying that these require much larger system sizes.
This is consistent with the prediction from the linear analysis above that the optimal wavelength (or the distance between discommensurations in the nonlinear regime) grows very abruptly as $D_2\to0$ (see Fig.~\ref{fig:lambdaJD1p}).

\subsection{The parent $(\pi,\pi)$ manifold of the $p1x$ phase}
We now turn to the second most extended phase of the phase diagram of Fig.~\ref{fig:phasediag}, the $p1x$ phase. As it turns out, this phase is spawn out of another family of parent states, which we now analyse.

\begin{figure}[!h]
\includegraphics[width=0.75\linewidth,angle=0]{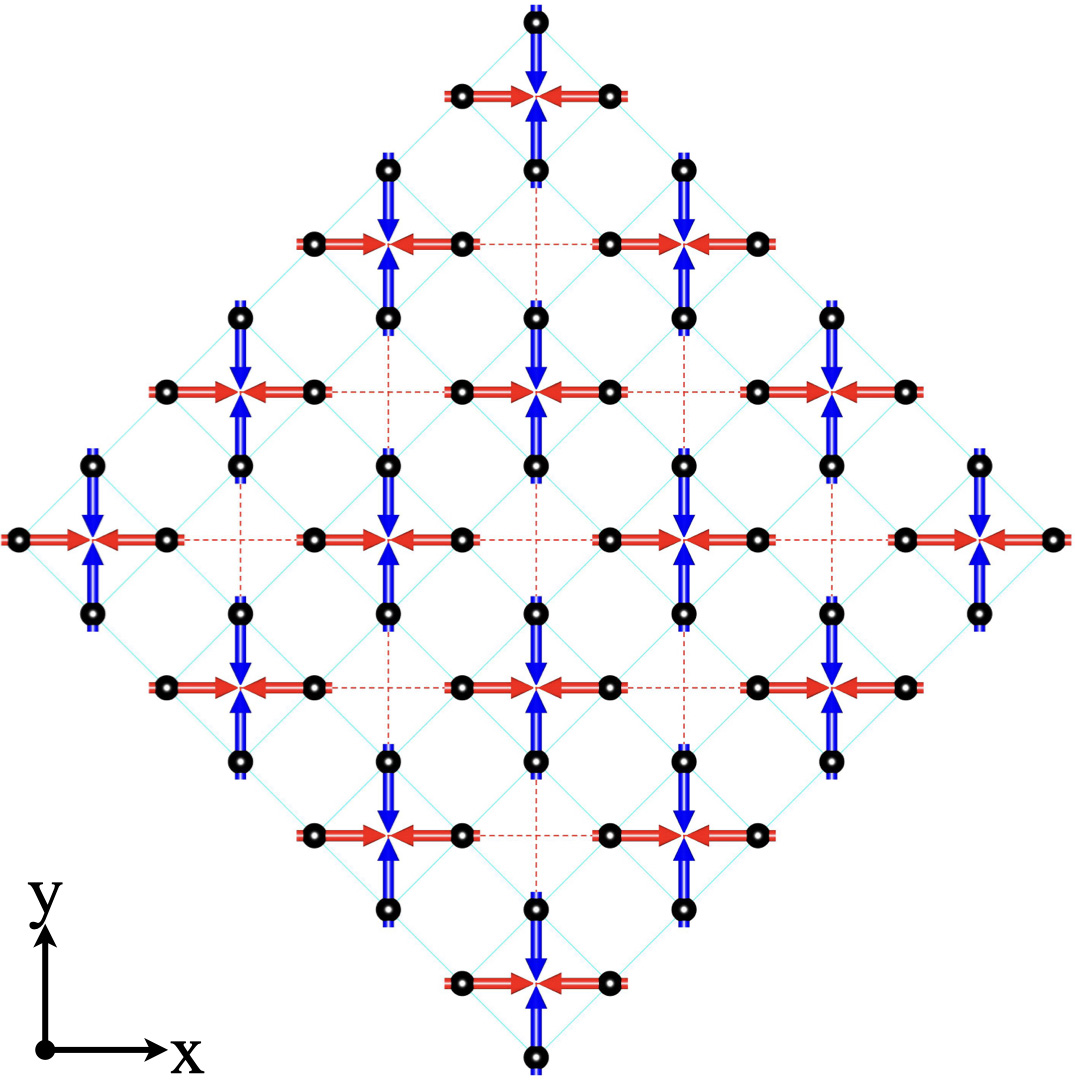}
\caption{(Color online) Tiling of the $D_{1,2}\!=\!0$ version of the cross state of the single tetrahedron problem. This state is a special member of the parent $(\pi,\pi)$ manifold of the $p1x$ phase of the lattice. Red and blue colors denote A- and B-sublattice spins respectively.}
\label{fig:AllInAllOut}
\end{figure}

We have mentioned above that the ground states of the single tetrahedron problem cannot be generally tiled on the lattice. 
Take, for example, the cross phase of Fig.~\ref{fig:single-tetra-phase}, which involves an all-in (or all-out) arrangement of the in-plane components of the spins and a uniform out-of-plane canting (which gives a nonzero $S_t^z$). 
The reason that this state cannot be tiled on the lattice is precisely the presence of this type of canting, as the sign of $S_t^z$ depends on whether the in-plane components arrange in the all-in or the all-out configuration (see Fig.~\ref{fig:single-tetra-phase}).
However, in the limit of $D_{1,2}\to0$, the out-of-plane spin components in the cross state vanish, and the cross state can then be tiled with ordering wavevector $(\pi,\pi)$, leading to the coplanar all-in/all-out configuration of Fig.~\ref{fig:AllInAllOut}. 

As it turns out, this state is only one special member of a manifold of $(\pi,\pi)$ states, which arise from the coplanar all-in/all-out member by adding a $(\pi,\pi)$ modulated out-of plane components. This corresponds to a vanishing $S_t^z$ for all tetrahedra $t$, and therefore does not cause any extra Heisenberg energy cost. 
The resulting one-parameter family of $(\pi,\pi)$ states is described by 
\be\label{eq:SRparentAllinAllout}
\renewcommand\arraystretch{1.5}
\begin{array}{l}
{\bf S}_{{\bf R},\text{A}}=\cos({\bf K}\!\cdot\!{\bf R}+\varphi_0) {\bf x}+
\sin({\bf K}\!\cdot\!{\bf R}+\varphi_0) {\bf z}
,\\
{\bf S}_{{\bf R},\text{B}}=-\cos({\bf K}\!\cdot\!{\bf R}+\varphi_0) {\bf y}+
\sin({\bf K}\!\cdot\!{\bf R}+\varphi_0) {\bf z}
\end{array}
\ee
where ${\bf K}=(\pi,\pi)$ is the ordering wavevector and $\varphi_0$ is a free parameter that quantifies the admixure of out-of-plane and in-plane components.

Let us summarise the main features of these parent states. 
First, the A spins are collinear [they point along $\pm (\cos\varphi_0 {\bf x}+\sin\varphi_0{\bf z})$] and the same is true for the B spins [which point along $\pm (-\cos\varphi_0 {\bf y}+\sin\varphi_0{\bf z})$]. Second, the angle between A and B spins depends on $\varphi_0$. In particular, the two sublattices turn orthogonal to each other for $\varphi_0\!=\!0$, see Fig.~\ref{fig:AllInAllOut}. 
Third, the manifold contains the following three symmetry-resolved fields only
\be\label{eq:miparentAllInAllout}
\renewcommand\arraystretch{1.5}
\begin{array}{c}
m_{\mc{A}_{2a}}({\bf R})=(-1)^{n_1+n_2} \cos\varphi_0\,,\\
m_{\mc{E}_{b},2}({\bf R})=-m_{\mc{E}_{c},2}({\bf R})=(-1)^{n_1+n_2} \sin\varphi_0/\sqrt{2}\,,
\end{array}
\ee
where $n_{1,2}$ are the integers labeling ${\bf R}\!=\!n_1 {\bf a}_1+n_2{\bf a}_2$. 
The corresponding mean squared values over all lattice sites are 
\be\label{eq:miparentAllInAlloutSquared}
\renewcommand\arraystretch{1.5}
\begin{array}{c}
\langle m_{\mc{A}_{2a}}^2\rangle=\cos^2\varphi_0
,\\
\langle m_{\mc{E}_{b},2}^2\rangle=\sin^2\varphi_0/2,~~
\langle m_{\mc{E}_{c},2}^2\rangle=\sin^2\varphi_0/2,
\end{array}
\ee
which depend on the particular choice of $\varphi_0$, unlike the case for the parent $JD_1^\pm$ states (the difference is due to the different ordering wavevectors).

\begin{figure*}[!t]
\includegraphics[width=\textwidth]{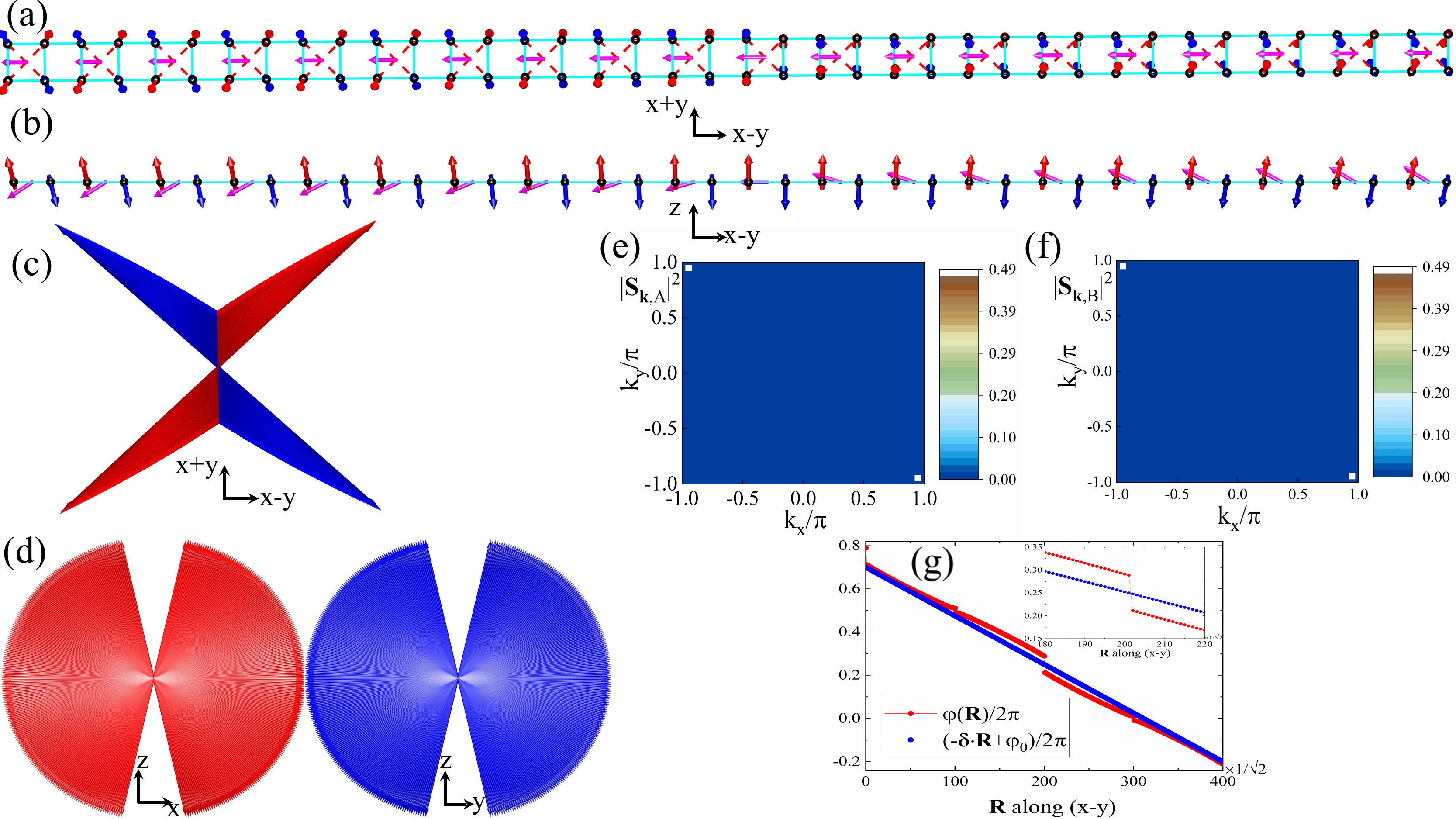}
\caption{(Color online) Key aspects of the $p1x$ phase for a representative point $(D_1,D_2)=(0.01,\sqrt{2}D_{1})$, as found on a finite-size cluster with $N=160000$ spins and spanning vectors $\mathbf{T}_{1}=200(\mathbf{x}-\mathbf{y})$ and $\mathbf{T}_{2}=200(\mathbf{x}+\mathbf{y})$.
(a-b) $(x-y,x+y)$ and $(x-y,z)$ views of $1/5$ of a modulation period. Red and blue arrows show A- and B-sublattice spins, respectively, whereas magenta arrows at the center of each tetrahedron show the total spin ${\bf S}_t$.
(c) $(x-y,x+y)$ view of common origin plot of A (red) and B (blue) spins.
(d) $xz$-view of A spins and $yz$ view of B spins.
(e-f) Structure factors $|{\bf S}_{\mathbf{k},A}|^2$ and $|{\bf S}_{\mathbf{k},B}|^2$.  
(g) Evolution of the angle $\varphi(\mathbf{R})$ of Eq.~(\ref{eq:p1x-ansatz}) with ${\bf R}$, as extracted from the numerical data, and its linear approximation, Eq.~(\ref{eq:phivsRp1x}).
}\label{fig:p1x-panel}
\end{figure*}

\subsection{The $p1x$ phase} \label{sec:p1x-state}

\subsubsection{Numerical results}
\begin{figure}[!t]
\includegraphics[width=0.9\linewidth,angle=0]{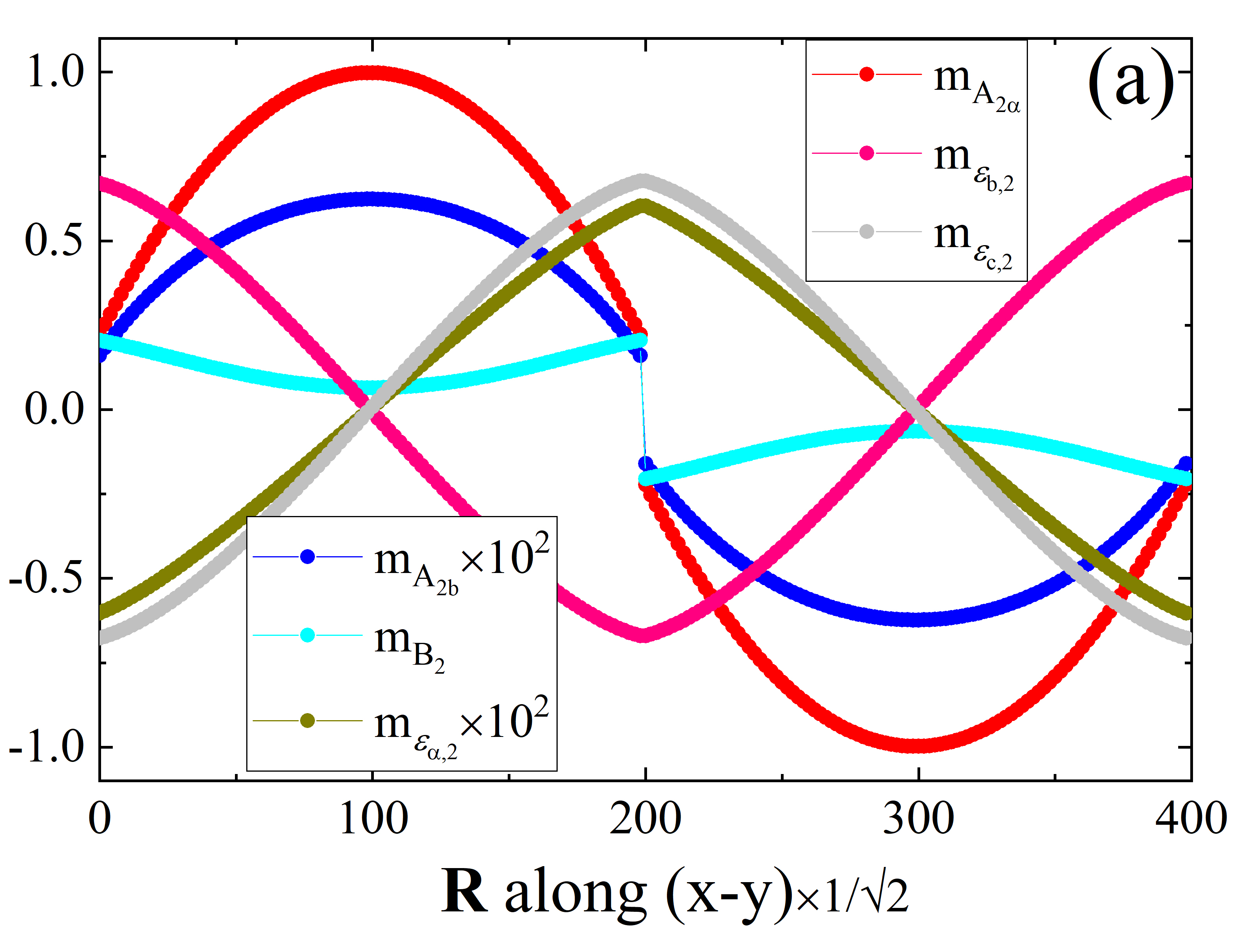}
\includegraphics[width=0.9\linewidth,angle=0]{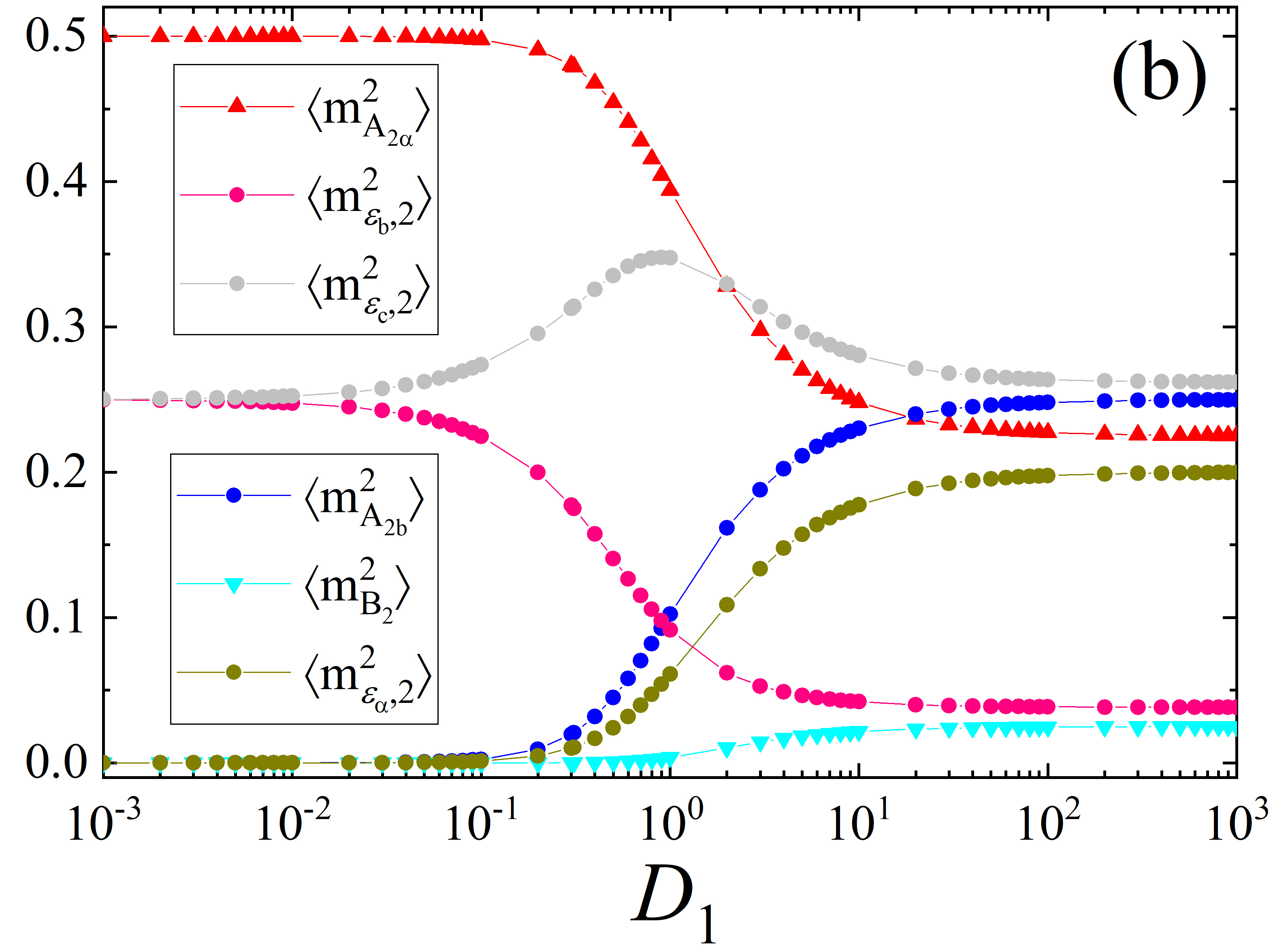}
\caption{(Color online) (a) Evolution of the symmetry-resolved fields in the $p1x$ state of Fig.~\ref{fig:p1x-panel} with ${\bf R}$ (on every second lattice site along the propagation direction), as found by numerics. 
(b) Evolution of the mean squared values $\langle m_i^2\rangle$ over all lattice sites along the line $D_2=\sqrt{2} D_{1}$, as given by the linear ansatz predictions of Eq.~(\ref{eq:p1xAnsatzfieldsquared}) with $\delta$ and $\xi$ determined by a minimization of Eq.~(\ref{eq:p1x-Euc})].
}\label{fig:mip1x}
\end{figure}

We now turn to the $p1x$ phase, depicted by green color in Fig.~\ref{fig:phasediag}. We shall focus on the positive $D_{1,2}$ region, as the negative one can be obtained by the duality transformation $\mc{M}_{xy}$.
Figure~\ref{fig:p1x-panel} summarizes the main features for a representative point $(D_1,D_2)=(0.01,0.0141)$, as found numerically on a cluster with $N=160000$ spins, with spanning vectors ${\bf T}_1\!=\!200({\bf x}+{\bf y})$ and ${\bf T}_2\!=\!200({\bf x}-{\bf y})$. This cluster accommodates the closest commensurate approximant of the phase at this parameter point. The main features are as follows:

{\cbl{\it i) Local spin structure:}}
Figures~\ref{fig:p1x-panel}\,(a) and (b) show the (x-y,x+y) and (x-y,z) view, respectively, of 1/5 of the period of the state, along the propagation direction ${\bf x}$-${\bf y}$. 
As before, red and blue depict the directions of the A- and B-sublattice spins, respectively, while the larger, magenta arrows drawn at the centers of the tetrahedra depict the direction of the total moment ${\bf S}_t$. 
As seen in Fig.~\ref{fig:p1x-panel}\,(a), the local structure matches that of of the parent $(\pi,\pi)$ manifold, with the characteristic all-in/all-out pattern of the in-plane components and the N\'eel-like pattern of the out-of-plane components. 
The local similarity with the parent $(\pi,\pi)$ manifold is also seen in the symmetry-resolved fields contained in the $p1x$ state. Indeed, as shown in Fig.~\ref{fig:mip1x}, the dominant fields are precisely the ones of the parent $(\pi,\pi)$ manifold of Eq.~(\ref{eq:miparentAllInAllout}), see further discussion in Sec.~\ref{sec:LinearAnsatzp1x}.

{\cbl{\it ii) Global two-plane \& `Butterfly' structure:}}
The eigenvalues of the spin inertial tensors $\mc{I}^{\text{A}/\text{B}}$ 
for the state of Fig.~\ref{fig:p1x-panel} are 
\be
(\lambda^\mu_{1},\lambda^\mu_{2},\lambda^\mu_{3}) = (0.006, 0.604, 0.390)\,,
\ee
for both $\mu=$A, B. This means that the spins of each sublattice rotate in specific planes. 
The eigenvectors of $\mc{I}^{\text{A}/\text{B}}$ take the general form of Eq.~(\ref{eq:evecsIAIBTwistedJD1p}) with $\chi\simeq0.46\pi$, close to $\pi/2$. So, the plane of the A-spins is close to the $xz$-plane and the plane of the B-spins is close to the $yz$-plane. 
This is consistent with the orientation of the spins in the cross phase of the single-tetrahedron problem.
The almost planar structure of each sublattice is also corroborated by the common-origin plots in Fig.~\ref{fig:p1x-panel}\,(c-d), which show the rotation of the A spins in the $xz$ plane, and of the B spins in the $yz$ plane.

While the spins of each sublattice seem to form {\it co-rotating spirals} in Fig.~\ref{fig:p1x-panel}\,(b), the spirals are, in fact, not entirely perfect helices. Indeed, as shown in Fig.~\ref{fig:p1x-panel}\,(c-d), the spins form a peculiar  `butterfly' structure with two main features: i) the $x$ (respectively, $y$) component of the A (B) spins, has a larger weight compared to the $z$ component. This is also reflected in the deviation of $\lambda_2/\lambda_3$ from one. 
ii) The angle of rotation $\varphi({\bf R})$ is not perfectly linear, but shows a jump somewhere in the middle of the period, see Fig.~\ref{fig:p1x-panel}\,(g).
These jumps will be discussed further in Sec.~\ref{sec:solitonsp1x}.

{\cbl{\it iii) Ferromagnetic canting:}}
As shown by the magenta arrows in Figs.~\ref{fig:p1x-panel}\,(a-b), the $p1x$ phase, unlike the parent $(\pi,\pi)$ states, features a nonzero ferromagnetic moment $S_t^z$ in each tetrahedron, which is inevitably present due to frustration (as is the case for the cross state of the single tetrahedron). 
Specifically, in one period of the modulation, the in-plane components of the spins rotate between the all-in and the all-out pattern. In doing so, the total spin ${\bf S}_t$, rotates from $+{\bf z}$ to $-{\bf z}$, consistent with the correlations between in-plane and out-of-plane components featured by the cross state of the single-tetrahedron problem. 
We note further that, unlike the twisted-$JD_1^\pm$ phases, where ${\bf S}_t$ features a density-wave-like modulation, here the total moment shows a helical-like rotation in the $(x-y,z)$ plane.

{\cbl{\it iv) 1D character of the modulation:}}
As shown in Figs.~ \ref{fig:p1x-panel}\,(e-f), the sublattice spin structure factors $|{\bf S}_{{\bf k},A}|^2$ and $|{\bf S}_{{\bf k},B}|^2$ feature two dominant peaks (carrying 98\% of the total intensity), centered at $\pm(\pi\!-\!\delta, -\pi\!+\!\delta)$ with $\delta$ small.
This is consistent with the phase of Fig.~\ref{fig:p1x-panel} being an incommensurate 1D modulation that is spawn out of the parent $(\pi,\pi)$ manifold. 
Furthermore, as the $p1x$ state breaks the four-fold symmetry of the model, there is an additional classical ground state with ordering wavevector $(\pi\!-\!\delta,\pi\!-\!\delta)$, along the direction ${\bf x}\!+\!{\bf y}$.

\subsubsection{The linear ansatz approximation}\label{sec:LinearAnsatzp1x}
From the structure factors ${\bf S}_{{\bf k},A}$, ${\bf S}_{{\bf k},B}$ we can derive, using inverse Fourier Transform, the following approximate ansatz for the $p1x$ phase:
\be\label{eq:p1x-ansatz}
{\bf S}_{{\bf R},\mu}\!\simeq\!\cos\big({\bf K}\!\cdot\!{\bf R}\!+\!\varphi({\bf R})\big) {\bf v}_{\mu} \!+\! \sin\big({\bf K}\!\cdot\!{\bf R}\!+\!\varphi({\bf R})\big) {\bf z} \,,
\ee
where $\mathbf{v}_{A}$, $\mathbf{v}_{B}$ take the form 
\be\label{eq:twisted-AllinAllout-va-vb-vectors}
{\bf v}_{\text{A}}=\cos\frac{\xi}{2}~{\bf x} - \sin\frac{\xi}{2}~{\bf y}\,,~~
{\bf v}_{\text{B}}=\sin\frac{\xi}{2}~{\bf x} - \cos\frac{\xi}{2}~{\bf y}\,,
\ee
and the angle $\varphi({\bf R})$ is approximated by 
\be\label{eq:phivsRp1x}
\varphi({\bf R})\simeq-\bs{\delta}\cdot{\bf R}+\varphi_{0}\,.
\ee

According to this ansatz, the $p1x$ phase features the three fields of the parent $(\pi,\pi)$ manifold of Eq.~(\ref{eq:miparentAllInAllout}), as well as three additional, (but much weaker for realistic values of $D_{1,2}$) fields: 
\be\label{eq:TwistedAllinAlloutfields}
\renewcommand\arraystretch{1.5}
\!\!\!\begin{array}{l}
m_{\mc{A}_{2a}}({\bf R})=\cos\frac{\delta}{2} \cos\frac{\xi}{2}~c_{\bf R}\,,
\\
m_{\mc{E}_b,2}({\bf R})=\frac{1}{\sqrt{2}}   [-\cos\frac{\delta}{2}+\sin\frac{\delta}{2}\cos\frac{2\xi+\pi}{4}]~s_{\bf R}\,,
\\
m_{\mc{E}_c,2}({\bf R})=\frac{1}{\sqrt{2}} [ \cos\frac{\delta}{2}+\sin\frac{\delta}{2} \cos\frac{2\xi+\pi}{4}]~s_{\bf R}\,,\\
\hline
m_{\mc{A}_{2b}}({\bf R})= \sin\frac{\delta}{2}~c_{\bf R}\,,
\\
m_{\mc{B}_2}({\bf R})=-\cos\frac{\delta}{2}\sin\frac{\xi}{2}~c_{\bf R}\,,
\\
m_{\mc{E}_a,2}({\bf R})=\sin\frac{\delta}{2} \cos\frac{2\xi-\pi}{4}~s_{\bf R}\,,
\end{array}
\ee
where, as before, ${\bf R}=n_1 {\bf a}_1+ n_2 {\bf a}_2$ specifies the position of the A$_1$ site of the tetrahedron (see site labeling convention in Fig.~\ref{fig:checkerboard-lattice-b}), and we have introduced the quantities 
\be 
\renewcommand\arraystretch{1.5}
\!\!\!\begin{array}{c}
c_{\bf R}=(-1)^{n_1+n_2}\cos\gamma_{\bf R},~~ 
s_{\bf R}=(-1)^{n_1+n_2}\sin\gamma_{\bf R},\\
\gamma_{\bf R}=\delta (1/2+n_1-n_2)-\varphi_0\,.
\end{array}
\ee
The mean squared values of these fields over the lattice are 
\be\label{eq:p1xAnsatzfieldsquared}
\renewcommand\arraystretch{1.5}
\begin{array}{l}
\langle m_{\mc{A}_{2a}}^2\rangle = \frac{1}{2}\cos^2\frac{\delta}{2} \cos^2\frac{\xi}{2},\\
\langle m_{\mc{E}_{b},2}^2\rangle = \frac{1}{4}[-\cos\frac{\delta}{2}+\sin\frac{\delta}{2}\cos\frac{2\xi+\pi}{4}]^2\,,\\
\langle m_{\mc{E}_{c},2}^2\rangle = \frac{1}{4}[\cos\frac{\delta}{2}+\sin\frac{\delta}{2}\cos\frac{2\xi+\pi}{4}]^2\,,\\
\hline
\langle m_{\mc{A}_{2b}}^2\rangle = \frac{1}{2}\sin^2\frac{\delta}{2},\\
\langle m_{\mc{B}_{2}}^2\rangle = \frac{1}{2}\cos^2\frac{\delta}{2} \sin^2\frac{\xi}{2},\\
\langle m_{\mc{E}_{a},2}^2\rangle = \frac{1}{2}\sin^2\frac{\delta}{2} \cos^2\frac{2\xi-\pi}{4}\,.
\end{array}
\ee
Their evolution along the line $D_2\!=\!\sqrt{2}D_1$ is shown in Fig.~\ref{fig:mip1x}\,(b), where $\delta$ and $\xi$ have been determined by minimizing the linear ansatz prediction for the energy given in Eq.~(\ref{eq:p1x-Euc}) below. Note that for $\delta\to0$ and $\xi\to0$, these values tend to the averages of Eq.~(\ref{eq:miparentAllInAlloutSquared}) over all possible $\varphi_0$, 
\be\label{eq:p1xAnsatzfieldsquaredlimit}
\renewcommand\arraystretch{1.5}
\begin{array}{l}
\langle m_{\mc{A}_{2a}}^2\rangle \to \frac{1}{2}\,,~
\langle m_{\mc{E}_{b},2}^2\rangle \to \frac{1}{4}\,,~
\langle m_{\mc{E}_{c},2}^2\rangle \to \frac{1}{4}\,,\\
\langle m_{\mc{A}_{2b}}^2\rangle \to 0\,,~
\langle m_{\mc{B}_{2}}^2\rangle \to 0\,,~
\langle m_{\mc{E}_{a},2}^2\rangle \to 0\,,
\end{array}
\ee
since the $p1x$ state effectively combines all members of the parent $(\pi,\pi)$ manifold.

Additionally, we see that, for realistic values of $D_{1,2}$, the dominant symmetry-resolved fields are the ones of the parent $(\pi,\pi)$ manifold, namely $m_{\mc{A}_{2a}}$, $m_{\mc{E}_{b},2}$ and $m_{\mc{E}_{c},2}$. The latter two are, in particular, almost opposite to each other, consistent with Eq.~(\ref{eq:miparentAllInAllout}).
Among the weaker fields, $m_{\mc{A}_{2b}}$ (and $m_{\mc{B}_2}$) seems to track the behaviour of $m_{\mc{A}_{2a}}$, which can be traced back to the cross-coupling term $\propto d_+ m_{\mc{A}_{2b}} m_{\mc{A}_{2b}}$ in Eq.~(\ref{eq:crosscouplings}).
Likewise, the field $m_{\mc{E}_{a},2}$ seems to track the behaviour of $m_{\mc{E}_{c},2}\approx -m_{\mc{E}_{b},2}$, which can be traced back to the cross-coupling term $\propto d_+ {\bf m}_{\mc{E}_{a}}\cdot({\bf m}_{\mc{E}_{b}}-{\bf m}_{\mc{E}_{c}})$  in Eq.~(\ref{eq:crosscouplings}). Both $m_{\mc{A}_{2b}}$ and $m_{\mc{E}_{a},2}$ [which are the two fields that cost Heisenberg energy, see Eq.~(\ref{eq:crosscouplings})] remain very weak, however, up to unrealistically large values of DM interactions, as shown in Fig.~\ref{fig:mip1x}\,(b). This is consistent with the total moment of each tetrahedron being proportional to the small quantity $\sin(\delta/2)$:
\be
\begin{array}{l}
{\bf S}_t({\bf R})=
 4 \sin\frac{\delta}{2}
 \Big[
 \cos\frac{2\xi-\pi}{4} s_{\bf R}~\frac{{\bf x}-{\bf y}}{\sqrt{2}}
+ c_{\bf R}~{\bf z}
\Big]\,.
\end{array}
\ee
This equation further shows that the total moment rotates in the $(x-y,z)$ plane, again in agreement with the numerical results, see Fig.~\ref{fig:AllInAllOut}\,(b).

{\cbl{\it Obtaining estimates for $\delta$ and $\xi$:}} 
As we did for the linear ansatz of the twisted $JD_1^+$ state, we can determine analytically the energy per unit cell, $E_{uc}$, in the thermodynamic limit by summing the energy contribution of each tetrahedron over ${\bf R}$, and then use the resulting expressions to obtain estimates for $\delta$ and $\xi$. 
We find 
\begin{align}
\label{eq:p1x-Euc}
E_{uc}&= 1-\cos \delta(3+\sin \xi) + \sin \xi  \notag \\
& - \sin \delta \Big\{\sqrt{2}D_{1}\big(\cos\frac{\xi}{2}+\sin \frac{\xi}{2} \big) + 2 D_{2}\cos \frac{\xi}{2}\Big\}, 
\end{align}
which, as in the twisted $JD_1^+$ phase, does not depend on $\varphi_0$, because the latter is a free parameter characterising a global shift of the IC phase along the propagation direction.

\begin{figure}[!t]
\subfloat{\includegraphics[width=0.85\linewidth]{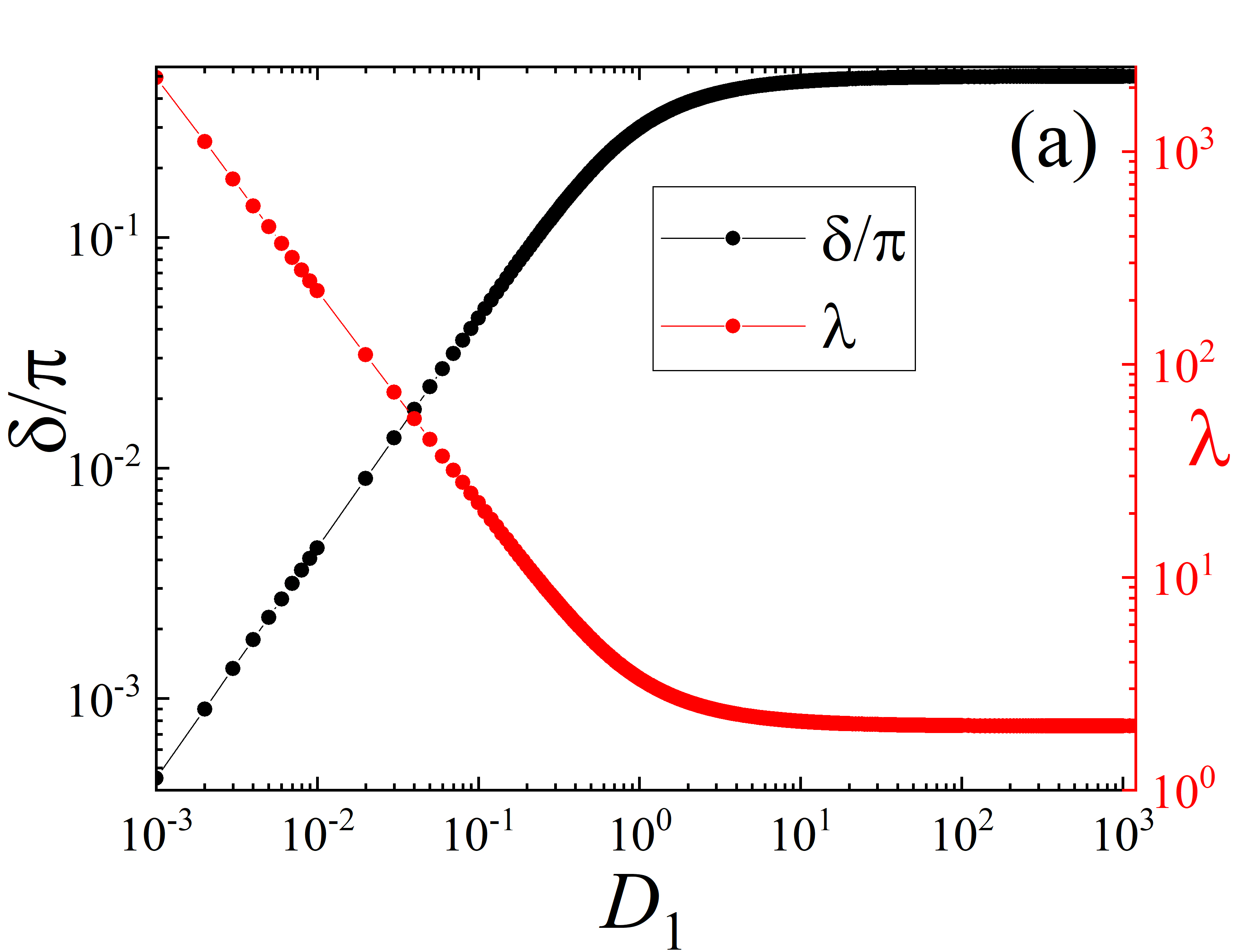}\label{fig:lambdap1xa}}\\
\subfloat{\includegraphics[width=0.85\linewidth]{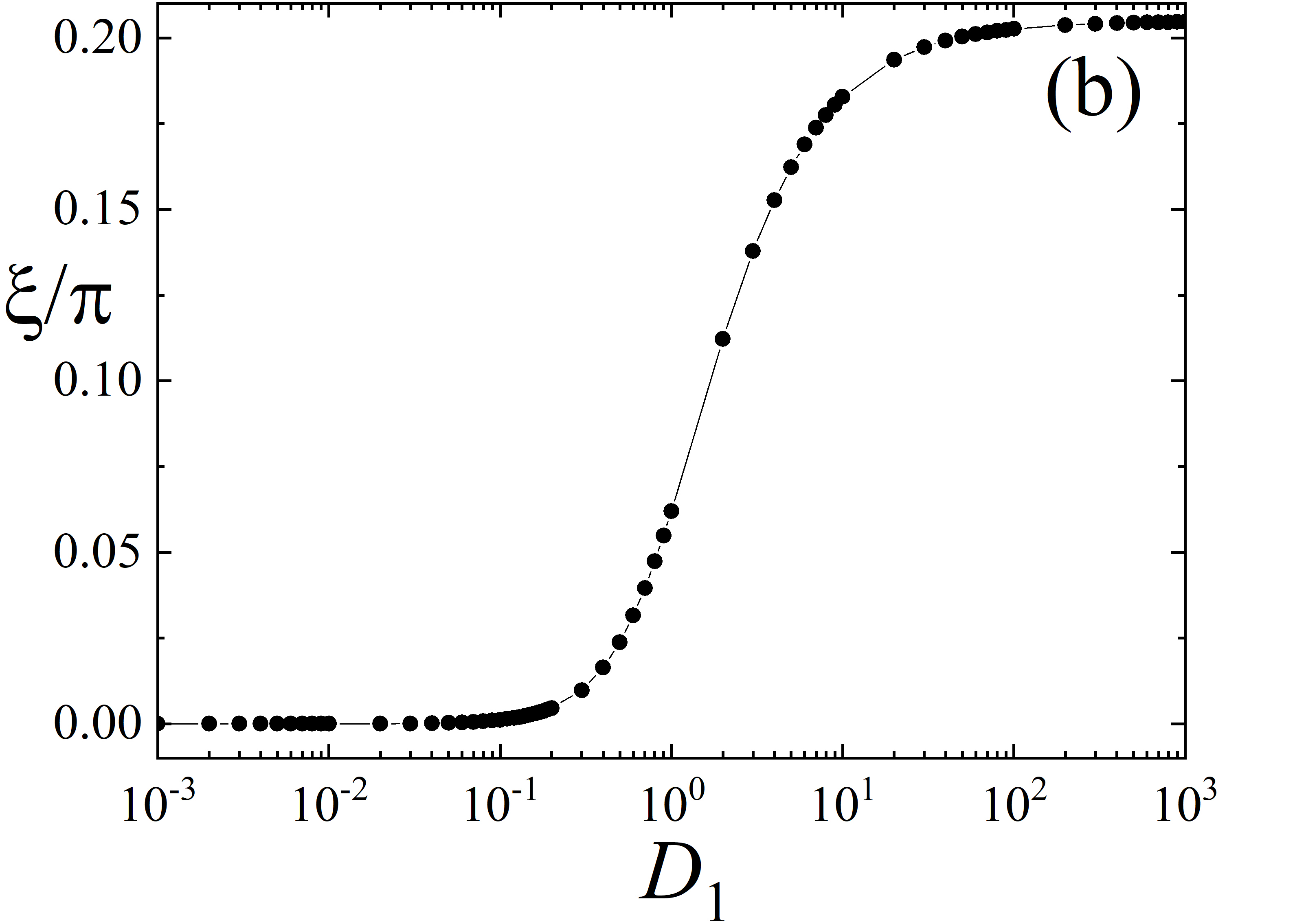}\label{fig:lambdap1xb}}
\caption{(Color online) Evolution of $\delta$ (a),  the corresponding cluster size $\lambda$ (a) and the angle $\xi$ (b) with $D_1$ along the line $D_{2}=\sqrt{2} D_{1}$. }\label{fig:lambdap1x}
\end{figure}

A numerical minimization of $E_{uc}$ with respect to $\xi$ and $\delta$ delivers the plots shown in Fig.~\ref{fig:lambdap1x}, as a function of $D_1$, along the line $D_2=\sqrt{2}D_1$, which is within the $p1x$ phase. 
As expected, $\delta$ drops (and correspondingly, the optimal wavelength $\lambda$ grows) as we approach the isotropic limit, which is consistent with the commensurate-incommensurate transition scenario discussed in Sec.~\ref{sec:solitonsp1x}.
Moreover, the angle $\xi$ of the linear ansatz, which quantifies the deviation from the parent $(\pi,\pi)$ manifold, drops to zero at the isotropic limit.

In order to address the limitations of the linear ansatz of the $p1x$ phase, we first note that the minimum energy $E_{uc}$ obtained by the linear ansatz is slightly higher than the one obtained from numerics. 
This stems from the fact that the linear ansatz misses the abrupt jumps seen in Fig.~\ref{fig:p1x-panel}\,(g) and in Fig.~\ref{fig:mip1x}\,(a). It also misses the weak non-coplanarity evidenced in the common origin plots of Fig.~\ref{fig:p1x-panel}\,(c-d) and the deviation of $\lambda_2/\lambda_3$ from one. 
Finally, as in the case of the twisted $JD_1^\pm$ phases, the linear ansatz does not satisfy local stability exactly, and misses the (much weaker) higher harmonics of the spin structure factors.

\subsubsection{C-I transition for the $p1x$ phase}
\label{sec:solitonsp1x}
Let us return to the peculiar butterfly structure of the two spin sublattices in the $p1x$ phase, and, in particular,  to the abrupt jumps that are evident in Fig.~\ref{fig:p1x-panel}\,(g) as well as in the evolution of the symmetry-resolved fields in Fig.~\ref{fig:mip1x}\,(a).  

Similarly to the discussion of Sec.~\ref{sec:solitonsJD1p}, it is again expected that these jumps tend to become more and more pronounced as we approach the isotropic limit.
In this scenario, the linear ansatz would break down and the angle $\varphi$ would show a piece-wise constant  behaviour of a soliton lattice. The value of $\varphi$ would be constant inside `domains' of (any of) the parent $(\pi,\pi)$ states, separated by abrupt jumps that play the role of `domain walls' (or `discommensurations') with a negative energy density $\propto d_+$. The distance between the domain walls would then diverge in the isotropic limit $d_+=0$, leaving a single domain of the parent phase.
As in the corresponding discussion in Sec.~\ref{sec:solitonsJD1p}, checking this scenario explicitly requires studying cluster sizes that are far too large for an efficient minimization. This relies on the linear ansatz prediction that the optimal wavelength (or the distance between discommensurations in the nonlinear regime) grows very fast on approaching the isotropic limit, see Fig.~\ref{fig:lambdaJD1p}\,(a).

\section{Full lattice model,  II: Multi-domain phases} \label{sec:full-model-phasesII}
We now turn to the multi-domain phases $p11$, $p31$,  $p11+p31$ and $p13$, depicted by red, blue, light grey and green colors respectively in Fig.~\ref{fig:phasediag}. 
These phases appear around the special line $D_2\!=\!\sqrt{2}D_1$ and are sandwiched between the more extended, 1D modulated phases discussed above. 
Among these, the stability region of $p11$ is much more more extended than the remaining three phases, and even more extended compared to the IC $p1x$ phase. 

\begin{figure*}[!t]
\includegraphics[width=\textwidth]{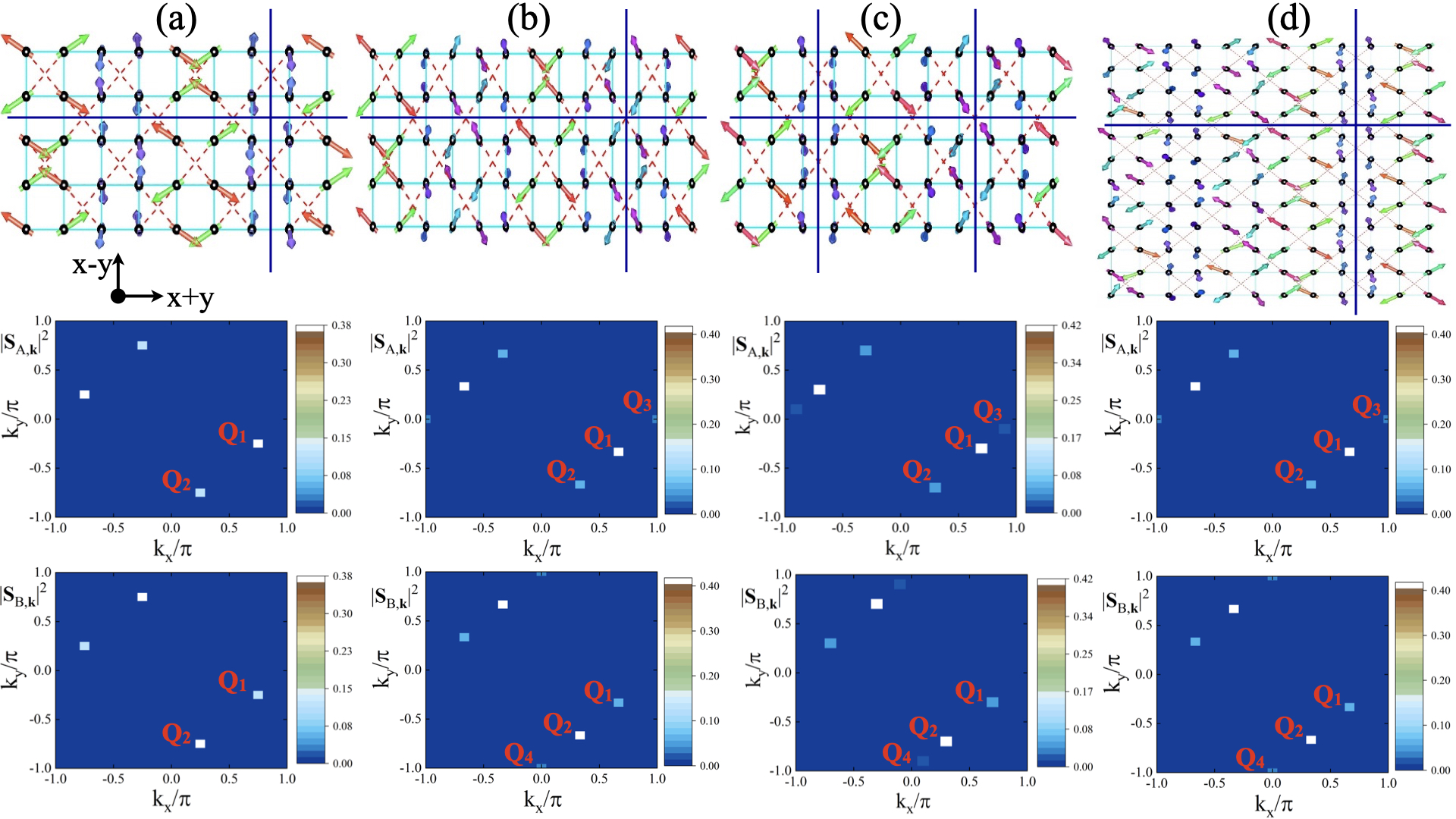} 
\caption{(Color online) Local view of various multi-domain phases of the model (top row), and the corresponding static spin structure factors of the two sublattices,  $|\mathbf{S}_{\mathbf{k},A}|^{2}$ and $|\mathbf{S}_{\mathbf{k},B}|^{2}$ (bottom row): 
(a) $p11$ phase at the representative point $(D_1,D_2)=(0.3,0.25)$, (b) $p31$ phase at $(D_1,D_2)=(0.2,0.45)$, (c) $p11+p31$ phase at $(D_1,D_2)=(0.2,0.38)$, and (d) $p13$ phase at $(D_1,D_2)=(0.5,0.5)$.  Lines in the top row show the $(x\pm y,z)$ reflection planes discussed in the text.}\label{fig:p-phases-panel} 
\end{figure*}

Moreover, the multi-domain states break the $C_4$ rotational symmetry, but retain the two reflection symmetries $\sigma_{x\pm y,z}$, through the planes $(x\pm y,z)$   (crossing the centers of a subset of tetrahedra, see, e.g., lines in the upper panels of Fig.~\ref{fig:p-phases-panel}).
This is reflected in the characteristic spin structure factor patterns as discussed below.

{\cbl (i) Multi-domain structure:} Figure~\ref{fig:p-phases-panel} depicts the spin profiles and corresponding static spin structure factor of some representative points within each of these phases. 
According to these profiles, these phases show two types of alternating domains that somewhat resemble the structure of the `twisted $JD_1^\pm$' (spins shown in nearly blue colors) and the all-in/all-out arrangement (spins shown in red and green colours), with variable widths. For example,  the `p11' phase (which is the most extended among these phases) features domains with widths $\ell_1\!=\!1$ and $\ell_2\!=\!1$, whereas the narrower `p31' phase has $\ell_1\!=\!3$ and $\ell_2\!=\!1$.
Unlike the 1D modulated phases, therefore, here there are two different modulations: a) the characteristic one-dimensional modulation (along the direction ${\bf x}$-${\bf y}$ in this case) within each domain, and b) the additional modulation in the orthogonal direction (${\bf x}$+${\bf y}$) with the alternation between the two types of domains. 
{\cbl (ii) Static spin structure factors:} It is noteworthy, that while our numerical results show that the multidomain states are commensurate with large magnetic unit cells, the static spin structure factors show a small number of Bragg peaks, suggesting that these phases can be approximated by the use of a few  harmonics.
These characteristic aspects can be summarised as follows.

{\it (a) $p11$ phase:} This phase features a 16-site magnetic unit cell, with spanning vectors 
\be
{\bf T}_1=(a,3a),~~
{\bf T}_2=(-3a,-a)\,,
\ee
and corresponding reciprocal vectors
\be
{\bf G}_1=(-\frac{\pi}{4a},\frac{3\pi}{4a}),~~
{\bf G}_2=(-\frac{3\pi}{4a},\frac{\pi}{4a})\,.
\ee
The static spin structure factors show four main Bragg peaks 
[$\pm{\bf Q}_1$ and $\pm{\bf Q}_2$ in panel (a)], 
with approximate intensities: \be\label{eq:p11-intensities}
\renewcommand\arraystretch{1.25}
\begin{array}{|c|c|c|c|}
\hline
\nu&{\bf Q}_\nu&
|{\bf S}_{\mathbf{Q},A}|^{2} &
|{\bf S}_{{\bf Q},B}|^{2} \\
\hline
1&
{\bf G}_1&
0.38& 0.12 
\\
2&
{\bf G}_2&
0.12&0.38\\
\hline
\end{array}
\ee
which obey the characteristic relations 
\be\label{eq:SofQAvsB1}
\renewcommand\arraystretch{1.25}
\begin{array}{ll}
|S_{{\bf Q}_1,A}|^2\!=\!|S_{{\bf Q}_2,B}|^2\,,
& |S_{{\bf Q}_2,A}|^2\!=\!|S_{{\bf Q}_1,B}|^2\,,
\end{array}
\ee
due to the symmetries $\sigma_{x\pm y,z}$ mentioned above (which swap the two sublattices and map ${\bf Q}$ to ${\bf Q}'\!=\!\bs{\sigma}_{x\pm y,z}\!\cdot\!{\bf Q}$).

{\it (b) $p31$ phase:} This phase features a 24-site magnetic unit cell, with spanning vectors
\be
{\bf T}_1=(2a,-2a),~~
{\bf T}_2=(4a,2a)\,,
\ee
and corresponding reciprocal vectors 
\be
{\bf G}_1=(\frac{\pi}{3a},-\frac{2\pi}{3a}),~~
{\bf G}_2=(\frac{\pi}{3a},\frac{\pi}{3a})\,.
\ee
The static spin structure factors reveal eight (in total) main Bragg peaks ($\pm{\bf Q}_\nu$, $\nu=1-4$) with approximate intensities:
\be\label{eq:p31-intensities}
\renewcommand\arraystretch{1.25}
\begin{array}{|c|c|c|c|}
\hline
\nu &{\bf Q}_\nu&
|{\bf S}_{\mathbf{Q},A}|^{2} &
|{\bf S}_{{\bf Q},B}|^{2} \\
\hline
1&\mathbf{G}_{1}+\mathbf{G}_{2}& 0.41 & 0.06 
\\
2&
\mathbf{G}_{2}&
0.06 & 0.41
\\
3&
\mathbf{G}_{1}+2\mathbf{G}_{2}&
0.03 & 0
\\
4&\mathbf{G}_{1}+\mathbf{G}_{2}&
0&0.03\\
\hline
\end{array}~~
\ee
which obey the characteristic relations 
\be\label{eq:SofQAvsB2}
\renewcommand\arraystretch{1.25}
\begin{array}{ll}
|S_{{\bf Q}_1,A}|^2\!=\!|S_{{\bf Q}_2,B}|^2\,,
& |S_{{\bf Q}_2,A}|^2\!=\!|S_{{\bf Q}_1,B}|^2\,,
\\
|S_{{\bf Q}_3,A}|^2\!=\!|S_{{\bf Q}_4,B}|^2\,, 
& |S_{{\bf Q}_4,A}|^2\!=\!|S_{{\bf Q}_3,B}|^2\,,
\end{array}
\ee
due to the $\sigma_{x\pm y,z}$ symmetries.

{\it (c) $p11$+$p31$ phase:} This phase features a 40-site magnetic unit cell, with spanning vectors
\be
{\bf T}_1=(2a,-2a),~~
{\bf T}_2=(7a,3a)
\ee
and corresponding reciprocal vectors
\be
{\bf G}_1=(\frac{3\pi}{10a},-\frac{7\pi}{10a}),~~
{\bf G}_2=(\frac{\pi}{5a},\frac{\pi}{5a})
\ee
As in $p31$, the spin structure factors reveal eight main Bragg peaks ($\pm{\bf Q}_\nu$, $\nu=1-4$) with approximate intensities:
\be
\label{eq:p11p31-intensities}
\renewcommand\arraystretch{1.25}
\begin{array}{|c|c|c|c|}
\hline
\nu &{\bf Q}_\nu&
|{\bf S}_{\mathbf{Q},A}|^{2} &
|{\bf S}_{{\bf Q},B}|^{2} \\
\hline
1&
2\mathbf{G}_{1}+\mathbf{G}_{2}&
0.42&0.05
\\
2&
\mathbf{G}_{2}&
0.05&0.42
\\
3&
3\mathbf{G}_{1}+\mathbf{G}_{2}&
0.02 & 0
\\
4&
-\mathbf{G}_{1}+\mathbf{G}_{2}&
0&0.02\\
\hline
\end{array}
\ee
which obey Eq.~(\ref{eq:SofQAvsB2}) above, due to the $\sigma_{x\pm y,z}$ symmetries.

We also note that here the main (visible) peaks give approximately 98\% of the total intensity, with the remaining 2\% goes into a number of much weaker peaks, that are not visible in the static structure factor profiles. 

{\it (d) $p13$ phase:} As shown in Fig.~\ref{fig:phasediag}, the $p13$ phase appears for relatively large values of $D_{1,2}/J$ and is therefore less likely to be realised.
This phase features a very large magnetic unit cell, with 48 sites and spanning vectors
\be
{\bf T}_1=(2a,-2a),~~
{\bf T}_2=(6a,6a)
\ee
and corresponding reciprocal vectors
\be
{\bf G}_1=(\frac{\pi}{2a},-\frac{\pi}{2a}),~~
{\bf G}_2=(\frac{\pi}{6a},\frac{\pi}{6a})
\ee
The static spin structure factors reveal eight main Bragg peaks ($\pm{\bf Q}_\nu$, $\nu=1-4$) with approximate intensities
\be\label{eq:p13-intensities}
\renewcommand\arraystretch{1.25}
\begin{array}{|c|c|c|c|}
\hline
\nu&{\bf Q}_\nu&
|{\bf S}_{\mathbf{Q},A}|^{2} &
|{\bf S}_{{\bf Q},B}|^{2} \\
\hline
1&
\mathbf{G}_{1}+2\mathbf{G}_{2}&
0.33 &0.10
\\
2&
\mathbf{G}_{1}-2\mathbf{G}_{2}&
0.10 &0.33
\\
3&
\mathbf{G}_{1}+3\mathbf{G}_{2}&
0.02 &0
\\
4&
\mathbf{G}_{1}-3\mathbf{G}_{2}&
0&0.02\\
\hline
\end{array}
\ee
which again obey Eq.~(\ref{eq:SofQAvsB2}) above, due to the $\sigma_{x\pm y,z}$ symmetries.

Here, the missing intensity corresponding to the much weaker Bragg peaks that are not visible in the structure factor profiles amounts to 10\%.

\begin{figure}[!t]
\includegraphics[width=0.95\linewidth]{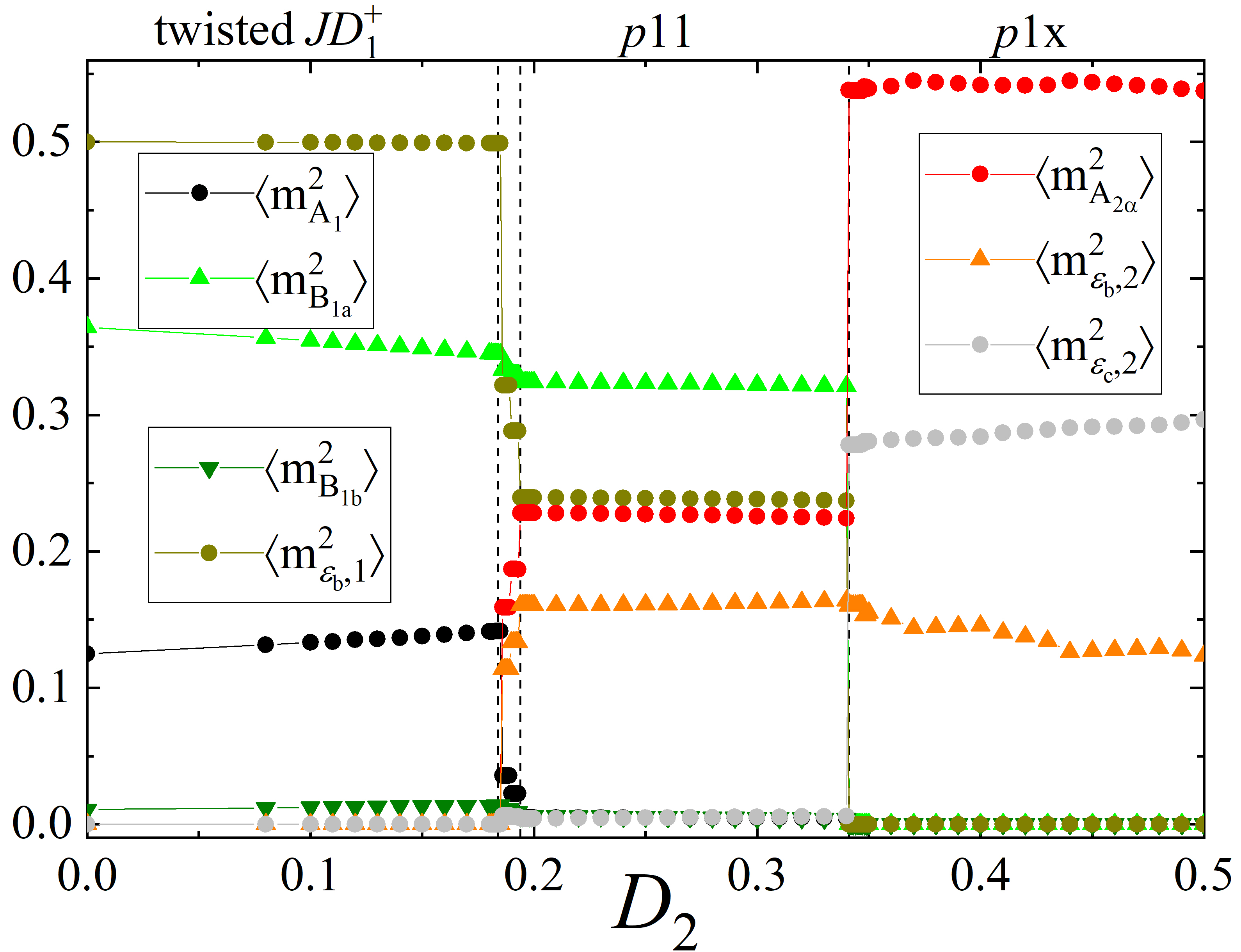}\label{fig:phase-ops}
\caption{(Color online) Evolution of the mean squared symmetry-resolved fields $m^{2}_{\mathcal{A}_{1}}$, $m^{2}_{\mathcal{B}_{1a}}$, $m^{2}_{\mathcal{B}_{1b}}$, and $m^{2}_{\mc{E}_b,1}$ with $D_2$ for $D_1\!=\!0.3$, as the system crosses from the `twisted $JD_1^\pm$' to the $p11$ and then to the $p1x$ phase.}\label{fig:phase-trans1}
\end{figure}

{\cbl (iii) Symmetry-resolved fields:}
The resemblance of the two types of domains with the local structure of the `twisted $JD_1^\pm$' and the all-in/all-out pattern is reflected in the analysis of the symmetry-resolved fields. 
Figure~\ref{fig:phase-trans1} shows, for example, the evolution of the mean squared symmetry-resolved fields $m^{2}_{\mathcal{A}_{1}}$, $m^{2}_{\mathcal{B}_{1a}}$, $m^{2}_{\mathcal{B}_{1b}}$, and $m^{2}_{\mc{E}_{b},1}$ across a representative cut in the phase diagram, as the system crosses from the `twisted $JD_1^\pm$' to the $p11$ and then to the $p1x$ phase.
We see that the $p11$ phase shows an appreciable weight for the symmetry-resolved fields $m_{\mc{E}_b,1}$ and $m_{\mc{B}_{1a}}$, which are among the dominant fields of the twisted $JD_1^\pm$ phase, as well as the fields $m_{\mc{A}_{2a}}$ and $m_{\mc{E}_b,2}$ which are among the dominant fields of the $p1x$ phase. The $p11$ phase contains other fields as well, whose weights are however much weaker. Among these are, for example, the fields $m_{\mc{B}_2}$ and $m_{\mc{A}_{2b}}$ (the latter is one the two fields that cost Heisenberg energy).

\section{Summary and discussion} \label{sec:summary}

In this work, we have presented a novel mechanism of generating multi-$Q$ magnetic states via the synergy of robust frustration and the presence of chiral interactions. 
We have elucidated how this synergy arises in the planar pyrochlore antiferromagnet, one of the paradigmatic models of frustration in two spatial dimensions, within a minimal model with $C_{4v}$ symmetry. In this model, the chiral interactions arise from the in-plane DM anisotropy allowed by the $C_{4v}$ symmetry, whereas the robust frustration is identified around the special line $D_2=\sqrt{2}D_1$ of the parameter space. 
Our combined analytical and numerical results demonstrate how these two ingredients --frustration and chiral interactions-- give rise to a rich phase diagram, with incommensurate (IC) one-dimensional modulated phases, as well as a cascade of commensurate, multi-domain phases with large magnetic unit cells. 

For the extended IC phases (twisted-$JD_1^\pm$ and $p1x$), we have identified their parent phases and elucidated the respective commensurate-incommensurate (C-I) transitions as the mechanism that drives these phases. We provided linear ansatz approximations which work well in the majority of the stability region of the IC phases, except sufficiently close to the C-I transitions, where non-linearities become more pronounced. In the latter regions, the linear ansatze provided clear insights on the way the wavelength of the IC modulation diverges on approaching the C-I transitions, which in turn was used to select appropriate finite-size clusters in our numerics.
Finally, we analysed the main local and global aspects of these phases in detail, using a number of diagnostic tools, including spin structure factors and symmetry-resolved fields. 

A similar in-depth analysis is given for the multi-domain phases. Here, the spin profiles show spatially intertwined domains of the underlying competing phases of the highly frustrated point, whereas the spin structure factors reveal a characteristic multi-$Q$ pattern, with most of the intensity concentrated on 4-6 main Bragg peaks.
The overall structure of the multi-domain phases suggests that the relevant parameter region around the special line $D_2=\sqrt{2}D_1$ may actually consist of many more such phases, where the evolution of the relative widths of the two domains is such that the system interpolates between the twisted $JD_1^\pm$ (where $\ell_1\to\infty$ and $\ell_2\to 0$) and the $p1x$ phase (where $\ell_1\to0$ and $\ell_2\to \infty$). Our numerical results, however, do not reveal phases with $\ell_{1,2}$ larger than 3, meaning that, if present, the stability region of such phases are too narrow to be identified numerically. 

We should comment here that, while we have focused on the classical limit, we anticipate that much of the phase diagram will broadly survive in the presence of quantum fluctuations, because these are generally mitigated by anisotropic terms in the Hamiltonian. 
An exception is the neighborhood of the isotropic point, where quantum effects are believed (see e.g., \cite{BrenigHonecker2002,Fouet2003,Moessner2004,Chan2011,Bishop2012} and refs.~therein) to select a valence bond crystal with a nonzero excitation gap. One would then expect that this quantum phase shall remain stable under weak enough anisotropies, and even more so along the special line $D_2\!=\!\sqrt{2}D_1$, where the DM anisotropy does not influence the energy at the mean-field level to leading order in $D/J$. 

Finally, as mentioned in the introduction, this study can be relevant to a number of layered checkerboard antiferromagnets, but also a series of 3D pyrochlores, which can naturally satisfy the requirement of perfect frustration (in the absence of anisotropy) by symmetry, without fine tuning. 
The distinctive fingerprints
of the various phases reported here can be used for the experimental detection of these phases by various probes, including, for example, neutron diffraction and local spectroscopic probes such as NMR and $\mu{SR}$. The latter, in particular, can detect the wide distributions of local magnetic moments associated with the characteristic nonlinearities that are inherently present in both the incommensurate and the commensurate multi-domain phases, as analyzed above. In addition, the thorough understanding of the effects of DM interaction and the distinct phases, is particularly important when these influence or drive other non-magnetic phases, e.g. in multiferroic materials \cite{Mostovoy,Betouras}.

\vspace*{0.2cm}
\begin{center}{\bf Acknowledgements}\end{center}
We thank Natalia Perkins, Yang Yang, Alexander Tsirlin, Johannes Richter and Damian Farnell for fruitful discussions. This work has been supported by the Engineering and Physical Sciences Research Council Grants No. EP/V038281/1, EP/P002811/1 and EP/T034351/1. We acknowledge the use of the Lovelace HPC service at Loughborough University.

\vspace*{0.3cm}
\appendix

\begin{center}
{\bf APPENDIX}
\end{center}
\vspace*{-0.25cm}

\section{Insights from the single tetrahedron problem} \label{sec:single-tetra-app}

\subsection{Coupling matrix}
The Hamiltonian of the single tetrahedron problem (see Fig.~\ref{fig:checkerboard-lattice-b})) can be written in the compact form 
\be\label{eq:single-tetra-LT-Ham}
\mc{H} = \frac{1}{2} \mathbfcal{S}^{T} \cdot \boldsymbol \Lambda \cdot \mathbfcal{S}\,
\ee
where $\mathbfcal{S}$ is the twelve-dimensional vector 
\be
\!\!\!\!\mc{S}^T\!\!=\!\!( A_1^x, A_1^y, A_1^z, A_2^x, A_2^y, A_2^z, B_1^x, B_1^y, B_1^z, B_2^x, B_2^y, B_2^z)
\label{eq:single-tetra-LT-spinor-1}
\ee
(we use the notation of Fig.~\ref{fig:checkerboard-lattice-b}, where the four spins of the tetrahedron are labeled as ${\bf A}_1$, ${\bf A}_2$, ${\bf B}_1$ and ${\bf B}_2$), and $\boldsymbol \Lambda$ is the 
matrix 
\small
\be
\bs{\Lambda}\!\!=\!\! \left[
\setlength\arraycolsep{0.1pt}
\begin{array}{ccc|ccc|ccc|ccc}
0& 0& 0& 1& 0& -D_2& 1& 0& -\widetilde{D}_1 & 1&  0 & -\widetilde{D}_1\\
0&  0& 0& 0& 1& 0& 0& 1& -\widetilde{D}_1 & 0& 1& \widetilde{D}_1\\
0& 0& 0& D_2& 0& 1& \widetilde{D}_1& \widetilde{D}_1& 1& \widetilde{D}_1& -\widetilde{D}_1& 1\\
\hline
1&   0& D_2& 0& 0& 0& 1& 0& \widetilde{D}_1& 1& 0& \widetilde{D}_1\\
0& 1& 0& 0&  0& 0& 0& 1& -\widetilde{D}_1& 0& 1& \widetilde{D}_1\\
-D_2& 0& 1& 0& 0&   0& -\widetilde{D}_1& \widetilde{D}_1& 1& -\widetilde{D}_1&-\widetilde{D}_1&   1\\
\hline
1& 0& \widetilde{D}_1& 1& 0& -\widetilde{D}_1& 0& 0& 0& 1& 0& 0\\
0&   1& \widetilde{D}_1& 0& 1& \widetilde{D}_1& 0& 0& 0& 0& 1&   D_2\\
-\widetilde{D}_1& -\widetilde{D}_1& 1& \widetilde{D}_1& -\widetilde{D}_1& 1&   0& 0& 0& 0& -D_2& 1\\
\hline
1& 0& \widetilde{D}_1& 1& 0& -\widetilde{D}_1& 1& 0&   0& 0& 0& 0\\
0& 1& -\widetilde{D}_1& 0& 1& -\widetilde{D}_1& 0& 1& -D_2&   0& 0& 0\\
-\widetilde{D}_1& \widetilde{D}_1& 1& \widetilde{D}_1& \widetilde{D}_1& 1& 0& D_2& 1& 0& 0& 0
\end{array}
\right]\nonumber
\ee
\normalsize
with $\tilde{D}_{1} \equiv D_{1}/\sqrt{2}$.

\subsection{Symmetry-resolved fields}\label{sec:symmetry-analysis}

Let $\bs{\Gamma}$ be the representation of $C_{4\mathrm{v}}$ in the space of all 12$\times$1 vectors $\mathbfcal{S}$ of Eq.~(\ref{eq:single-tetra-LT-spinor-1}). This representation is reducible and can be decomposed into a direct sum of the irreducible representations (irreps) $\bs{\Gamma}_{l}$ of $C_{4\mathrm{v}}$  
\be\label{eq:irrep-decomp1}
\bs{\Gamma}= \sum_l^\oplus
a_{l} \bs{\Gamma}_{l}\,
\ee
where the index $l$ runs over the six irreps of $C_{4\mathrm{v}}$ (which include the four one-dimensional irreps  $\mc{A}_{1}$, $\mc{A}_{2}$, $\mc{B}_{1}$ and $\mc{B}_{2}$, and the  two-dimensional irrep $\mc{E}$) and the multiplicities $a_{l}$ can be found using the formula
\be\label{eq:irrep-decomp2}
a_{l}= \frac{1}{N_{G}}\sum_{g\epsilon C_{4\mathrm{v}}} \chi_{l}^\ast(g) X(g)\,
\ee
where $\chi_{l}(g)$ is the character of the $l$-th irrep, $X(g)=\text{Tr}[\bs{\Gamma}(g)]$ and $N_{G}=8$. 
We find
\be
\bs{\Gamma}= \mc{A}_1 \oplus 2\mc{A}_2 \oplus 2 \mc{B}_1 \oplus \mc{B}_2 \oplus 3 \mc{E}\,
\ee
Next, we search for appropriate symmetrized basis vectors ${\bf v}_i$ and associated `order parameter' fields $m_i$. To this end, we take $\mathbfcal{S}$ and apply the projection operators
\be\label{eq:irrep-decomp3}
{\bf P}_{l}= \frac{1}{N_{G}}\sum_{g\epsilon C_{4\mathrm{v}}} \chi_{l}^\ast(g)~\bs{\Gamma}(g)\,
\ee
We find the following (normalized) basis vectors
\be
\begin{aligned}
\label{eq:irrep-basis-vecs}
\mathbf{v}_{\mc{A}_{1}}&=(0, 1, 0, 0, -1, 0, 1, 0, 0, -1, 0, 0)^{T}/2 \\
\mathbf{v}_{\mc{A}_{2a}}&=(1, 0, 0, -1, 0, 0, 0, -1, 0, 0, 1, 0)^{T}/2 \\
\mathbf{v}_{\mc{A}_{2b}}&=(0, 0, 1, 0, 0, 1, 0, 0, 1, 0, 0, 1)^{T}/2\\
\mathbf{v}_{\mc{B}_{1a}}&=(1, 0, -1, -1, 0, -1, 0, 1, 1, 0, -1, 1)^{T}/\sqrt{8} \\
\mathbf{v}_{\mc{B}_{1b}}&=(1, 0, 1, -1, 0, 1, 0, 1, -1, 0, -1, -1)^{T}/\sqrt{8} \\
\mathbf{v}_{\mc{B}_{2}}&=(0, 1, 0, 0, -1, 0, -1, 0, 0, 1, 0, 0)^{T}/2 \\
\mathbf{v}_{\mc{E}_a,1}&=(1, 1, 0, 1, 1, 0, 1, 1, 0, 1, 1, 0)^{T}/\sqrt{8} \\
\mathbf{v}_{\mc{E}_a,2}&=(1, -1, 0, 1, -1, 0, 1, -1, 0, 1, -1, 0)^{T}/\sqrt{8} \\
\mathbf{v}_{\mc{E}_{b'},1}&=(1, -1, 0, 1, -1, 0, -1, 1, 0, -1, 1, 0)^{T}/\sqrt{8}\\
\mathbf{v}_{\mc{E}_{b'},2}&=(1, 1, 0, 1, 1, 0, -1, -1, 0, -1, -1, 0)^{T}/\sqrt{8}\\
\mathbf{v}_{\mc{E}_{c'},1}&=(0, 0, 1, 0, 0, -1, 0, 0, -1, 0, 0, 1)^{T}/2\\
\mathbf{v}_{\mc{E}_{c'},2}&=(0, 0, 1, 0, 0, -1, 0, 0, 1, 0, 0, -1)^{T}/2
\end{aligned}
\ee
We note that the basis vectors corresponding to the fields defined in Eq.~(\ref{eq:bpcp}) are combinations of $\mathbf{v}_{\mc{E}_{b'}}$ and $\mathbf{v}_{\mc{E}_{c'}}$, 
\be
\mathbf{v}_{\mc{E}_{b},j}=\frac{\mathbf{v}_{\mc{E}_{b'},j}+\mathbf{v}_{\mc{E}_{c'},j}}{\sqrt{2}}\,,~~~
\mathbf{v}_{\mc{E}_{c},j}=\frac{\mathbf{v}_{\mc{E}_{b'},j}-\mathbf{v}_{\mc{E}_{c'},j}}{\sqrt{2}}\,
\ee
where $j=1,2$. 

The basis vectors ${\bf v}_i$ can be used to decompose the general configuration $\bs{\mc{S}}$ as in Eq.~(\ref{eq:spinor-in-basis}), with fields $m_i$ given by Eq.~(\ref{eq:order-params}). The latter can, in turn, be used to re-write the single-tetrahedron Hamiltonian as in Eq.~(\ref{eq:crosscouplings}).

\subsection{Luttinger-Tisza method for a single tetrahedron}\label{app:LTsingleTetrahedron}
Here we discuss the classical ground state phase diagram of the single-tetrahedron problem. To this end, we use the Luttinger-Tisza method~\cite{Luttinger, Bertraut,Litvin,Kaplan}, which,  in this problem, is able to deliver the exact classical ground states. 
This method consists in replacing the problem of minimizing the classical energy under the $N$ `hard' spin length constraints ${\bf S}^{2}_{i}\!=\! S^2$, $i=1$-$N$ (here $N\!=\!4$) with the much easier problem of minimizing the classical energy under a single, `soft' constraint $\sum_{i} {\bf S}^{2}_{i}\!=\!N S^2$. 
This problem amounts to minimizing  $\mc{F}\!=\!\mc{H}\!-\! \frac{1}{2}\lambda \big( \sum_{i} {\bf S}^{2}_{i}-4 S^{2}\big)$, with $\mc{H}$ given by Eq.~(\ref{eq:single-tetra-LT-Ham}). This leads to the eigenvalue problem of the coupling matrix $\bs{\Lambda}$,
\be\label{eq:single-tetra-LT-eigen}
\bs{\Lambda}\cdot \mathbfcal{S}=\lambda \mathbfcal{S}~
\ee
One then checks whether the eigenvector(s) associated with the lowest eigenvalue of $\bs{\Lambda}$ can be used to construct one or more configurations that satisfy the hard constraints. If the answer is positive then these configurations must belong to the classical ground state manifold, because the domain ${\it D}_s$ of configurations that satisfy the soft constraint encompasses the domain ${\it D}_h$ of configurations that satisfy the hard constraints~\cite{Kaplan}.

Denoting by $\lambda_n$ and ${\bf V}_n$ the eigenvalues and eigenvectors of $\bs{\Lambda}$, respectively, and then decomposing 
\be
\bs{\Lambda}=\sum_n \lambda_n {\bf V}_n {\bf V}_n^\dagger~
\ee
leads to the following energy bound 
\bea
&&\mc{H} = \frac{1}{2}\sum_n \lambda_n (\mathbfcal{S}^T\cdot{\bf V}_n)({\bf V}_n^\dagger\cdot\mathbfcal{S}) \nonumber\\
\Rightarrow &&
\mc{H}\geq
 \frac{1}{2}\lambda_{\text{min}}\sum_n  (\mathbfcal{S}^T\cdot{\bf V}_n)({\bf V}_n^\dagger\cdot\mathbfcal{S})
= \frac{1}{2} \lambda_{\text{min}} \underbrace{\mathbfcal{S}^T\!\cdot\!\mathbfcal{S}}_{4S^2} \nonumber\\
\Rightarrow &&
\mc{H}\geq 2 \lambda_{\text{min}}S^2\,
\eea
In the present problem the eigenvector corresponding to the minimum eigenvalue $\lambda_{\text{min}}$ satisfies the spin length constraints and therefore the LT method is able to deliver the exact ground states. 
Specifically, the (non-normalised) eigenvector with the minimum eigenvalue $\lambda_{\text{min}}$ takes the following form in each of the three phases of Fig.~\ref{fig:PhaseDiagrams}\,(a):
\be\label{eq:single-tetra-LT-spinor-2}  
\renewcommand{\arraystretch}{1.25}
\!\!\!\!
\begin{array}{cl}
\alpha:&\mathbfcal{S}^T\!=\!\big(1, 0, -1, -1,0,-1, 0,1,1,0,-1,1\big)
\\
\beta:&\mathbfcal{S}^T\!=\!\big(1, 0, 1, -1,0,1, 0,1,-1,0,-1,-1\big)
\\
\text{cross}:&\mathbfcal{S}^T\!=\!\big(c,0,s,-c,0,s,0,-c,s,0,c,s \big)
\end{array}~~
\ee
where $c\!=\!\cos\phi$, $s\!=\!\sin\phi$, $\tan(2\phi)\!=\!(\sqrt{2}D_1\!+\!D_2)/(2J)$ [see Eq.~(\ref{eq:tan2phi})], and $\tan\phi=\frac{-1+\sqrt{1+\tan^2(2\phi)}}{\tan(2\phi)}$.
The corresponding energies per site $E/N=\frac{1}{2}\lambda_{\text{min}}$ (where $N\!=\!4$ here) are:
\be
\renewcommand{\arraystretch}{1.25}
\!\!\!\!\begin{array}{cl}
\alpha:&E_\alpha/N\!=\!-J/2\!-\!(\sqrt{2}D_1\!-\!D_2)/2
\\
\beta:&E_\beta/N\!=\!-J/2\!+\!(\sqrt{2}D_1\!-\!D_2)/2
\\
\text{cross}:&E_{\text{cross}}/N\!=\!
\big(J\!-\!\sqrt{4J^2\!+\! (\sqrt{2}D_1\!+\!D_2)^2}\big)/2
\end{array}
\ee

{\cbl{\it Special line $D_2=\sqrt{2}D_1$:}} This is a special line inside the cross phase. Along this line we have
\be
\tan(2\phi)=\frac{\sqrt{2}D_1}{J}, ~~
\tan\phi = \frac{\sqrt{2} D_1}{J + \sqrt{J^2 + 2 D_1^2}}
\ee
and
\be
E/N=\big(J - 2 \sqrt{J^2 + 2 D_1^2}\big)/2\,
\ee

{\cbl{\it Purely anisotropic model:}} The classical ground state phase diagram of the purely anisotropic model ($J_1\!=\!J_2\!=\!0$) is occupied by the cross phase, with  $\phi\!=\!\pi/4$ for $\sqrt{2}D_1+D_2\!>\!0$, and $\phi\!=\!-\pi/4$ for $\sqrt{2}D_1+D_2<0$, with energy per site $E/N\!=\!-|\sqrt{2}D_1+D_2|/2$, and total moment ${\bf S}_t=\pm 2\sqrt{2}~{\bf z}$.

\section{Full model: Luttinger-Tisza method} \label{App:LT-full-model}
Let us label the spin sites by $i\mapsto ({\bf r}, \mu)$, where $\mu$ is the sublattice index, A or B, and ${\bf r}$ is the {\it physical} position of the spin in the lattice. The latter is related to the corresponding position ${\bf R}$ of the underlying square Bravais lattice by
\be
{\bf r} = {\bf R}+\bs{\rho}_\mu\,
\ee
where $\bs{\rho}_A\!=\!0$ and $\bs{\rho}_B\!=\!({\bf a}_1+{\bf a}_2)/2$.
To work in momentum space we introduce
\be\label{eq:LT-2}
\renewcommand{\arraystretch}{1.25}
\begin{array}{l}
{\bf S}_{{\bf r},\mu}=
\sum_{{\bf k}} e^{\mi {\bf k} \cdot {\bf r}} {\bf S}_{{\bf k},\mu}\,
\\
{\bf S}_{{\bf k},\mu} =
\frac{1}{N/2}\sum_{{\bf R}} e^{-\mi {\bf k} \cdot ({\bf R}+\bs{\rho}_\mu)} {\bf S}_{{\bf R}+\bs{\rho}_\mu,\mu}\,
\end{array}
\ee
where ${\bf k}$ belongs to the first Brillouin Zone of the underlying Bravais (square) lattice. 
Using this definition, the structure factor `sum rule' takes the form
\be
\sum\nolimits_{\bf k} |{\bf S}_{{\bf k},\mu}|^2=
S^2\,,~~ \mu=A, B\,
\ee
where $S$ is the spin length ($S\!=\!1$ in our simulations).

Next, we rewrite the energy in the compact form
\begin{align}
\label{eq:LT-4}
\mc{H}/N&= \frac{1}{2} \sum_{{\bf k}} \mc{S}^{\dagger}_{{\bf k}} \cdot \bf{\Lambda}_{{\bf k}} \cdot \mc{S}_{{\bf k}}\,
\end{align}
where the 6$\times$1 vector $\mc{S}_{{\bf k}}$ is given by
\begin{align}
\label{eq:LT-3}
\mc{S}_{{\bf k}}&=
\begin{pmatrix}
S^{x}_{{\bf k},A} & S^{y}_{{\bf k},A} & S^{z}_{{\bf k},A} & S^{x}_{{\bf k},B} & S^{y}_{{\bf k},B} & S^{z}_{{\bf k},B} \end{pmatrix}^{T}\,
\end{align}
and the 6$\times$6 coupling matrix $\bf{\Lambda}_{{\bf k}}$ is given by
\begin{align}
\label{eq:LambdaFullLattice}
\bf{\Lambda}_{{\bf k}}&= \begin{pmatrix}
f_{1,{\bf k}} & 0 & f_{4,{\bf k}}& f_{3,{\bf k}} &0 & f_{6,{\bf k}} \\
0 & f_{1,{\bf k}} & 0 &0 & f_{3,{\bf k}} & f_{7,{\bf k}} \\
 f^{*}_{4,{\bf k}} & 0 & f_{1,{\bf k}}& f^{*}_{6,{\bf k}} & f^{*}_{7,{\bf k}} & f_{3,{\bf k}} \\
f_{3,{\bf k}} & 0 &f_{6,{\bf k}} & f_{2,{\bf k}} &0 & 0 \\
0 & f_{3,{\bf k}} & f_{7,{\bf k}} & 0& f_{2,{\bf k}} & f_{5,{\bf k}} \\
f^{*}_{6,{\bf k}} & f^{*}_{7,{\bf k}}  &f_{3,{\bf k}} & 0 & f^{*}_{5,{\bf k}} & f_{2,{\bf k}}
\end{pmatrix}\,
\end{align}
with
\be
\begin{aligned}
&f_{1,{\bf k}}=2 J_{2} \cos k_x
\\
&f_{2,{\bf k}}=2 J_{2} \cos k_y  \label{eq:LambdaMat2}
\\
&f_{3,{\bf k}}= 4 J_{1} \cos (k_{x}/2) \cos (k_{y}/2)
\\
&f_{4,{\bf k}}=  -2\mi D_{2} \sin k_{x}
\\
&f_{5,{\bf k}}= -2\mi D_{2} \sin k_{y}
\\
&f_{6,{\bf k}}= -2 \sqrt{2}\mi D_{1} \sin (k_{x}/2) \cos (k_{y}/2) \\
&f_{7,{\bf k}}= -2 \sqrt{2}\mi D_{1} \cos (k_{x}/2) \sin (k_{y}/2)
\end{aligned}
\ee

Denoting the eigenvalues and eigenvectors of $\bs{\Lambda}_{{\bf k}}$ by $\lambda_{{\bf k},\nu}$ and ${\bf V}_{{\bf k}, \nu}$, respectively, the eigenvalue equation reads 
\begin{align}
\label{eq:LT-5}
\bs{\Lambda}_{{\bf k}} \cdot{\bf V}_{{\bf k}, \nu}&= \lambda_{{\bf k},\nu} {\bf V}_{{\bf k}, \nu}
\end{align}
where $\nu\!=\!1$-$6$.
In analogy to the single tetrahedron case, we then decompose $\bf{\Lambda}_{{\bf k}}=\sum_{\nu}\lambda_{{\bf k},\nu} V_{{\bf k}, \nu} V^{\dagger}_{{\bf k}, \nu}$
and rewrite:
\bea
\mc{H} &=& \frac{1}{2}\sum_{{\bf k},\nu} \lambda_{{\bf k},\nu} (\mathbfcal{S}_{\bf k}^\dagger\cdot{\bf V}_{{\bf k},\nu})({\bf V}_{{\bf k},\nu}^\dagger\cdot\mathbfcal{S}_{\bf k}) \nonumber\\
\Rightarrow
\mc{H}&\geq&
 \frac{1}{2}\lambda_{\text{min}}\sum_{{\bf k},\nu}  (\mathbfcal{S}_{\bf k}^\dagger\cdot{\bf V}_{{\bf k},\nu})({\bf V}_{{\bf k},\nu}^\dagger\cdot\mathbfcal{S}_{\bf k})
 \nonumber\\
&=& \frac{1}{2} \lambda_{\text{min}} \underbrace{\sum_{\bf k}\mathbfcal{S}_{\bf k}^\dagger \cdot \mathbfcal{S}_{\bf k}}_{N S^2} \nonumber\\
\Rightarrow
\mc{H} &\geq& \frac{N}{2}\lambda_{\text{min}}S^2\,
\eea
This inequality allows us to approach from below the minimum of the energy by using the modes at the
special points $({\bf k}^{*}, \nu^{*})$ which correspond to the minimum eigenvalue $\lambda_{\text{min}}$.

\subsection{Special line $D_2=\sqrt{2}D_1$}\label{sec:special-line-D2-sq2D1}
This line is rather special because the leading effect of the DM perturbation to the manifold of the lowest three flat modes of the isotropic point vanishes identically. To see this, let us denote by ${\bf V}^0_{1,{\bf k}}$, ${\bf V}^0_{2,{\bf k}}$ and ${\bf V}^0_{3,{\bf k}}$ the three (lowest and degenerate) flat bands of the coupling matrix $\bs{\Lambda}_{\bf k}^0$ of the isotropic point. Then, the leading effect of the DM perturbation $\bs{\Lambda}_{\bf k}^{\text{DM}}$ can be obtained by diagonalizing its matrix representation inside the unperturbed degenerate basis, namely
\vspace{0.2cm}
 \begin{widetext}
 \be
 \bs{\Lambda}_{\bf k}^{\text{DM}}=\frac{4 i(\sqrt{2}D_1-D_2)}{2+\cos{k_x}+\cos{k_y}}
 \left(\begin{array}{ccc}
 0 & 0 & \cos^2\frac{k_y}{2}\sin{k_x} \\
 0&0&\cos^2\frac{k_x}{2}\sin{k_y} \\
 -\cos^2\frac{k_y}{2}\sin{k_x}& -\cos^2\frac{k_x}{2}\sin{k_y} & 0
 \end{array}
 \right)_{\big\{ {\bf V}^0_{1,{\bf k}}, {\bf V}^0_{2,{\bf k}}, {\bf V}^0_{3,{\bf k}}\big\}}\,
 \ee
 \end{widetext}
Hence, $\bs{\Lambda}_{\bf k}^{\text{DM}}$ vanishes identically along the special line $D_2\!=\!\sqrt{2}D_1$ and as a result, the three flat bands remain completely flat up to first order in $D_{1,2}$. 

The region around $(\pi,\pi)$ needs special care. At this point, all six modes are degenerate and one needs to consider the $\bs{\Lambda}_{\bf k}^{\text{DM}}$ matrix in the full eigenbasis. And while, exactly at this point, the DM coupling matrix vanishes identically, the near degeneracy of the six branches in the neighborhood of $(\pi,\pi)$ implies that the effect of the DM perturbation will be strongest in this region of momentum space. In turn, this suggests that the new minimum of the energy will likely reside in the neighborhood of $(\pi,\pi)$. This is indeed confirmed by our numerics. 

\subsection{Special line $D_2=0$: The $JD_1^\pm$ manifolds} \label{App:LTJD1pm}
In this case, the effect of the DM perturbation on the coupling matrix gives rise to a first order lifting of the flat bands and a minimum at the points  $(\pm\pi/2,\pm\pi/2)$. Furthermore, as we show below, there is no mixing to the excited bands at all orders, which explains why the $D_1$ coupling selects configurations with total spin ${\bf S}_t=0$ for each tetrahedron.

The line $D_2\!=\!0$ is the only part of the phase diagram of the full lattice problem where the LT method works. 
Along this line, the lowest eigenvalue of the coupling matrix $\bs{\Lambda}_{{\bf k}}$ is
\be
\lambda_{\text{min}}=-2J-2|D_1|\,
\ee
and this occurs at the four wavevectors $\pm{\bf Q}_0$ and $\pm {\bf Q}_0'$, where ${\bf Q}_0\!=\!(\pi/2,-\pi/2)$ and ${\bf Q}_0'\!=\!(\pi/2,\pi/2)$. 
The corresponding eigenvectors are 
\be\label{eq:JD1-LT2}
\renewcommand{\arraystretch}{1.25}
\!\!\!\!\begin{array}{ll}
{\bf k}\!=\!{\bf Q}_0:~&V_{{\bf Q}_{0}}= 
\big(1, -1, \mi\sqrt{2}\sigma, -1, 1, -\mi \sqrt{2}\sigma \big)^{T}\\
{\bf k}\!=\!-{\bf Q}_0:~&V_{-{\bf Q}_{0}}=V_{{\bf Q}_{0}}^\ast\\
{\bf k}\!=\!{\bf Q}_0':~&V_{{\bf Q}_{0}'}= 
\big(1, 1, \mi\sqrt{2}\sigma, -1, -1, -\mi \sqrt{2}\sigma \big)^{T}\\
{\bf k}\!=\!-{\bf Q}_0':~&V_{-{\bf Q}_{0}'}=V_{{\bf Q}_{0}'}^\ast
\end{array}
\ee
where $\sigma\!=\!\mathrm{sign}(D_1)$. 
Using these eigenvectors, we can construct two types of solutions, one propagating along ${\bf Q}_0$ and one along ${\bf Q}_0'$. 
Considering the former, we can approach the minimum energy by taking 
\be
\mc{S}_{\bf k} = \left\{
\begin{array}{ll}
c {\bf V}_{{\bf Q}_0}, & \text{if}~{\bf k}={\bf Q}_0\\
c^\ast {\bf V}_{{\bf Q}_0}^\ast, & \text{if}~{\bf k}=-{\bf Q}_0\\
0, & \text{otherwise}\,
\end{array}\right.
\ee
The inverse Fourier transform provides the configuration in real space
\bea
\left(
\begin{array}{c}
{\bf S}_{{\bf r},A}\\
{\bf S}_{{\bf r},B}
\end{array}
\right) &=& c ~e^{i {\bf Q}_0\cdot {\bf r}} {\bf V}_{{\bf Q}_0}
+c^\ast ~e^{-i {\bf Q}_0\cdot {\bf r}} {\bf V}_{{\bf Q}_0}^\ast
\nonumber\\
&=& \frac{1}{\sqrt{2}}
\left(\begin{array}{c}
\cos({\bf Q}_0\cdot{\bf r}+\varphi)\\
-\cos({\bf Q}_0\cdot{\bf r}+\varphi)\\
-\sqrt{2}\sigma\sin({\bf Q}_0\cdot{\bf r}+\varphi)\\
-\cos({\bf Q}_0\cdot{\bf r}+\varphi)\\
\cos({\bf Q}_0\cdot{\bf r}+\varphi)\\
\sqrt{2}\sigma\sin({\bf Q}_0\cdot{\bf r}+\varphi)\\
\end{array}\right)\,
\eea
where we have taken $c=|c| e^{i\varphi}$, with $|c|=\frac{1}{\sqrt{2}}$ in order to satisfy the spin length constraints. 
Similarly, the configuration that propagates along ${\bf Q}_0'$ takes the form
\bea
\left(\begin{array}{c}
{\bf S}_{{\bf r},A}\\
{\bf S}_{{\bf r},B}
\end{array}
\right)
&=& \frac{1}{\sqrt{2}}
\left(\begin{array}{c}
\cos({\bf Q}'_0\cdot{\bf r}+\varphi)\\
\cos({\bf Q}'_0\cdot{\bf r}+\varphi)\\
-\sqrt{2}\sigma\sin({\bf Q}'_0\cdot{\bf r}+\varphi)\\
-\cos({\bf Q}'_0\cdot{\bf r}+\varphi)\\
-\cos({\bf Q}'_0\cdot{\bf r}+\varphi)\\
\sqrt{2}\sigma\sin({\bf Q}'_0\cdot{\bf r}+\varphi)\\
\end{array}\right)
\eea
This state can arise from the ${\bf Q}_0=(\pi/2,-\pi/2)$ state by a $C_4$ rotation in the combined spin-orbit space, which maps ${\bf Q}_0\cdot{\bf r}\mapsto{\bf Q}_0\cdot{\bf r}'={\bf Q}_0'\cdot{\bf r}$ in real space, and $(S^x,S^y,S^z)\mapsto(-S_y,S^x,S^z)$ in spin space. Here ${\bf r}\!=\!(x,y,0)$, ${\bf r}'\!=\!(-y,x,0)$.

The above two types of solutions are one-dimensional modulations with spins on the $(x-y,z)$ or $(x+y,z)$ planes, respectively, and rotating by 90$^\circ$ as we travel along the corresponding propagation directions (${\bf x}-{\bf y}$ and ${\bf x}+{\bf y}$, respectively). One of these states is shown in Fig.~\ref{fig:JD1p}. 
Finally, both types of configurations satisfy ${\bf S}_t\!=\!0$, meaning that they belong to the GS manifold of the isotropic point.

\section{Full model: Iterative numerical minimization method} \label{app:num-method}
In the numerical work we used finite-size clusters with periodic boundary conditions (PBC).
More specifically, we considered tetragonal clusters with spanning vectors $\mathbf{t}_1\!=\!n(\mathbf{a}_{1}\!-\!\mathbf{a}_{2})$ and $\mathbf{t}_2\!=\!n(\mathbf{a}_{1}\!+\!\mathbf{a}_{2})$, where $\mathbf{a}_{1}\!=\!(a,0)$ and $\mathbf{a}_{1}\!=\!(0,a)$, and PBC along $\mathbf{t}_1$ and $\mathbf{t}_2$. 
We have also explored other types of clusters, but the above have been found to be sufficient given the nature of the ground states.  
The numerical approach is a variation of the iterative method discussed, e.g., in Refs.~\cite{LapaHenley2012,SklanHenley2012}.
Starting from an initial random configuration, one chooses at random, in each iteration, a site 
$i$, calculates the local mean field ${\bf h}_i$ exerted by its neighbours, and then aligns the direction of ${\bf S}_i$ along ${\bf h}_i$.
This step can then be repeated a number of times $N_{rep}$, and an energy convergence criterion (along with the constraint that the local torque $\boldsymbol\tau_{i}\!=\!{\bf h}_{i}\times{\bf S}_{i}$ is sufficiently small at every site) can be used to terminate the process. 

While the above approach can be successful in unfrustrated systems, it does not {\it a priori} work well in systems with many, and nearly degenerate local energy minima. Indeed, as the above iterative procedure relies on single-spin updates, the system can be easily trapped in one of the local energy minima.
We have mitigated this problem to a large extent by: 
i) by running large ensembles of iterative minimizations with different initial random states and selecting the ones with the lowest energies,
ii) incorporating random `shake-offs' of the configurations to allow the system to escape a local minimum, and iii) studying systematically different cluster sizes, with $n$ up to 18 ($N\!=\!1296$ sites).

In most regions of the phase diagram, this systematic procedure does alleviate the problems associated with the multiple local energy minima and delivers accurate predictions for the ground states.
However, as explained in Sec.~\ref{sec:LinearAnsatzTwistedJD1p}, the iterative approach becomes inefficient in the vicinity of commensurate-incommensurate transitions, where the wavelength grows significantly. In these regions, the use of the linear ansatz approximations discussed in the text has been crucial, as it has allowed: i) to choose appropriately large cluster sizes (see e.g., Fig.~\ref{fig:p1x-panel} where $n\!=\!200$) to work with, and ii) to start the algorithm with appropriate initial states (instead of random ones) that closely reproduce the local spin correlations of the actual ground state. With such good `variational' states as input, the convergence to the actual ground state for each given cluster becomes efficient, and the optimal ground state among different cluster sizes can be identified. A similar combination of iterative (single- or multi-spin) configuration updates with appropriately chosen  variational inputs has been used successfully in Refs.~\cite{Z2vortices,Z2vortices2}.

\begin{figure}[!t]
\includegraphics[width=0.85\linewidth]{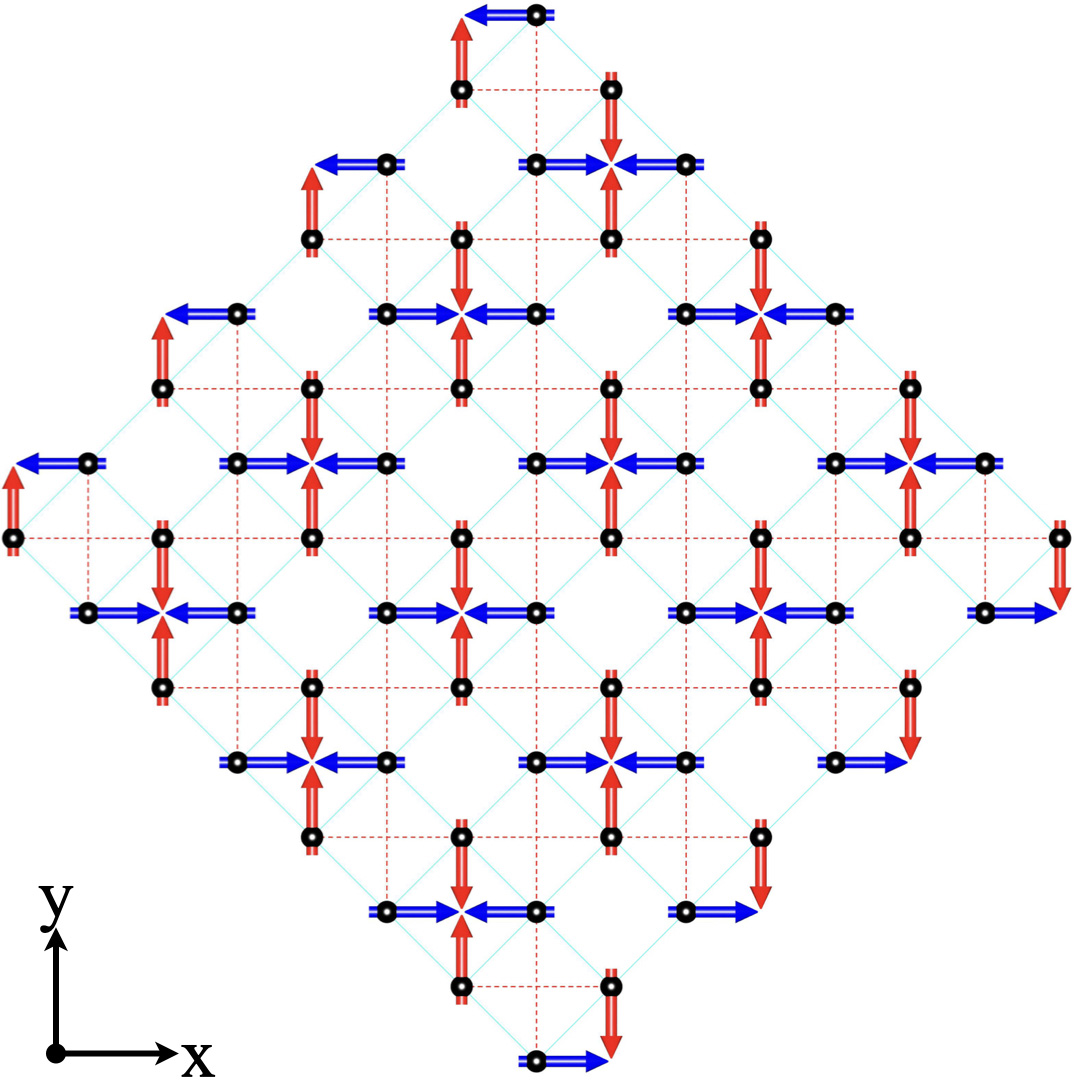}
\caption{
One of the classical ground states of (\ref{eq:single-tetra-DM-z-spins}) with $\phi\!=\!\pi/2$. 
}
\label{fig:DM_z_tiling_example}
\end{figure}

\section{Effect of out-of-plane DM interactions}
\label{sec:DM-along-z}
As mentioned in the introduction, the $C_{4\mathrm{v}}$ symmetry of the system allows for a uniform, out-of-plane component of the nearest-neighbour DM vectors, along the  $\mathbf{z}$ axis, which we write as (following the site labelling of Fig.~\ref{fig:checkerboard-lattice-b}) 
\be
\label{eq:DM-interactions-along-z}
 \mathbf{D}_{A_{1},B_{2}}'=\mathbf{D}_{B_{2},A_{2}}'=\mathbf{D}_{A_{2},B_{1}}'=\mathbf{D}_{B_{1},A_{1}}'= D_{1}' \mathbf{z}   
\ee
Here we examine the effect of these couplings, in the absence of the in-plane DM components studied in the paper. 
To this end, let us first examine a single tetrahedron. 
The classical ground state arises simply by saturating the lower bound of the energy contributions from $D_1'$ by simply placing the spins on the $xy$-plane at right angles between nearest neighbours. Importantly, this can be done at no energy cost for the isotropic couplings as can be seen explicitly below.
Alternatively, the classical minimum can also be found by the LT method. Indeed, the $\bs{\Lambda}$ matrix of Eq.~(\ref{eq:single-tetra-LT-Ham}) takes the form
\be
\bs{\Lambda}\!=\!\left[
\begin{array}{cccc}
{\bf 0}_3 & {\bf 1}_3 & {\bf L}_1 & {\bf L}_2 \\ 
{\bf 1}_3 & {\bf 0}_3 & {\bf L}_2 & {\bf L}_1 \\ 
{\bf L}_2 & {\bf L}_1 & {\bf 0}_3 & {\bf 1}_3 \\ 
{\bf L}_1 & {\bf L}_2 & {\bf 1}_3 & {\bf 0}_3 
\end{array}\right]
\ee
where ${\bf 0}_3$ and ${\bf 1}_3$ are the $3\times3$ zero and identity matrices, and
\be
{\bf L}_1\!=\!\left[\begin{array}{ccc}
1& -D_1'& 0\\ 
D_1' & 1 & 0\\
0 & 0 & 1
\end{array}\right],~~
{\bf L}_2\!=\!\left[\begin{array}{ccc}
1& D_1'& 0\\ 
-D_1' & 1 & 0\\
0 & 0 & 1
\end{array}\right]
\ee
The minimum eigenvalue of $\bs{\Lambda}$ is given by 
\be
\lambda_{\text{min}}=-1-2|D_1'|\,
\ee
which is always two-fold degenerate with eigenvectors:
\be
\renewcommand{\arraystretch}{1.25}
\begin{array}{l}
v_1^T=(1, 0, 0, -1, 0, 0, 0, \sigma, 0, 0, -\sigma, 0)\,\\
v_2^T=(0, 1, 0, 0, -1, 0, -\sigma, 0, 0, \sigma, 0, 0)\,
\end{array}
\ee
where $\sigma\!\equiv\!\text{sign}(D_1')$. Taking linear combinations of these two orthogonal solutions gives a one-parameter family of coplanar states 
\be\label{eq:single-tetra-DM-z-spins}
\renewcommand{\arraystretch}{1.25}
\begin{array}{ll}
{\bf A}_1=\cos\phi {\bf x} +\sin\phi {\bf y}, 
& {\bf A}_2=-{\bf A}_1\\
{\bf B}_1=\sigma (-\sin\phi{\bf x}+\cos \phi{\bf y}), & {\bf B}_2=-{\bf B}_1\\
\end{array}
\ee
where NN spins are at right angles with respect to each other, and the total energy is given by $E=2\lambda_{\text{min}}=-2(1+2|D_1'|)$.
The degeneracy with respect to the angle $\phi$ arises from the U(1) rotational symmetry around the $z$ axis for this particular problem. The above single-tetrahedron configurations can be tiled on the whole lattice leading to an orthogonal configuration with ordering wavevector ${\bf Q}\!=\!(\pi,\pi)$ (see example in Fig.~\ref{fig:DM_z_tiling_example} with  $\phi\!=\!\pi/2$).


\begin{thebibliography}{71}%
\makeatletter
\providecommand \@ifxundefined [1]{%
 \@ifx{#1\undefined}
}%
\providecommand \@ifnum [1]{%
 \ifnum #1\expandafter \@firstoftwo
 \else \expandafter \@secondoftwo
 \fi
}%
\providecommand \@ifx [1]{%
 \ifx #1\expandafter \@firstoftwo
 \else \expandafter \@secondoftwo
 \fi
}%
\providecommand \natexlab [1]{#1}%
\providecommand \enquote  [1]{``#1''}%
\providecommand \bibnamefont  [1]{#1}%
\providecommand \bibfnamefont [1]{#1}%
\providecommand \citenamefont [1]{#1}%
\providecommand \href@noop [0]{\@secondoftwo}%
\providecommand \href [0]{\begingroup \@sanitize@url \@href}%
\providecommand \@href[1]{\@@startlink{#1}\@@href}%
\providecommand \@@href[1]{\endgroup#1\@@endlink}%
\providecommand \@sanitize@url [0]{\catcode `\\12\catcode `\$12\catcode
  `\&12\catcode `\#12\catcode `\^12\catcode `\_12\catcode `\%12\relax}%
\providecommand \@@startlink[1]{}%
\providecommand \@@endlink[0]{}%
\providecommand \url  [0]{\begingroup\@sanitize@url \@url }%
\providecommand \@url [1]{\endgroup\@href {#1}{\urlprefix }}%
\providecommand \urlprefix  [0]{URL }%
\providecommand \Eprint [0]{\href }%
\providecommand \doibase [0]{https://doi.org/}%
\providecommand \selectlanguage [0]{\@gobble}%
\providecommand \bibinfo  [0]{\@secondoftwo}%
\providecommand \bibfield  [0]{\@secondoftwo}%
\providecommand \translation [1]{[#1]}%
\providecommand \BibitemOpen [0]{}%
\providecommand \bibitemStop [0]{}%
\providecommand \bibitemNoStop [0]{.\EOS\space}%
\providecommand \EOS [0]{\spacefactor3000\relax}%
\providecommand \BibitemShut  [1]{\csname bibitem#1\endcsname}%
\let\auto@bib@innerbib\@empty
\bibitem [{HFM(2011)}]{HFMBook}%
  \BibitemOpen
  \href@noop {} {\emph {\bibinfo {title} {{Introduction to Frustrated
  Magnetism: Materials, Experiments, Theory}}}}\ (\bibinfo  {publisher}
  {Springer Series in Solid-State Sciences},\ \bibinfo {address} {Berlin},\
  \bibinfo {year} {2011})\BibitemShut {NoStop}%
\bibitem [{Die(2013)}]{DiepBook}%
  \BibitemOpen
  \href@noop {} {\emph {\bibinfo {title} {{Frustrated Spin Systems}}}},\
  \bibinfo {edition} {2nd}\ ed.\ (\bibinfo  {publisher} {World Scientific},\
  \bibinfo {year} {2013})\BibitemShut {NoStop}%
\bibitem [{\citenamefont {P.}(2001)}]{RamirezBook}%
  \BibitemOpen
  \bibfield  {author} {\bibinfo {author} {\bibfnamefont {R.~A.}\ \bibnamefont
  {P.}},\ }in\ \href@noop {} {\emph {\bibinfo {booktitle} {{Handbook of
  Magnetic Materials}}}},\ Vol.~\bibinfo {volume} {13},\ \bibinfo {editor}
  {edited by\ \bibinfo {editor} {\bibfnamefont {B.~K.~H.}\ \bibnamefont {J.}}}\
  (\bibinfo  {publisher} {Elservier},\ \bibinfo {address} {Amsterdam},\
  \bibinfo {year} {2001})\ Chap.~\bibinfo {chapter} {4}, pp.\ \bibinfo {pages}
  {423--520}\BibitemShut {NoStop}%
\bibitem [{\citenamefont {Anderson}(1973)}]{Anderson1973}%
  \BibitemOpen
  \bibfield  {author} {\bibinfo {author} {\bibfnamefont {P.}~\bibnamefont
  {Anderson}},\ }\href
  {http://www.sciencedirect.com/science/article/pii/0025540873901670}
  {\bibfield  {journal} {\bibinfo  {journal} {Mat. Res. Bull}\ }\textbf
  {\bibinfo {volume} {8}},\ \bibinfo {pages} {153 } (\bibinfo {year}
  {1973})}\BibitemShut {NoStop}%
\bibitem [{\citenamefont {Balents}(2010)}]{Balents2010}%
  \BibitemOpen
  \bibfield  {author} {\bibinfo {author} {\bibfnamefont {L.}~\bibnamefont
  {Balents}},\ }\href {http://dx.doi.org/10.1038/nature08917} {\bibfield
  {journal} {\bibinfo  {journal} {Nature (London)}\ }\textbf {\bibinfo {volume}
  {464}} (\bibinfo {year} {2010})}\BibitemShut {NoStop}%
\bibitem [{\citenamefont {Savary}\ and\ \citenamefont
  {Balents}(2017)}]{Savary2016}%
  \BibitemOpen
  \bibfield  {author} {\bibinfo {author} {\bibfnamefont {L.}~\bibnamefont
  {Savary}}\ and\ \bibinfo {author} {\bibfnamefont {L.}~\bibnamefont
  {Balents}},\ }\href {http://stacks.iop.org/0034-4885/80/i=1/a=016502}
  {\bibfield  {journal} {\bibinfo  {journal} {Rep. Prog. Phys.}\ }\textbf
  {\bibinfo {volume} {80}},\ \bibinfo {pages} {016502} (\bibinfo {year}
  {2017})}\BibitemShut {NoStop}%
\bibitem [{\citenamefont {Trebst}(2017)}]{Trebst2017}%
  \BibitemOpen
  \bibfield  {author} {\bibinfo {author} {\bibfnamefont {S.}~\bibnamefont
  {Trebst}},\ }\href {http://arxiv.org/abs/1701.07056} {\bibfield  {journal}
  {\bibinfo  {journal} {arXiv:1701.07056}\ } (\bibinfo {year}
  {2017})}\BibitemShut {NoStop}%
\bibitem [{\citenamefont {Zhou}\ \emph {et~al.}(2017)\citenamefont {Zhou},
  \citenamefont {Kanoda},\ and\ \citenamefont {Ng}}]{Zhou2017}%
  \BibitemOpen
  \bibfield  {author} {\bibinfo {author} {\bibfnamefont {Y.}~\bibnamefont
  {Zhou}}, \bibinfo {author} {\bibfnamefont {K.}~\bibnamefont {Kanoda}},\ and\
  \bibinfo {author} {\bibfnamefont {T.-K.}\ \bibnamefont {Ng}},\ }\href
  {https://doi.org/10.1103/RevModPhys.89.025003} {\bibfield  {journal}
  {\bibinfo  {journal} {Rev. Mod. Phys.}\ }\textbf {\bibinfo {volume} {89}},\
  \bibinfo {pages} {025003} (\bibinfo {year} {2017})}\BibitemShut {NoStop}%
\bibitem [{\citenamefont {Winter}\ \emph {et~al.}(2017)\citenamefont {Winter},
  \citenamefont {Tsirlin}, \citenamefont {Daghofer}, \citenamefont {van~den
  Brink}, \citenamefont {Singh}, \citenamefont {Gegenwart},\ and\ \citenamefont
  {Valenti}}]{Winter2017}%
  \BibitemOpen
  \bibfield  {author} {\bibinfo {author} {\bibfnamefont {S.~M.}\ \bibnamefont
  {Winter}}, \bibinfo {author} {\bibfnamefont {A.~A.}\ \bibnamefont {Tsirlin}},
  \bibinfo {author} {\bibfnamefont {M.}~\bibnamefont {Daghofer}}, \bibinfo
  {author} {\bibfnamefont {J.}~\bibnamefont {van~den Brink}}, \bibinfo {author}
  {\bibfnamefont {Y.}~\bibnamefont {Singh}}, \bibinfo {author} {\bibfnamefont
  {P.}~\bibnamefont {Gegenwart}},\ and\ \bibinfo {author} {\bibfnamefont
  {R.}~\bibnamefont {Valenti}},\ }\href
  {http://iopscience.iop.org/article/10.1088/1361-648X/aa8cf5} {\bibfield
  {journal} {\bibinfo  {journal} {J. Phys.: Condens. Matter}\ }\textbf
  {\bibinfo {volume} {29}},\ \bibinfo {pages} {493002} (\bibinfo {year}
  {2017})}\BibitemShut {NoStop}%
\bibitem [{\citenamefont {Hermanns}\ \emph {et~al.}(2018)\citenamefont
  {Hermanns}, \citenamefont {Kimchi},\ and\ \citenamefont
  {Knolle}}]{Hermanns2017}%
  \BibitemOpen
  \bibfield  {author} {\bibinfo {author} {\bibfnamefont {M.}~\bibnamefont
  {Hermanns}}, \bibinfo {author} {\bibfnamefont {I.}~\bibnamefont {Kimchi}},\
  and\ \bibinfo {author} {\bibfnamefont {J.}~\bibnamefont {Knolle}},\ }\href
  {https://doi.org/10.1146/annurev-conmatphys-033117-053934} {\bibfield
  {journal} {\bibinfo  {journal} {Annual Review of Condensed Matter Physics}\
  }\textbf {\bibinfo {volume} {9}},\ \bibinfo {pages} {17} (\bibinfo {year}
  {2018})},\ \Eprint
  {https://arxiv.org/abs/https://doi.org/10.1146/annurev-conmatphys-033117-053934}
  {https://doi.org/10.1146/annurev-conmatphys-033117-053934} \BibitemShut
  {NoStop}%
\bibitem [{\citenamefont {Knolle}\ and\ \citenamefont
  {Moessner}(2019)}]{Knolle2019}%
  \BibitemOpen
  \bibfield  {author} {\bibinfo {author} {\bibfnamefont {J.}~\bibnamefont
  {Knolle}}\ and\ \bibinfo {author} {\bibfnamefont {R.}~\bibnamefont
  {Moessner}},\ }\href
  {https://doi.org/10.1146/annurev-conmatphys-031218-013401} {\bibfield
  {journal} {\bibinfo  {journal} {Annual Review of Condensed Matter Physics}\
  }\textbf {\bibinfo {volume} {10}},\ \bibinfo {pages} {451} (\bibinfo {year}
  {2019})},\ \Eprint
  {https://arxiv.org/abs/https://doi.org/10.1146/annurev-conmatphys-031218-013401}
  {https://doi.org/10.1146/annurev-conmatphys-031218-013401} \BibitemShut
  {NoStop}%
\bibitem [{\citenamefont {Dzyaloshinskii}(1964)}]{Dz64}%
  \BibitemOpen
  \bibfield  {author} {\bibinfo {author} {\bibfnamefont {I.}~\bibnamefont
  {Dzyaloshinskii}},\ }\href
  {http://www.jetp.ac.ru/cgi-bin/e/index/e/19/4/p960?a=list} {\bibfield
  {journal} {\bibinfo  {journal} {Sov.\ Phys.\ JETP}\ }\textbf {\bibinfo
  {volume} {19}},\ \bibinfo {pages} {960} (\bibinfo {year} {1964})}\BibitemShut
  {NoStop}%
\bibitem [{\citenamefont {Izyumov}\ and\ \citenamefont
  {Syromyatnikov}(1990)}]{IZYUMOV}%
  \BibitemOpen
  \bibfield  {author} {\bibinfo {author} {\bibfnamefont {A.}~\bibnamefont
  {Izyumov}}\ and\ \bibinfo {author} {\bibfnamefont {V.}~\bibnamefont
  {Syromyatnikov}},\ }\href@noop {} {\emph {\bibinfo {title} {{Phase
  Transitions and Crystal Symmetry}}}}\ (\bibinfo  {publisher} {Kluwer Academic
  Publishers},\ \bibinfo {year} {1990})\BibitemShut {NoStop}%
\bibitem [{\citenamefont {Bogdanov}\ and\ \citenamefont
  {Yablonskii}(1989)}]{Bogdanov1989}%
  \BibitemOpen
  \bibfield  {author} {\bibinfo {author} {\bibfnamefont {A.~N.}\ \bibnamefont
  {Bogdanov}}\ and\ \bibinfo {author} {\bibfnamefont {D.~A.}\ \bibnamefont
  {Yablonskii}},\ }\href
  {http://www.jetp.ac.ru/cgi-bin/e/index/e/68/1/p101?a=list} {\bibfield
  {journal} {\bibinfo  {journal} {Sov. Phys. JETP}\ }\textbf {\bibinfo {volume}
  {68}},\ \bibinfo {pages} {101} (\bibinfo {year} {1989})}\BibitemShut
  {NoStop}%
\bibitem [{\citenamefont {Bogdanov}\ and\ \citenamefont
  {Hubert}(1994)}]{Bogdanov1994}%
  \BibitemOpen
  \bibfield  {author} {\bibinfo {author} {\bibfnamefont {A.~N.}\ \bibnamefont
  {Bogdanov}}\ and\ \bibinfo {author} {\bibfnamefont {A.}~\bibnamefont
  {Hubert}},\ }\href@noop {} {\bibfield  {journal} {\bibinfo  {journal} {J.
  Magn. Magn. Mater.}\ }\textbf {\bibinfo {volume} {138}},\ \bibinfo {pages}
  {255} (\bibinfo {year} {1994})}\BibitemShut {NoStop}%
\bibitem [{\citenamefont {R\"o{\ss}ler}\ \emph {et~al.}(2006)\citenamefont
  {R\"o{\ss}ler}, \citenamefont {Bogdanov},\ and\ \citenamefont
  {Pfleiderer}}]{Roessler2006}%
  \BibitemOpen
  \bibfield  {author} {\bibinfo {author} {\bibfnamefont {U.~K.}\ \bibnamefont
  {R\"o{\ss}ler}}, \bibinfo {author} {\bibfnamefont {A.~N.}\ \bibnamefont
  {Bogdanov}},\ and\ \bibinfo {author} {\bibfnamefont {C.}~\bibnamefont
  {Pfleiderer}},\ }\href {https://doi.org/doi:10.1038/nature05056} {\bibfield
  {journal} {\bibinfo  {journal} {Nature}\ }\textbf {\bibinfo {volume} {442}},\
  \bibinfo {pages} {797} (\bibinfo {year} {2006})}\BibitemShut {NoStop}%
\bibitem [{\citenamefont {Janson}\ \emph {et~al.}(2014)\citenamefont {Janson},
  \citenamefont {Rousochatzakis}, \citenamefont {Tsirlin}, \citenamefont
  {Belesi}, \citenamefont {Leonov}, \citenamefont {R\"o{\ss}ler}, \citenamefont
  {van~den Brink},\ and\ \citenamefont {Rosner}}]{Oleg2014}%
  \BibitemOpen
  \bibfield  {author} {\bibinfo {author} {\bibfnamefont {O.}~\bibnamefont
  {Janson}}, \bibinfo {author} {\bibfnamefont {I.}~\bibnamefont
  {Rousochatzakis}}, \bibinfo {author} {\bibfnamefont {A.~A.}\ \bibnamefont
  {Tsirlin}}, \bibinfo {author} {\bibfnamefont {M.}~\bibnamefont {Belesi}},
  \bibinfo {author} {\bibfnamefont {A.~A.}\ \bibnamefont {Leonov}}, \bibinfo
  {author} {\bibfnamefont {U.~K.}\ \bibnamefont {R\"o{\ss}ler}}, \bibinfo
  {author} {\bibfnamefont {J.}~\bibnamefont {van~den Brink}},\ and\ \bibinfo
  {author} {\bibfnamefont {H.}~\bibnamefont {Rosner}},\ }\href
  {https://doi.org/10.1038/ncomms6376} {\bibfield  {journal} {\bibinfo
  {journal} {Nature Communications}\ }\textbf {\bibinfo {volume} {5}},\
  \bibinfo {pages} {5376} (\bibinfo {year} {2014})}\BibitemShut {NoStop}%
\bibitem [{\citenamefont {M{\"{u}}hlbauer}\ \emph {et~al.}(2009)\citenamefont
  {M{\"{u}}hlbauer}, \citenamefont {Binz}, \citenamefont {Jonietz},
  \citenamefont {Pfleiderer}, \citenamefont {Rosch}, \citenamefont {Neubauer},
  \citenamefont {Georgii},\ and\ \citenamefont {B{\"{o}}ni}}]{muehlbauer2009}%
  \BibitemOpen
  \bibfield  {author} {\bibinfo {author} {\bibfnamefont {S.}~\bibnamefont
  {M{\"{u}}hlbauer}}, \bibinfo {author} {\bibfnamefont {B.}~\bibnamefont
  {Binz}}, \bibinfo {author} {\bibfnamefont {F.}~\bibnamefont {Jonietz}},
  \bibinfo {author} {\bibfnamefont {C.}~\bibnamefont {Pfleiderer}}, \bibinfo
  {author} {\bibfnamefont {A.}~\bibnamefont {Rosch}}, \bibinfo {author}
  {\bibfnamefont {A.}~\bibnamefont {Neubauer}}, \bibinfo {author}
  {\bibfnamefont {R.}~\bibnamefont {Georgii}},\ and\ \bibinfo {author}
  {\bibfnamefont {P.}~\bibnamefont {B{\"{o}}ni}},\ }\href
  {https://doi.org/10.1126/science.1166767} {\bibfield  {journal} {\bibinfo
  {journal} {Science}\ }\textbf {\bibinfo {volume} {323}},\ \bibinfo {pages}
  {915} (\bibinfo {year} {2009})}\BibitemShut {NoStop}%
\bibitem [{\citenamefont {Yu}\ \emph {et~al.}(2010)\citenamefont {Yu},
  \citenamefont {Onose}, \citenamefont {Kanazawa}, \citenamefont {Park},
  \citenamefont {Han}, \citenamefont {Matsui}, \citenamefont {Nagaosa},\ and\
  \citenamefont {Tokura}}]{yu2010}%
  \BibitemOpen
  \bibfield  {author} {\bibinfo {author} {\bibfnamefont {X.~Z.}\ \bibnamefont
  {Yu}}, \bibinfo {author} {\bibfnamefont {Y.}~\bibnamefont {Onose}}, \bibinfo
  {author} {\bibfnamefont {N.}~\bibnamefont {Kanazawa}}, \bibinfo {author}
  {\bibfnamefont {J.~H.}\ \bibnamefont {Park}}, \bibinfo {author}
  {\bibfnamefont {J.~H.}\ \bibnamefont {Han}}, \bibinfo {author} {\bibfnamefont
  {Y.}~\bibnamefont {Matsui}}, \bibinfo {author} {\bibfnamefont
  {N.}~\bibnamefont {Nagaosa}},\ and\ \bibinfo {author} {\bibfnamefont
  {Y.}~\bibnamefont {Tokura}},\ }\href {https://doi.org/10.1038/nature09124}
  {\bibfield  {journal} {\bibinfo  {journal} {Nature}\ }\textbf {\bibinfo
  {volume} {465}},\ \bibinfo {pages} {901} (\bibinfo {year}
  {2010})}\BibitemShut {NoStop}%
\bibitem [{\citenamefont {Tonomura}\ \emph {et~al.}(2012)\citenamefont
  {Tonomura}, \citenamefont {Yu}, \citenamefont {Yanagisawa}, \citenamefont
  {Matsuda}, \citenamefont {Onose}, \citenamefont {Kanazawa}, \citenamefont
  {Park},\ and\ \citenamefont {Tokura}}]{tonomura2012}%
  \BibitemOpen
  \bibfield  {author} {\bibinfo {author} {\bibfnamefont {A.}~\bibnamefont
  {Tonomura}}, \bibinfo {author} {\bibfnamefont {X.}~\bibnamefont {Yu}},
  \bibinfo {author} {\bibfnamefont {K.}~\bibnamefont {Yanagisawa}}, \bibinfo
  {author} {\bibfnamefont {T.}~\bibnamefont {Matsuda}}, \bibinfo {author}
  {\bibfnamefont {Y.}~\bibnamefont {Onose}}, \bibinfo {author} {\bibfnamefont
  {N.}~\bibnamefont {Kanazawa}}, \bibinfo {author} {\bibfnamefont {H.~S.}\
  \bibnamefont {Park}},\ and\ \bibinfo {author} {\bibfnamefont
  {Y.}~\bibnamefont {Tokura}},\ }\href {https://doi.org/10.1021/nl300073m}
  {\bibfield  {journal} {\bibinfo  {journal} {Nano Letters}\ }\textbf {\bibinfo
  {volume} {12}},\ \bibinfo {pages} {1673} (\bibinfo {year}
  {2012})}\BibitemShut {NoStop}%
\bibitem [{\citenamefont {Seki}\ \emph {et~al.}(2012)\citenamefont {Seki},
  \citenamefont {Yu}, \citenamefont {Ishiwata},\ and\ \citenamefont
  {Tokura}}]{Seki2012}%
  \BibitemOpen
  \bibfield  {author} {\bibinfo {author} {\bibfnamefont {S.}~\bibnamefont
  {Seki}}, \bibinfo {author} {\bibfnamefont {X.~Z.}\ \bibnamefont {Yu}},
  \bibinfo {author} {\bibfnamefont {S.}~\bibnamefont {Ishiwata}},\ and\
  \bibinfo {author} {\bibfnamefont {Y.}~\bibnamefont {Tokura}},\ }\href
  {https://doi.org/10.1126/science.1214143} {\bibfield  {journal} {\bibinfo
  {journal} {Science}\ }\textbf {\bibinfo {volume} {336}},\ \bibinfo {pages}
  {198} (\bibinfo {year} {2012})}\BibitemShut {NoStop}%
\bibitem [{\citenamefont {Shiba}\ and\ \citenamefont
  {Suzuki}(1983)}]{Shiba:1983js}%
  \BibitemOpen
  \bibfield  {author} {\bibinfo {author} {\bibfnamefont {H.}~\bibnamefont
  {Shiba}}\ and\ \bibinfo {author} {\bibfnamefont {N.}~\bibnamefont {Suzuki}},\
  }\href {https://doi.org/10.1143/JPSJ.52.1382} {\bibfield  {journal} {\bibinfo
   {journal} {J. Phys. Soc. Jpn.}\ }\textbf {\bibinfo {volume} {52}},\ \bibinfo
  {pages} {1382} (\bibinfo {year} {1983})}\BibitemShut {NoStop}%
\bibitem [{\citenamefont {M\"uhlbauer}\ \emph {et~al.}(2011)\citenamefont
  {M\"uhlbauer}, \citenamefont {Gvasaliya}, \citenamefont {Pomjakushina},\ and\
  \citenamefont {Zheludev}}]{Muhlbauer2011}%
  \BibitemOpen
  \bibfield  {author} {\bibinfo {author} {\bibfnamefont {S.}~\bibnamefont
  {M\"uhlbauer}}, \bibinfo {author} {\bibfnamefont {S.~N.}\ \bibnamefont
  {Gvasaliya}}, \bibinfo {author} {\bibfnamefont {E.}~\bibnamefont
  {Pomjakushina}},\ and\ \bibinfo {author} {\bibfnamefont {A.}~\bibnamefont
  {Zheludev}},\ }\href {https://doi.org/10.1103/PhysRevB.84.180406} {\bibfield
  {journal} {\bibinfo  {journal} {Phys. Rev. B}\ }\textbf {\bibinfo {volume}
  {84}},\ \bibinfo {pages} {180406} (\bibinfo {year} {2011})}\BibitemShut
  {NoStop}%
\bibitem [{\citenamefont {Artyukhin}\ \emph {et~al.}(2012)\citenamefont
  {Artyukhin}, \citenamefont {Mostovoy}, \citenamefont {Jensen}, \citenamefont
  {Le}, \citenamefont {Prokes}, \citenamefont {de~Paula}, \citenamefont
  {Bordallo}, \citenamefont {Maljuk}, \citenamefont {Landsgesell},
  \citenamefont {Ryll}, \citenamefont {Klemke}, \citenamefont {Paeckel},
  \citenamefont {Kiefer}, \citenamefont {Lefmann}, \citenamefont {Kuhn},\ and\
  \citenamefont {Argyriou}}]{Yukawa2012}%
  \BibitemOpen
  \bibfield  {author} {\bibinfo {author} {\bibfnamefont {S.}~\bibnamefont
  {Artyukhin}}, \bibinfo {author} {\bibfnamefont {M.}~\bibnamefont {Mostovoy}},
  \bibinfo {author} {\bibfnamefont {N.~P.}\ \bibnamefont {Jensen}}, \bibinfo
  {author} {\bibfnamefont {D.}~\bibnamefont {Le}}, \bibinfo {author}
  {\bibfnamefont {K.}~\bibnamefont {Prokes}}, \bibinfo {author} {\bibfnamefont
  {V.~G.}\ \bibnamefont {de~Paula}}, \bibinfo {author} {\bibfnamefont {H.~N.}\
  \bibnamefont {Bordallo}}, \bibinfo {author} {\bibfnamefont {A.}~\bibnamefont
  {Maljuk}}, \bibinfo {author} {\bibfnamefont {S.}~\bibnamefont {Landsgesell}},
  \bibinfo {author} {\bibfnamefont {H.}~\bibnamefont {Ryll}}, \bibinfo {author}
  {\bibfnamefont {B.}~\bibnamefont {Klemke}}, \bibinfo {author} {\bibfnamefont
  {S.}~\bibnamefont {Paeckel}}, \bibinfo {author} {\bibfnamefont
  {K.}~\bibnamefont {Kiefer}}, \bibinfo {author} {\bibfnamefont
  {K.}~\bibnamefont {Lefmann}}, \bibinfo {author} {\bibfnamefont {L.~T.}\
  \bibnamefont {Kuhn}},\ and\ \bibinfo {author} {\bibfnamefont {D.~N.}\
  \bibnamefont {Argyriou}},\ }\href {https://doi.org/10.1038/nmat3358}
  {\bibfield  {journal} {\bibinfo  {journal} {Nat. Mater.}\ }\textbf {\bibinfo
  {volume} {11}},\ \bibinfo {pages} {694} (\bibinfo {year} {2012})}\BibitemShut
  {NoStop}%
\bibitem [{\citenamefont {Okubo}\ \emph {et~al.}(2011)\citenamefont {Okubo},
  \citenamefont {Nguyen},\ and\ \citenamefont {Kawamura}}]{Okubo2011}%
  \BibitemOpen
  \bibfield  {author} {\bibinfo {author} {\bibfnamefont {T.}~\bibnamefont
  {Okubo}}, \bibinfo {author} {\bibfnamefont {T.~H.}\ \bibnamefont {Nguyen}},\
  and\ \bibinfo {author} {\bibfnamefont {H.}~\bibnamefont {Kawamura}},\ }\href
  {https://doi.org/10.1103/PhysRevB.84.144432} {\bibfield  {journal} {\bibinfo
  {journal} {Phys. Rev. B}\ }\textbf {\bibinfo {volume} {84}},\ \bibinfo
  {pages} {144432} (\bibinfo {year} {2011})}\BibitemShut {NoStop}%
\bibitem [{\citenamefont {Okubo}\ \emph {et~al.}(2012)\citenamefont {Okubo},
  \citenamefont {Chung},\ and\ \citenamefont {Kawamura}}]{Okubo2012}%
  \BibitemOpen
  \bibfield  {author} {\bibinfo {author} {\bibfnamefont {T.}~\bibnamefont
  {Okubo}}, \bibinfo {author} {\bibfnamefont {S.}~\bibnamefont {Chung}},\ and\
  \bibinfo {author} {\bibfnamefont {H.}~\bibnamefont {Kawamura}},\ }\href
  {https://doi.org/10.1103/PhysRevLett.108.017206} {\bibfield  {journal}
  {\bibinfo  {journal} {Phys. Rev. Lett.}\ }\textbf {\bibinfo {volume} {108}},\
  \bibinfo {pages} {017206} (\bibinfo {year} {2012})}\BibitemShut {NoStop}%
\bibitem [{\citenamefont {Kamiya}\ and\ \citenamefont
  {Batista}(2014)}]{Kamiya2014}%
  \BibitemOpen
  \bibfield  {author} {\bibinfo {author} {\bibfnamefont {Y.}~\bibnamefont
  {Kamiya}}\ and\ \bibinfo {author} {\bibfnamefont {C.~D.}\ \bibnamefont
  {Batista}},\ }\href {https://doi.org/10.1103/PhysRevX.4.011023} {\bibfield
  {journal} {\bibinfo  {journal} {Phys. Rev. X}\ }\textbf {\bibinfo {volume}
  {4}},\ \bibinfo {pages} {011023} (\bibinfo {year} {2014})}\BibitemShut
  {NoStop}%
\bibitem [{\citenamefont {Rousochatzakis}\ \emph {et~al.}(2016)\citenamefont
  {Rousochatzakis}, \citenamefont {R\"ossler}, \citenamefont {van~den Brink},\
  and\ \citenamefont {Daghofer}}]{Z2vortices}%
  \BibitemOpen
  \bibfield  {author} {\bibinfo {author} {\bibfnamefont {I.}~\bibnamefont
  {Rousochatzakis}}, \bibinfo {author} {\bibfnamefont {U.~K.}\ \bibnamefont
  {R\"ossler}}, \bibinfo {author} {\bibfnamefont {J.}~\bibnamefont {van~den
  Brink}},\ and\ \bibinfo {author} {\bibfnamefont {M.}~\bibnamefont
  {Daghofer}},\ }\href {https://doi.org/10.1103/PhysRevB.93.104417} {\bibfield
  {journal} {\bibinfo  {journal} {Phys. Rev. B}\ }\textbf {\bibinfo {volume}
  {93}},\ \bibinfo {pages} {104417} (\bibinfo {year} {2016})}\BibitemShut
  {NoStop}%
\bibitem [{\citenamefont {Dzyaloshinsky}(1958)}]{Dz1958}%
  \BibitemOpen
  \bibfield  {author} {\bibinfo {author} {\bibfnamefont {I.}~\bibnamefont
  {Dzyaloshinsky}},\ }\href
  {https://doi.org/https://doi.org/10.1016/0022-3697(58)90076-3} {\bibfield
  {journal} {\bibinfo  {journal} {Journal of Physics and Chemistry of Solids}\
  }\textbf {\bibinfo {volume} {4}},\ \bibinfo {pages} {241} (\bibinfo {year}
  {1958})}\BibitemShut {NoStop}%
\bibitem [{\citenamefont {Moriya}(1960)}]{Moriya1960}%
  \BibitemOpen
  \bibfield  {author} {\bibinfo {author} {\bibfnamefont {T.}~\bibnamefont
  {Moriya}},\ }\href {https://doi.org/10.1103/PhysRevLett.4.228} {\bibfield
  {journal} {\bibinfo  {journal} {Phys. Rev. Lett.}\ }\textbf {\bibinfo
  {volume} {4}},\ \bibinfo {pages} {228} (\bibinfo {year} {1960})}\BibitemShut
  {NoStop}%
\bibitem [{\citenamefont {Moessner}\ and\ \citenamefont
  {Chalker}(1998)}]{MoessnerChalker1998}%
  \BibitemOpen
  \bibfield  {author} {\bibinfo {author} {\bibfnamefont {R.}~\bibnamefont
  {Moessner}}\ and\ \bibinfo {author} {\bibfnamefont {J.~T.}\ \bibnamefont
  {Chalker}},\ }\href {https://doi.org/10.1103/PhysRevB.58.12049} {\bibfield
  {journal} {\bibinfo  {journal} {Phys. Rev. B}\ }\textbf {\bibinfo {volume}
  {58}},\ \bibinfo {pages} {12049} (\bibinfo {year} {1998})}\BibitemShut
  {NoStop}%
\bibitem [{\citenamefont {Singh}\ \emph {et~al.}(1998)\citenamefont {Singh},
  \citenamefont {Starykh},\ and\ \citenamefont {Freitas}}]{Singh1998}%
  \BibitemOpen
  \bibfield  {author} {\bibinfo {author} {\bibfnamefont {R.~R.~P.}\
  \bibnamefont {Singh}}, \bibinfo {author} {\bibfnamefont {O.~A.}\ \bibnamefont
  {Starykh}},\ and\ \bibinfo {author} {\bibfnamefont {P.~J.}\ \bibnamefont
  {Freitas}},\ }\href {https://doi.org/10.1063/1.367682} {\bibfield  {journal}
  {\bibinfo  {journal} {Journal of Applied Physics}\ }\textbf {\bibinfo
  {volume} {83}},\ \bibinfo {pages} {7387} (\bibinfo {year} {1998})},\ \Eprint
  {https://arxiv.org/abs/https://doi.org/10.1063/1.367682}
  {https://doi.org/10.1063/1.367682} \BibitemShut {NoStop}%
\bibitem [{\citenamefont {Lieb}\ and\ \citenamefont {Schupp}(1999)}]{Lieb1999}%
  \BibitemOpen
  \bibfield  {author} {\bibinfo {author} {\bibfnamefont {E.~H.}\ \bibnamefont
  {Lieb}}\ and\ \bibinfo {author} {\bibfnamefont {P.}~\bibnamefont {Schupp}},\
  }\href {https://doi.org/10.1103/PhysRevLett.83.5362} {\bibfield  {journal}
  {\bibinfo  {journal} {Phys. Rev. Lett.}\ }\textbf {\bibinfo {volume} {83}},\
  \bibinfo {pages} {5362} (\bibinfo {year} {1999})}\BibitemShut {NoStop}%
\bibitem [{\citenamefont {Elhajal}\ \emph {et~al.}(2001)\citenamefont
  {Elhajal}, \citenamefont {Canals},\ and\ \citenamefont
  {Lacroix}}]{Elhajal2001}%
  \BibitemOpen
  \bibfield  {author} {\bibinfo {author} {\bibfnamefont {M.}~\bibnamefont
  {Elhajal}}, \bibinfo {author} {\bibfnamefont {B.}~\bibnamefont {Canals}},\
  and\ \bibinfo {author} {\bibfnamefont {C.}~\bibnamefont {Lacroix}},\ }\href
  {https://doi.org/10.1139/p01-107} {\bibfield  {journal} {\bibinfo  {journal}
  {Canadian Journal of Physics}\ }\textbf {\bibinfo {volume} {79}},\ \bibinfo
  {pages} {1353} (\bibinfo {year} {2001})},\ \Eprint
  {https://arxiv.org/abs/https://doi.org/10.1139/p01-107}
  {https://doi.org/10.1139/p01-107} \BibitemShut {NoStop}%
\bibitem [{\citenamefont {{Canals, B.}}\ and\ \citenamefont {{Garanin, D.
  A.}}(2002)}]{CanalsGaranin2002}%
  \BibitemOpen
  \bibfield  {author} {\bibinfo {author} {\bibnamefont {{Canals, B.}}}\ and\
  \bibinfo {author} {\bibnamefont {{Garanin, D. A.}}},\ }\href
  {https://doi.org/10.1140/epjb/e20020112} {\bibfield  {journal} {\bibinfo
  {journal} {Eur. Phys. J. B}\ }\textbf {\bibinfo {volume} {26}},\ \bibinfo
  {pages} {439} (\bibinfo {year} {2002})}\BibitemShut {NoStop}%
\bibitem [{\citenamefont {Canals}(2002)}]{Canals2002}%
  \BibitemOpen
  \bibfield  {author} {\bibinfo {author} {\bibfnamefont {B.}~\bibnamefont
  {Canals}},\ }\href {https://doi.org/10.1103/PhysRevB.65.184408} {\bibfield
  {journal} {\bibinfo  {journal} {Phys. Rev. B}\ }\textbf {\bibinfo {volume}
  {65}},\ \bibinfo {pages} {184408} (\bibinfo {year} {2002})}\BibitemShut
  {NoStop}%
\bibitem [{\citenamefont {Brenig}\ and\ \citenamefont
  {Honecker}(2002)}]{BrenigHonecker2002}%
  \BibitemOpen
  \bibfield  {author} {\bibinfo {author} {\bibfnamefont {W.}~\bibnamefont
  {Brenig}}\ and\ \bibinfo {author} {\bibfnamefont {A.}~\bibnamefont
  {Honecker}},\ }\href {https://doi.org/10.1103/PhysRevB.65.140407} {\bibfield
  {journal} {\bibinfo  {journal} {Phys. Rev. B}\ }\textbf {\bibinfo {volume}
  {65}},\ \bibinfo {pages} {140407} (\bibinfo {year} {2002})}\BibitemShut
  {NoStop}%
\bibitem [{\citenamefont {Fouet}\ \emph {et~al.}(2003)\citenamefont {Fouet},
  \citenamefont {Mambrini}, \citenamefont {Sindzingre},\ and\ \citenamefont
  {Lhuillier}}]{Fouet2003}%
  \BibitemOpen
  \bibfield  {author} {\bibinfo {author} {\bibfnamefont {J.-B.}\ \bibnamefont
  {Fouet}}, \bibinfo {author} {\bibfnamefont {M.}~\bibnamefont {Mambrini}},
  \bibinfo {author} {\bibfnamefont {P.}~\bibnamefont {Sindzingre}},\ and\
  \bibinfo {author} {\bibfnamefont {C.}~\bibnamefont {Lhuillier}},\ }\href
  {https://doi.org/10.1103/PhysRevB.67.054411} {\bibfield  {journal} {\bibinfo
  {journal} {Phys. Rev. B}\ }\textbf {\bibinfo {volume} {67}},\ \bibinfo
  {pages} {054411} (\bibinfo {year} {2003})}\BibitemShut {NoStop}%
\bibitem [{\citenamefont {Moessner}\ \emph {et~al.}(2004)\citenamefont
  {Moessner}, \citenamefont {Tchernyshyov},\ and\ \citenamefont
  {Sondhi}}]{Moessner2004}%
  \BibitemOpen
  \bibfield  {author} {\bibinfo {author} {\bibfnamefont {R.}~\bibnamefont
  {Moessner}}, \bibinfo {author} {\bibfnamefont {O.}~\bibnamefont
  {Tchernyshyov}},\ and\ \bibinfo {author} {\bibfnamefont {S.~L.}\ \bibnamefont
  {Sondhi}},\ }\href {https://doi.org/10.1023/B:JOSS.0000037247.54022.62}
  {\bibfield  {journal} {\bibinfo  {journal} {Journal of Statistical Physics}\
  }\textbf {\bibinfo {volume} {116}},\ \bibinfo {pages} {755} (\bibinfo {year}
  {2004})}\BibitemShut {NoStop}%
\bibitem [{\citenamefont {Bernier}\ \emph {et~al.}(2004)\citenamefont
  {Bernier}, \citenamefont {Chung}, \citenamefont {Kim},\ and\ \citenamefont
  {Sachdev}}]{Bernier2004}%
  \BibitemOpen
  \bibfield  {author} {\bibinfo {author} {\bibfnamefont {J.-S.}\ \bibnamefont
  {Bernier}}, \bibinfo {author} {\bibfnamefont {C.-H.}\ \bibnamefont {Chung}},
  \bibinfo {author} {\bibfnamefont {Y.~B.}\ \bibnamefont {Kim}},\ and\ \bibinfo
  {author} {\bibfnamefont {S.}~\bibnamefont {Sachdev}},\ }\href
  {https://doi.org/10.1103/PhysRevB.69.214427} {\bibfield  {journal} {\bibinfo
  {journal} {Phys. Rev. B}\ }\textbf {\bibinfo {volume} {69}},\ \bibinfo
  {pages} {214427} (\bibinfo {year} {2004})}\BibitemShut {NoStop}%
\bibitem [{\citenamefont {Chan}\ \emph {et~al.}(2011)\citenamefont {Chan},
  \citenamefont {Han},\ and\ \citenamefont {Duan}}]{Chan2011}%
  \BibitemOpen
  \bibfield  {author} {\bibinfo {author} {\bibfnamefont {Y.-H.}\ \bibnamefont
  {Chan}}, \bibinfo {author} {\bibfnamefont {Y.-J.}\ \bibnamefont {Han}},\ and\
  \bibinfo {author} {\bibfnamefont {L.-M.}\ \bibnamefont {Duan}},\ }\href
  {https://doi.org/10.1103/PhysRevB.84.224407} {\bibfield  {journal} {\bibinfo
  {journal} {Phys. Rev. B}\ }\textbf {\bibinfo {volume} {84}},\ \bibinfo
  {pages} {224407} (\bibinfo {year} {2011})}\BibitemShut {NoStop}%
\bibitem [{\citenamefont {Bishop}\ \emph {et~al.}(2012)\citenamefont {Bishop},
  \citenamefont {Li}, \citenamefont {Farnell}, \citenamefont {Richter},\ and\
  \citenamefont {Campbell}}]{Bishop2012}%
  \BibitemOpen
  \bibfield  {author} {\bibinfo {author} {\bibfnamefont {R.~F.}\ \bibnamefont
  {Bishop}}, \bibinfo {author} {\bibfnamefont {P.~H.~Y.}\ \bibnamefont {Li}},
  \bibinfo {author} {\bibfnamefont {D.~J.~J.}\ \bibnamefont {Farnell}},
  \bibinfo {author} {\bibfnamefont {J.}~\bibnamefont {Richter}},\ and\ \bibinfo
  {author} {\bibfnamefont {C.~E.}\ \bibnamefont {Campbell}},\ }\href
  {https://doi.org/10.1103/PhysRevB.85.205122} {\bibfield  {journal} {\bibinfo
  {journal} {Phys. Rev. B}\ }\textbf {\bibinfo {volume} {85}},\ \bibinfo
  {pages} {205122} (\bibinfo {year} {2012})}\BibitemShut {NoStop}%
\bibitem [{\citenamefont {Zhu}\ \emph {et~al.}(2010)\citenamefont {Zhu},
  \citenamefont {Yu}, \citenamefont {Wang}, \citenamefont {Zhao}, \citenamefont
  {Jones}, \citenamefont {Dai}, \citenamefont {Abrahams}, \citenamefont
  {Morosan}, \citenamefont {Fang},\ and\ \citenamefont {Si}}]{Si10}%
  \BibitemOpen
  \bibfield  {author} {\bibinfo {author} {\bibfnamefont {J.-X.}\ \bibnamefont
  {Zhu}}, \bibinfo {author} {\bibfnamefont {R.}~\bibnamefont {Yu}}, \bibinfo
  {author} {\bibfnamefont {H.}~\bibnamefont {Wang}}, \bibinfo {author}
  {\bibfnamefont {L.~L.}\ \bibnamefont {Zhao}}, \bibinfo {author}
  {\bibfnamefont {M.~D.}\ \bibnamefont {Jones}}, \bibinfo {author}
  {\bibfnamefont {J.}~\bibnamefont {Dai}}, \bibinfo {author} {\bibfnamefont
  {E.}~\bibnamefont {Abrahams}}, \bibinfo {author} {\bibfnamefont
  {E.}~\bibnamefont {Morosan}}, \bibinfo {author} {\bibfnamefont
  {M.}~\bibnamefont {Fang}},\ and\ \bibinfo {author} {\bibfnamefont
  {Q.}~\bibnamefont {Si}},\ }\href
  {https://doi.org/10.1103/PhysRevLett.104.216405} {\bibfield  {journal}
  {\bibinfo  {journal} {Phys. Rev. Lett.}\ }\textbf {\bibinfo {volume} {104}},\
  \bibinfo {pages} {216405} (\bibinfo {year} {2010})}\BibitemShut {NoStop}%
\bibitem [{\citenamefont {Freelon}\ \emph {et~al.}(2019)\citenamefont
  {Freelon}, \citenamefont {Yamani}, \citenamefont {Swainson}, \citenamefont
  {Flacau}, \citenamefont {Karki}, \citenamefont {Liu}, \citenamefont {Craco},
  \citenamefont {Laad}, \citenamefont {Wang}, \citenamefont {Chen},
  \citenamefont {Birgeneau},\ and\ \citenamefont {Fang}}]{Freelon19}%
  \BibitemOpen
  \bibfield  {author} {\bibinfo {author} {\bibfnamefont {B.}~\bibnamefont
  {Freelon}}, \bibinfo {author} {\bibfnamefont {Z.}~\bibnamefont {Yamani}},
  \bibinfo {author} {\bibfnamefont {I.}~\bibnamefont {Swainson}}, \bibinfo
  {author} {\bibfnamefont {R.}~\bibnamefont {Flacau}}, \bibinfo {author}
  {\bibfnamefont {B.}~\bibnamefont {Karki}}, \bibinfo {author} {\bibfnamefont
  {Y.~H.}\ \bibnamefont {Liu}}, \bibinfo {author} {\bibfnamefont
  {L.}~\bibnamefont {Craco}}, \bibinfo {author} {\bibfnamefont {M.~S.}\
  \bibnamefont {Laad}}, \bibinfo {author} {\bibfnamefont {M.}~\bibnamefont
  {Wang}}, \bibinfo {author} {\bibfnamefont {J.}~\bibnamefont {Chen}}, \bibinfo
  {author} {\bibfnamefont {R.~J.}\ \bibnamefont {Birgeneau}},\ and\ \bibinfo
  {author} {\bibfnamefont {M.}~\bibnamefont {Fang}},\ }\href
  {https://doi.org/10.1103/PhysRevB.99.024109} {\bibfield  {journal} {\bibinfo
  {journal} {Phys. Rev. B}\ }\textbf {\bibinfo {volume} {99}},\ \bibinfo
  {pages} {024109} (\bibinfo {year} {2019})}\BibitemShut {NoStop}%
\bibitem [{\citenamefont {Liu}\ \emph {et~al.}(2011)\citenamefont {Liu},
  \citenamefont {Zhang}, \citenamefont {Cheng}, \citenamefont {Luo},
  \citenamefont {Ying}, \citenamefont {Yan}, \citenamefont {Zhang},
  \citenamefont {Wang}, \citenamefont {Xiang}, \citenamefont {Ye},\ and\
  \citenamefont {Chen}}]{Chen1}%
  \BibitemOpen
  \bibfield  {author} {\bibinfo {author} {\bibfnamefont {R.~H.}\ \bibnamefont
  {Liu}}, \bibinfo {author} {\bibfnamefont {J.~S.}\ \bibnamefont {Zhang}},
  \bibinfo {author} {\bibfnamefont {P.}~\bibnamefont {Cheng}}, \bibinfo
  {author} {\bibfnamefont {X.~G.}\ \bibnamefont {Luo}}, \bibinfo {author}
  {\bibfnamefont {J.~J.}\ \bibnamefont {Ying}}, \bibinfo {author}
  {\bibfnamefont {Y.~J.}\ \bibnamefont {Yan}}, \bibinfo {author} {\bibfnamefont
  {M.}~\bibnamefont {Zhang}}, \bibinfo {author} {\bibfnamefont {A.~F.}\
  \bibnamefont {Wang}}, \bibinfo {author} {\bibfnamefont {Z.~J.}\ \bibnamefont
  {Xiang}}, \bibinfo {author} {\bibfnamefont {G.~J.}\ \bibnamefont {Ye}},\ and\
  \bibinfo {author} {\bibfnamefont {X.~H.}\ \bibnamefont {Chen}},\ }\href
  {https://doi.org/10.1103/PhysRevB.83.174450} {\bibfield  {journal} {\bibinfo
  {journal} {Phys. Rev. B}\ }\textbf {\bibinfo {volume} {83}},\ \bibinfo
  {pages} {174450} (\bibinfo {year} {2011})}\BibitemShut {NoStop}%
\bibitem [{\citenamefont {He}\ \emph {et~al.}(2011)\citenamefont {He},
  \citenamefont {Wang}, \citenamefont {Shi}, \citenamefont {Yang},
  \citenamefont {Li},\ and\ \citenamefont {Chen}}]{Chen2}%
  \BibitemOpen
  \bibfield  {author} {\bibinfo {author} {\bibfnamefont {J.~B.}\ \bibnamefont
  {He}}, \bibinfo {author} {\bibfnamefont {D.~M.}\ \bibnamefont {Wang}},
  \bibinfo {author} {\bibfnamefont {H.~L.}\ \bibnamefont {Shi}}, \bibinfo
  {author} {\bibfnamefont {H.~X.}\ \bibnamefont {Yang}}, \bibinfo {author}
  {\bibfnamefont {J.~Q.}\ \bibnamefont {Li}},\ and\ \bibinfo {author}
  {\bibfnamefont {G.~F.}\ \bibnamefont {Chen}},\ }\href
  {https://doi.org/10.1103/PhysRevB.84.205212} {\bibfield  {journal} {\bibinfo
  {journal} {Phys. Rev. B}\ }\textbf {\bibinfo {volume} {84}},\ \bibinfo
  {pages} {205212} (\bibinfo {year} {2011})}\BibitemShut {NoStop}%
\bibitem [{\citenamefont {Takeiri}\ \emph {et~al.}(2016)\citenamefont
  {Takeiri}, \citenamefont {Matsumoto}, \citenamefont {Yamamoto}, \citenamefont
  {Hayashi}, \citenamefont {Li}, \citenamefont {Tohyama}, \citenamefont
  {Tassel}, \citenamefont {Ritter}, \citenamefont {Narumi}, \citenamefont
  {Hagiwara},\ and\ \citenamefont {Kageyama}}]{Kageyama16}%
  \BibitemOpen
  \bibfield  {author} {\bibinfo {author} {\bibfnamefont {F.}~\bibnamefont
  {Takeiri}}, \bibinfo {author} {\bibfnamefont {Y.}~\bibnamefont {Matsumoto}},
  \bibinfo {author} {\bibfnamefont {T.}~\bibnamefont {Yamamoto}}, \bibinfo
  {author} {\bibfnamefont {N.}~\bibnamefont {Hayashi}}, \bibinfo {author}
  {\bibfnamefont {Z.}~\bibnamefont {Li}}, \bibinfo {author} {\bibfnamefont
  {T.}~\bibnamefont {Tohyama}}, \bibinfo {author} {\bibfnamefont
  {C.}~\bibnamefont {Tassel}}, \bibinfo {author} {\bibfnamefont
  {C.}~\bibnamefont {Ritter}}, \bibinfo {author} {\bibfnamefont
  {Y.}~\bibnamefont {Narumi}}, \bibinfo {author} {\bibfnamefont
  {M.}~\bibnamefont {Hagiwara}},\ and\ \bibinfo {author} {\bibfnamefont
  {H.}~\bibnamefont {Kageyama}},\ }\href
  {https://doi.org/10.1103/PhysRevB.94.184426} {\bibfield  {journal} {\bibinfo
  {journal} {Phys. Rev. B}\ }\textbf {\bibinfo {volume} {94}},\ \bibinfo
  {pages} {184426} (\bibinfo {year} {2016})}\BibitemShut {NoStop}%
\bibitem [{\citenamefont {Zhao}\ \emph {et~al.}(2013)\citenamefont {Zhao},
  \citenamefont {Wu}, \citenamefont {Wang}, \citenamefont {Hodges},
  \citenamefont {Broholm},\ and\ \citenamefont {Morosan}}]{Morosan13}%
  \BibitemOpen
  \bibfield  {author} {\bibinfo {author} {\bibfnamefont {L.~L.}\ \bibnamefont
  {Zhao}}, \bibinfo {author} {\bibfnamefont {S.}~\bibnamefont {Wu}}, \bibinfo
  {author} {\bibfnamefont {J.~K.}\ \bibnamefont {Wang}}, \bibinfo {author}
  {\bibfnamefont {J.~P.}\ \bibnamefont {Hodges}}, \bibinfo {author}
  {\bibfnamefont {C.}~\bibnamefont {Broholm}},\ and\ \bibinfo {author}
  {\bibfnamefont {E.}~\bibnamefont {Morosan}},\ }\href
  {https://doi.org/10.1103/PhysRevB.87.020406} {\bibfield  {journal} {\bibinfo
  {journal} {Phys. Rev. B}\ }\textbf {\bibinfo {volume} {87}},\ \bibinfo
  {pages} {020406} (\bibinfo {year} {2013})}\BibitemShut {NoStop}%
\bibitem [{\citenamefont {McCabe}\ \emph {et~al.}(2014)\citenamefont {McCabe},
  \citenamefont {Wills}, \citenamefont {Chapon}, \citenamefont {Manuel},\ and\
  \citenamefont {Evans}}]{McCabe14}%
  \BibitemOpen
  \bibfield  {author} {\bibinfo {author} {\bibfnamefont {E.~E.}\ \bibnamefont
  {McCabe}}, \bibinfo {author} {\bibfnamefont {A.~S.}\ \bibnamefont {Wills}},
  \bibinfo {author} {\bibfnamefont {L.}~\bibnamefont {Chapon}}, \bibinfo
  {author} {\bibfnamefont {P.}~\bibnamefont {Manuel}},\ and\ \bibinfo {author}
  {\bibfnamefont {J.~S.~O.}\ \bibnamefont {Evans}},\ }\href
  {https://doi.org/10.1103/PhysRevB.90.165111} {\bibfield  {journal} {\bibinfo
  {journal} {Phys. Rev. B}\ }\textbf {\bibinfo {volume} {90}},\ \bibinfo
  {pages} {165111} (\bibinfo {year} {2014})}\BibitemShut {NoStop}%
\bibitem [{\citenamefont {Oogarah}\ \emph {et~al.}(2017)\citenamefont
  {Oogarah}, \citenamefont {Stockdale}, \citenamefont {Stock}, \citenamefont
  {Evans}, \citenamefont {Wills}, \citenamefont {Taylor},\ and\ \citenamefont
  {McCabe}}]{McCabe17}%
  \BibitemOpen
  \bibfield  {author} {\bibinfo {author} {\bibfnamefont {R.~K.}\ \bibnamefont
  {Oogarah}}, \bibinfo {author} {\bibfnamefont {C.~P.~J.}\ \bibnamefont
  {Stockdale}}, \bibinfo {author} {\bibfnamefont {C.}~\bibnamefont {Stock}},
  \bibinfo {author} {\bibfnamefont {J.~S.~O.}\ \bibnamefont {Evans}}, \bibinfo
  {author} {\bibfnamefont {A.~S.}\ \bibnamefont {Wills}}, \bibinfo {author}
  {\bibfnamefont {J.~W.}\ \bibnamefont {Taylor}},\ and\ \bibinfo {author}
  {\bibfnamefont {E.~E.}\ \bibnamefont {McCabe}},\ }\href
  {https://doi.org/10.1103/PhysRevB.95.174441} {\bibfield  {journal} {\bibinfo
  {journal} {Phys. Rev. B}\ }\textbf {\bibinfo {volume} {95}},\ \bibinfo
  {pages} {174441} (\bibinfo {year} {2017})}\BibitemShut {NoStop}%
\bibitem [{\citenamefont {Elhajal}\ \emph {et~al.}(2005)\citenamefont
  {Elhajal}, \citenamefont {Canals}, \citenamefont {Sunyer},\ and\
  \citenamefont {Lacroix}}]{Elhajal05}%
  \BibitemOpen
  \bibfield  {author} {\bibinfo {author} {\bibfnamefont {M.}~\bibnamefont
  {Elhajal}}, \bibinfo {author} {\bibfnamefont {B.}~\bibnamefont {Canals}},
  \bibinfo {author} {\bibfnamefont {R.}~\bibnamefont {Sunyer}},\ and\ \bibinfo
  {author} {\bibfnamefont {C.}~\bibnamefont {Lacroix}},\ }\href
  {https://journals.aps.org/prb/abstract/10.1103/PhysRevB.71.094420} {\bibfield
   {journal} {\bibinfo  {journal} {Phys. Rev. B}\ }\textbf {\bibinfo {volume}
  {71}},\ \bibinfo {pages} {094420} (\bibinfo {year} {2005})}\BibitemShut
  {NoStop}%
\bibitem [{\citenamefont {Bogdanov}\ \emph {et~al.}(2013)\citenamefont
  {Bogdanov}, \citenamefont {Maurice}, \citenamefont {Rousochatzakis},
  \citenamefont {van~den Brink},\ and\ \citenamefont {Hozoi}}]{Bogdanov13}%
  \BibitemOpen
  \bibfield  {author} {\bibinfo {author} {\bibfnamefont {N.~A.}\ \bibnamefont
  {Bogdanov}}, \bibinfo {author} {\bibfnamefont {R.}~\bibnamefont {Maurice}},
  \bibinfo {author} {\bibfnamefont {I.}~\bibnamefont {Rousochatzakis}},
  \bibinfo {author} {\bibfnamefont {J.}~\bibnamefont {van~den Brink}},\ and\
  \bibinfo {author} {\bibfnamefont {L.}~\bibnamefont {Hozoi}},\ }\href
  {https://doi.org/10.1103/PhysRevLett.110.127206} {\bibfield  {journal}
  {\bibinfo  {journal} {Phys. Rev. Lett.}\ }\textbf {\bibinfo {volume} {110}},\
  \bibinfo {pages} {127206} (\bibinfo {year} {2013})}\BibitemShut {NoStop}%
\bibitem [{\citenamefont {Sohn}\ \emph {et~al.}(2017)\citenamefont {Sohn},
  \citenamefont {Kim}, \citenamefont {Sandilands}, \citenamefont {Hien},
  \citenamefont {Kim}, \citenamefont {Park}, \citenamefont {Kim}, \citenamefont
  {Moon}, \citenamefont {Yamaura}, \citenamefont {Hiroi},\ and\ \citenamefont
  {Noh}}]{Sohn17}%
  \BibitemOpen
  \bibfield  {author} {\bibinfo {author} {\bibfnamefont {C.~H.}\ \bibnamefont
  {Sohn}}, \bibinfo {author} {\bibfnamefont {C.~H.}\ \bibnamefont {Kim}},
  \bibinfo {author} {\bibfnamefont {L.~J.}\ \bibnamefont {Sandilands}},
  \bibinfo {author} {\bibfnamefont {N.~T.~M.}\ \bibnamefont {Hien}}, \bibinfo
  {author} {\bibfnamefont {S.~Y.}\ \bibnamefont {Kim}}, \bibinfo {author}
  {\bibfnamefont {H.~J.}\ \bibnamefont {Park}}, \bibinfo {author}
  {\bibfnamefont {K.~W.}\ \bibnamefont {Kim}}, \bibinfo {author} {\bibfnamefont
  {S.~J.}\ \bibnamefont {Moon}}, \bibinfo {author} {\bibfnamefont
  {J.}~\bibnamefont {Yamaura}}, \bibinfo {author} {\bibfnamefont
  {Z.}~\bibnamefont {Hiroi}},\ and\ \bibinfo {author} {\bibfnamefont {T.~W.}\
  \bibnamefont {Noh}},\ }\href {https://doi.org/10.1103/PhysRevLett.118.117201}
  {\bibfield  {journal} {\bibinfo  {journal} {Phys. Rev. Lett.}\ }\textbf
  {\bibinfo {volume} {118}},\ \bibinfo {pages} {117201} (\bibinfo {year}
  {2017})}\BibitemShut {NoStop}%
\bibitem [{\citenamefont {Zhang}\ \emph {et~al.}(2021)\citenamefont {Zhang},
  \citenamefont {Wang},\ and\ \citenamefont {Wu}}]{Zhang21}%
  \BibitemOpen
  \bibfield  {author} {\bibinfo {author} {\bibfnamefont {B.~H.}\ \bibnamefont
  {Zhang}}, \bibinfo {author} {\bibfnamefont {Z.}~\bibnamefont {Wang}},\ and\
  \bibinfo {author} {\bibfnamefont {R.~Q.}\ \bibnamefont {Wu}},\ }\href
  {https://doi.org/10.1103/PhysRevB.104.024411} {\bibfield  {journal} {\bibinfo
   {journal} {Phys. Rev. B}\ }\textbf {\bibinfo {volume} {104}},\ \bibinfo
  {pages} {024411} (\bibinfo {year} {2021})}\BibitemShut {NoStop}%
\bibitem [{\citenamefont {Amirabbasi}\ \emph {et~al.}(2019)\citenamefont
  {Amirabbasi}, \citenamefont {Rezaei}, \citenamefont {Alaei}, \citenamefont
  {Shahbazi},\ and\ \citenamefont {Akbarzadeh}}]{Amirabbasi19}%
  \BibitemOpen
  \bibfield  {author} {\bibinfo {author} {\bibfnamefont {M.}~\bibnamefont
  {Amirabbasi}}, \bibinfo {author} {\bibfnamefont {N.}~\bibnamefont {Rezaei}},
  \bibinfo {author} {\bibfnamefont {M.}~\bibnamefont {Alaei}}, \bibinfo
  {author} {\bibfnamefont {F.}~\bibnamefont {Shahbazi}},\ and\ \bibinfo
  {author} {\bibfnamefont {H.}~\bibnamefont {Akbarzadeh}},\ }\href
  {https://doi.org/10.1103/PhysRevB.100.054441} {\bibfield  {journal} {\bibinfo
   {journal} {Phys. Rev. B}\ }\textbf {\bibinfo {volume} {100}},\ \bibinfo
  {pages} {054441} (\bibinfo {year} {2019})}\BibitemShut {NoStop}%
\bibitem [{\citenamefont {Donnerer}\ \emph {et~al.}(2016)\citenamefont
  {Donnerer}, \citenamefont {Rahn}, \citenamefont {Sala}, \citenamefont {Vale},
  \citenamefont {Pincini}, \citenamefont {Strempfer}, \citenamefont {Krisch},
  \citenamefont {Prabhakaran}, \citenamefont {Boothroyd},\ and\ \citenamefont
  {McMorrow}}]{McMorrow16}%
  \BibitemOpen
  \bibfield  {author} {\bibinfo {author} {\bibfnamefont {C.}~\bibnamefont
  {Donnerer}}, \bibinfo {author} {\bibfnamefont {M.~C.}\ \bibnamefont {Rahn}},
  \bibinfo {author} {\bibfnamefont {M.~M.}\ \bibnamefont {Sala}}, \bibinfo
  {author} {\bibfnamefont {J.~G.}\ \bibnamefont {Vale}}, \bibinfo {author}
  {\bibfnamefont {D.}~\bibnamefont {Pincini}}, \bibinfo {author} {\bibfnamefont
  {J.}~\bibnamefont {Strempfer}}, \bibinfo {author} {\bibfnamefont
  {M.}~\bibnamefont {Krisch}}, \bibinfo {author} {\bibfnamefont
  {D.}~\bibnamefont {Prabhakaran}}, \bibinfo {author} {\bibfnamefont {A.~T.}\
  \bibnamefont {Boothroyd}},\ and\ \bibinfo {author} {\bibfnamefont {D.~F.}\
  \bibnamefont {McMorrow}},\ }\href
  {https://doi.org/10.1103/PhysRevLett.117.037201} {\bibfield  {journal}
  {\bibinfo  {journal} {Phys. Rev. Lett.}\ }\textbf {\bibinfo {volume} {117}},\
  \bibinfo {pages} {037201} (\bibinfo {year} {2016})}\BibitemShut {NoStop}%
\bibitem [{\citenamefont {Jaeseok~Son}\ and\ \citenamefont
  {Noh}(2019)}]{Son19}%
  \BibitemOpen
  \bibfield  {author} {\bibinfo {author} {\bibfnamefont {C.~H. K. H. C. S. Y.
  K. L. J. S. C. S. J.-G. P. S. J.~M.}\ \bibnamefont {Jaeseok~Son},
  \bibfnamefont {Byung Cheol~Park}}\ and\ \bibinfo {author} {\bibfnamefont
  {T.~W.}\ \bibnamefont {Noh}},\ }\href
  {https://doi.org/https://doi.org/10.1038/s41535-019-0157-0} {\bibfield
  {journal} {\bibinfo  {journal} {npj Quantum Mater.}\ }\textbf {\bibinfo
  {volume} {4}} (\bibinfo {year} {2019})}\BibitemShut {NoStop}%
\bibitem [{\citenamefont {Hwang}\ \emph {et~al.}(2020)\citenamefont {Hwang},
  \citenamefont {Trivedi},\ and\ \citenamefont {Randeria}}]{Hwang20}%
  \BibitemOpen
  \bibfield  {author} {\bibinfo {author} {\bibfnamefont {K.}~\bibnamefont
  {Hwang}}, \bibinfo {author} {\bibfnamefont {N.}~\bibnamefont {Trivedi}},\
  and\ \bibinfo {author} {\bibfnamefont {M.}~\bibnamefont {Randeria}},\ }\href
  {https://doi.org/10.1103/PhysRevLett.125.047203} {\bibfield  {journal}
  {\bibinfo  {journal} {Phys. Rev. Lett.}\ }\textbf {\bibinfo {volume} {125}},\
  \bibinfo {pages} {047203} (\bibinfo {year} {2020})}\BibitemShut {NoStop}%
\bibitem [{\citenamefont {Abrikosov}(1957)}]{Abrikosov1957}%
  \BibitemOpen
  \bibfield  {author} {\bibinfo {author} {\bibfnamefont {A.~A.}\ \bibnamefont
  {Abrikosov}},\ }\href@noop {} {\bibfield  {journal} {\bibinfo  {journal}
  {Sov. Phys. JETP}\ }\textbf {\bibinfo {volume} {5}},\ \bibinfo {pages} {1174}
  (\bibinfo {year} {1957})}\BibitemShut {NoStop}%
\bibitem [{\citenamefont {Wright}\ and\ \citenamefont
  {Mermin}(1989)}]{Wright89}%
  \BibitemOpen
  \bibfield  {author} {\bibinfo {author} {\bibfnamefont {D.~C.}\ \bibnamefont
  {Wright}}\ and\ \bibinfo {author} {\bibfnamefont {N.~D.}\ \bibnamefont
  {Mermin}},\ }\href {https://doi.org/10.1103/RevModPhys.61.385} {\bibfield
  {journal} {\bibinfo  {journal} {Rev. Mod. Phys.}\ }\textbf {\bibinfo {volume}
  {61}},\ \bibinfo {pages} {385} (\bibinfo {year} {1989})}\BibitemShut
  {NoStop}%
\bibitem [{\citenamefont {Bogdanov}(1995)}]{Bogdanov1995}%
  \BibitemOpen
  \bibfield  {author} {\bibinfo {author} {\bibfnamefont {A.}~\bibnamefont
  {Bogdanov}},\ }\href
  {http://www.jetpletters.ac.ru/ps/1214/article_18359.shtml} {\bibfield
  {journal} {\bibinfo  {journal} {JETP Lett.}\ }\textbf {\bibinfo {volume}
  {62}},\ \bibinfo {pages} {247} (\bibinfo {year} {1995})}\BibitemShut
  {NoStop}%
\bibitem [{\citenamefont {Matthew F.~Lapa}(2012)}]{LapaHenley2012}%
  \BibitemOpen
  \bibfield  {author} {\bibinfo {author} {\bibfnamefont {C.~L.~H.}\
  \bibnamefont {Matthew F.~Lapa}},\ }\href@noop {} {\  (\bibinfo {year}
  {2012})},\ \Eprint {https://arxiv.org/abs/1210.6810} {arXiv:1210.6810}
  \BibitemShut {NoStop}%
\bibitem [{\citenamefont {Sophia R.~Sklan}(2012)}]{SklanHenley2012}%
  \BibitemOpen
  \bibfield  {author} {\bibinfo {author} {\bibfnamefont {C.~L.~H.}\
  \bibnamefont {Sophia R.~Sklan}},\ }\href@noop {} {\  (\bibinfo {year}
  {2012})},\ \Eprint {https://arxiv.org/abs/1209.1381} {arXiv:1209.1381}
  \BibitemShut {NoStop}%
\bibitem [{\citenamefont {Henley}(1984)}]{Henley1984}%
  \BibitemOpen
  \bibfield  {author} {\bibinfo {author} {\bibfnamefont {C.~L.}\ \bibnamefont
  {Henley}},\ }\href@noop {} {\bibfield  {journal} {\bibinfo  {journal} {Annals
  of Phys.}\ }\textbf {\bibinfo {volume} {156}},\ \bibinfo {pages} {368}
  (\bibinfo {year} {1984})}\BibitemShut {NoStop}%
\bibitem [{\citenamefont {Mostovoy}(2006)}]{Mostovoy}%
  \BibitemOpen
  \bibfield  {author} {\bibinfo {author} {\bibfnamefont {M.}~\bibnamefont
  {Mostovoy}},\ }\href@noop {} {\bibfield  {journal} {\bibinfo  {journal}
  {Phys. Rev. Lett.}\ }\textbf {\bibinfo {volume} {96}},\ \bibinfo {pages}
  {067601} (\bibinfo {year} {2006})}\BibitemShut {NoStop}%
\bibitem [{\citenamefont {Tarkhany}\ \emph {et~al.}(2021)\citenamefont
  {Tarkhany}, \citenamefont {Discacciati},\ and\ \citenamefont
  {Betouras}}]{Betouras}%
  \BibitemOpen
  \bibfield  {author} {\bibinfo {author} {\bibfnamefont {A.~R.}\ \bibnamefont
  {Tarkhany}}, \bibinfo {author} {\bibfnamefont {M.}~\bibnamefont
  {Discacciati}},\ and\ \bibinfo {author} {\bibfnamefont {J.~J.}\ \bibnamefont
  {Betouras}},\ }\href@noop {} {\bibfield  {journal} {\bibinfo  {journal}
  {Phys. Rev. B}\ }\textbf {\bibinfo {volume} {103}},\ \bibinfo {pages}
  {205409} (\bibinfo {year} {2021})}\BibitemShut {NoStop}%
\bibitem [{\citenamefont {Luttinger}\ and\ \citenamefont
  {Tisza}(1946)}]{Luttinger}%
  \BibitemOpen
  \bibfield  {author} {\bibinfo {author} {\bibfnamefont {J.~M.}\ \bibnamefont
  {Luttinger}}\ and\ \bibinfo {author} {\bibfnamefont {L.}~\bibnamefont
  {Tisza}},\ }\href {https://doi.org/10.1103/PhysRev.70.954} {\bibfield
  {journal} {\bibinfo  {journal} {Phys. Rev.}\ }\textbf {\bibinfo {volume}
  {70}},\ \bibinfo {pages} {954} (\bibinfo {year} {1946})}\BibitemShut
  {NoStop}%
\bibitem [{\citenamefont {Bertaut}(1961)}]{Bertraut}%
  \BibitemOpen
  \bibfield  {author} {\bibinfo {author} {\bibfnamefont {E.}~\bibnamefont
  {Bertaut}},\ }\href
  {https://doi.org/https://doi.org/10.1016/0022-3697(61)90105-6} {\bibfield
  {journal} {\bibinfo  {journal} {Journal of Physics and Chemistry of Solids}\
  }\textbf {\bibinfo {volume} {21}},\ \bibinfo {pages} {256} (\bibinfo {year}
  {1961})}\BibitemShut {NoStop}%
\bibitem [{\citenamefont {Litvin}(1974)}]{Litvin}%
  \BibitemOpen
  \bibfield  {author} {\bibinfo {author} {\bibfnamefont {D.}~\bibnamefont
  {Litvin}},\ }\href
  {https://doi.org/https://doi.org/10.1016/0031-8914(74)90257-2} {\bibfield
  {journal} {\bibinfo  {journal} {Physica}\ }\textbf {\bibinfo {volume} {77}},\
  \bibinfo {pages} {205} (\bibinfo {year} {1974})}\BibitemShut {NoStop}%
\bibitem [{\citenamefont {Kaplan}\ and\ \citenamefont {Menyuk}(2007)}]{Kaplan}%
  \BibitemOpen
  \bibfield  {author} {\bibinfo {author} {\bibfnamefont {T.~A.}\ \bibnamefont
  {Kaplan}}\ and\ \bibinfo {author} {\bibfnamefont {N.}~\bibnamefont
  {Menyuk}},\ }\href
  {https://www.tandfonline.com/doi/abs/10.1080/14786430601080229} {\bibfield
  {journal} {\bibinfo  {journal} {Philos. Mag.}\ }\textbf {\bibinfo {volume}
  {87}},\ \bibinfo {pages} {3711} (\bibinfo {year} {2007})}\BibitemShut
  {NoStop}%
\bibitem [{\citenamefont {Li}\ \emph {et~al.}(2019)\citenamefont {Li},
  \citenamefont {Perkins},\ and\ \citenamefont {Rousochatzakis}}]{Z2vortices2}%
  \BibitemOpen
  \bibfield  {author} {\bibinfo {author} {\bibfnamefont {M.}~\bibnamefont
  {Li}}, \bibinfo {author} {\bibfnamefont {N.~B.}\ \bibnamefont {Perkins}},\
  and\ \bibinfo {author} {\bibfnamefont {I.}~\bibnamefont {Rousochatzakis}},\
  }\href {https://doi.org/10.1103/PhysRevResearch.1.013002} {\bibfield
  {journal} {\bibinfo  {journal} {Phys. Rev. Res.}\ }\textbf {\bibinfo {volume}
  {1}},\ \bibinfo {pages} {013002} (\bibinfo {year} {2019})}\BibitemShut
  {NoStop}%
\end{thebibliography}

%

\end{document}